\documentclass[a4paper,11pt]{article}
\pdfoutput=1 

\usepackage{jheppub} 
                     
\usepackage[utf8]{inputenc}
\usepackage{graphicx} 
\usepackage{grffile}
\usepackage{dcolumn}  
\usepackage{makecell}
\usepackage{colordvi}
\usepackage{color}
\usepackage{amssymb}
\usepackage{amsmath}
\usepackage{bm}
\usepackage{slashed}
\usepackage{enumerate}
\usepackage{url}
\usepackage{morefloats}
\usepackage{lineno}
\usepackage{subfig}
\usepackage{graphicx}
\usepackage[percent]{overpic}

\usepackage{natbib}

\usepackage[colorlinks=true,
urlcolor=blue,
linkcolor=blue,
citecolor=blue,
linktocpage=true,
pdfproducer=medialab,
pdfa=true,
anchorcolor=blue]{hyperref}

 \usepackage{footnotebackref}

\usepackage{tabularx,array}

\newcolumntype{C}{>{\centering\arraybackslash}X}
\usepackage{multirow}

\usepackage{lineno}

\usepackage{float}
\usepackage{xcolor}

\usepackage[title]{appendix}

\renewcommand{\thesection}{\arabic{section}}

\usepackage{changepage}
\usepackage{booktabs}

\input belle2sym.tex

\preprint{}
\makeatletter
\gdef\@fpheader{}
\makeatother

\begin{document}

\raggedbottom


\title{Analysis note: measurement of thrust and track energy-energy correlator in \boldmath $e^{+}e^{-}$ collisions at \boldmath $91.2$~GeV with DELPHI open data}

\author[a]{Jingyu Zhang,}
\author[b]{Tzu-An Sheng,}
\author[b]{Yu-Chen Chen,}
\author[b]{Hannah Bossi,}
\author[c]{Anthony Badea,}
\author[d]{Austin Baty,}
\author[b]{Chris McGinn,}
\author[b]{Yen-Jie Lee,}
\author[a]{Yi Chen}

\affiliation[a]{Vanderbilt University, Nashville, Tennessee, USA}
\affiliation[b]{Massachusetts Institute of Technology, Cambridge, Massachusetts, USA}
\affiliation[c]{Enrico Fermi Institute, University of Chicago, Chicago IL}
\affiliation[d]{University of Illinois Chicago, Chicago, Illinois, USA}%

\emailAdd{jingyu.zhang@cern.ch}
\emailAdd{tzu-an.sheng@cern.ch}
\emailAdd{janice\_c@mit.edu}
\emailAdd{hannah.bossi@cern.ch}
\emailAdd{anthony.badea@cern.ch}
\emailAdd{abaty@uic.edu}
\emailAdd{cfmcginn@mit.edu}
\emailAdd{yenjie@mit.edu}
\emailAdd{luna.chen@vanderbilt.edu}

\date{\today}

\abstract{
Recent theoretical developments, as well as experimental measurements at hadron collisions, have renewed interest in studying event shape variables in $e^{+}e^{-}$ collisions. We present a measurement of thrust and track-based energy-energy correlator in $e^+e^-$ collisions at center-of-mass energy of $91.2~\mathrm{GeV}$, using newly released open data from the DELPHI experiment. The event shapes, measured with unprecedented resolution and precision, are compared to various Monte Carlo and analytic predictions. Leveraging DELPHI’s unique detector geometry and reconstruction capabilities, the track energy-energy correlator measurement provides data with the highest angular resolution, offering critical inputs for precision tests of QCD in both collinear and back-to-back limits. This note presents the first physics analysis using DELPHI open data and establishes benchmarks necessary for future studies exploiting this legacy dataset.
}


\maketitle
\flushbottom

\section{Introduction}
\label{sec:Introduction}
The $e^{+}e^{-}$ collision environment provides the cleanest setting for studying Quantum Chromodynamics (QCD), as the colliding objects are fundamental particles, without the complications inherent to hadron collisions such as beam remnants and parton distribution functions.
Re-analysis of archived data from the Apparatus for LEp PHysics (ALEPH) experiment at the Large Electron Positron (LEP) collider has highlighted the potential of applying modern experimental techniques to legacy datasets~\cite{Chen:2021uws, Bossi:2025xsi, Badea:2025wzd, Badea:2019vey, Chen:2023njr}, enabling new precision tests of QCD at $e^+e^-$ collisions at center-of-mass energy $\sqrt{s}=91.2~\mathrm{GeV}$.
These measurements are specifically designed to test the latest advances in analytic QCD calculations with high-precision data—providing a fundamental probe of perturbative and non-perturbative dynamics—while also serving as a clean reference for models of jet substructure and the tuning of Monte Carlo (MC) generators used at hadron colliders.
Moreover, new studies of two particle correlations with LEP-1~\cite{Badea:2019vey} and LEP-2~\cite{Chen:2023njr} data reveal a potential excess in the highest multiplicity interval of LEP 2 data not seen in an archived \textsc{pythia} 6 MC sample from ALEPH, providing new insights into the presence of signatures typically attributed to hydrodynamical flow in heavy-ion collisions. 
In August 2024, the DELPHI (DEtector with Lepton, Photon and Hadron Identification) collaboration~\cite{DELPHI:1990cdc} released its entire dataset and software for open access, creating a unique opportunity to extend the current LEP re-analysis landscape.
Leveraging DELPHI's distinctive detector geometry and reconstruction capabilities, re-measurements of hadronic event shapes offer complementary results to recent ALEPH re-analysis efforts while providing important input to precision QCD studies with modern analysis techniques.

Hadronic event-shape observables are central to the understanding of the energy flow and radiative patterns of high-energy collider data and the underlying QCD dynamics.
Two-point energy correlation function, or energy-energy correlator (EEC), both theoretically developed~\cite{PhysRevLett.41.1585,Basham:1978zq,Basham:1979gh,Basham:1977iq} and experimentally measured~\cite{OPAL:1990reb, OPAL:1993pnw,DELPHI:1990sof,DELPHI:2003yqh,DELPHI:2000uri,SLD:1994idb}, are powerful tools for studying energy flow in $e^{+}e^{-}$ collisions.
This class of observables is sensitive to universal QCD parameters such as the strong coupling constant $\alpha_s$, the Collins-Soper kernel~\cite{Collins:1981uw,Collins:1984kg}, and the $\rm \Omega_{1}$ parameter in the non-perturbative power corrections~\cite{Dokshitzer:1995qm, Korchemsky:1999kt}. 
In the original proposals, the EEC was measured using all particles in the event, resulting in limited angular resolution at very small or very large opening angles due to the calorimeter for neutral particles.
Thanks to recent theoretical development of the track function formalism~\cite{Chang:2013rca,Chang:2013iba}, charged-particle-based observables are now theoretically well-understood in QCD. 
This calls for a measurement of track EEC to significantly enhance the resolution and precision in the very small and very large opening angles in $e^{+}e^{-}$ collisions. 

Another widely studied event shape variable is thrust~\cite{PhysRevLett.39.1587, Georgi:1977sf, ALEPH:2003obs, DELPHI:2003yqh, DELPHI:2000uri, DELPHI:1999vbd}, which quantifies how compatible an event is with two back-to-back objects. 
The thrust distribution has been extensively used to extract the strong coupling constant $\alpha_s(m_Z)$ with high precision, to constrain non-perturbative shape functions, and to test and refine models of hadronization~\cite{Abbate:2010xh, Davison:2009wzs, Hartgring:2013jma}. 
These capabilities make thrust a fundamental observable in both theoretical and experimental analyses of $e^{+}e^{-}$ data.
New theoretical results have re-emphasized the importance of the LEP thrust measurements for multiple research directions. 
Efforts to determine $\alpha_s(m_Z)$ from $e^+e^-$ thrust data have yielded precise theoretical predictions that lie several standard deviations below the world average~\cite{Benitez:2024nav}. 
The analysis reports $\alpha_s(m_Z) = 0.1136 \pm 0.0012$, a value significantly lower and incompatible with the 2023 PDG world average of $\alpha_s(m_Z) = 0.1180 \pm 0.0009$. 
The experimental world average has also evolved, with the 2023 recalculation excluding $e^{+}e^{-}$ event-shape observables due to MC-estimated hadronization corrections. 
In parallel, recent studies have highlighted the importance of logarithmic moments of event-shape variables, such as thrust, in constraining both perturbative and non-perturbative QCD effects~\cite{Assi:2025ibi}. 
Based on these results, in addition to others, it is well motivated to re-examine the thrust from LEP data, with an emphasis on the regime corresponding to jet cores in the dijet peak, which was not measured with high precision by the LEP collaborations and is most sensitive to non-perturbative effects.

In this note, we present a measurement of thrust and track EEC, using $e^{+}e^{-}$ collision data at $\sqrt{s} = 91.2$~GeV at LEP-1; collected by the DELPHI collaboration~\cite{DELPHI:1990cdc}.
A key focus of this analysis is a rigorous evaluation of systematic uncertainties. This is performed to modern standards, incorporating detailed detector effects studies, advanced unfolding techniques, and comparisons to multiple Monte Carlo (MC) generators.
In Section~\ref{sec:def}, a detailed definition of the EEC and thrust will be presented. 
In Section~\ref{sec:delphi} and~\ref{sec:sample}, the experimental setup, including details of the data and MC samples, will be described.
The details of the EEC analysis, including the study of detector effects, unfolding, and systematic uncertainties, are described in Section \ref{sec:analysis}. 
The details of the thrust analysis are provided in a separate section~\ref{sec:thrust}. 
The fully-corrected results are presented in Section \ref{sec:results}. 
Finally, an outlook is provided for future work.  

\section{Definition of the observables}\label{sec:def}
\subsection{Energy-energy correlator}\label{sec:eec}
Different measurements of the energy correlator may have slightly different definitions. The definition of the EEC employed in this work has similar form as the results previously measured in the LEP experiments and is expressed as:
\begin{equation}\label{eq:ENC}
    \text{EEC}(\theta_{\rm L}) = \sum^{\rm n}_{\rm i \; > \; j} \int d\sigma \frac{E_{\rm i}E_{\rm j}}{E^{2}}\delta(\theta_{\rm L} - \theta_{\rm i j})
\end{equation}
Here, $E$ is the total center-of-mass energy (sqrt(s)), and $E_{\rm i}$ is the energy of particle i. The sum runs over all unique pairs of charged particles (i,j) in a final state, and $\theta_{\rm ij}$ is the opening angle between them.
The term $d\sigma$ represents the differential cross-section for producing a specific hadronic final state. The integral over $d\sigma$ therefore implies an average over all possible hadronic events, with each event weighted by its quantum mechanical probability of occurring.
 The definition of the EEC provided here differs from some other approaches in a few ways. Firstly, the EEC in hadron collisions is typically expressed as a function of the distance variable $\Delta R = \sqrt{\Delta y^{2} + \Delta\phi^{2}}$ instead of the opening angle $\theta_{\rm L}$. Secondly, all charged particles in a given event are used in this definition as opposed to the definitions used in jet substructure measurements that only utilize particles within reconstructed jets. As will be discussed in detail later, this allows not only the collinear limit of QCD to be studied but also the back-to-back region. Thirdly, in this definition, correlations between two particles are only counted once, i.e., ij correlations and ji correlations are not double-counted as in previous measurements such as that in Ref.~\cite{OPAL:1990reb}. Lastly, as mentioned before, only charged particles are considered in this analysis in order to fully exploit the excellent angular resolution for charged tracks in DELPHI (See Section \ref{sec:sample} for more details). The set of charged particles used in this analysis includes electrons, muons, and charged hadrons considered stable within the detector volume (e.g., pions, kaons, protons). Charged hadrons originating from the decay of V0 particles, such as $K^0_S \to \pi^+\pi^-$, are also included. 
This definition inherently minimizes the impact of quantum electrodynamics (QED) final-state radiation (FSR), as the probability of a photon converting to an $e^{+}e^{-}$ pair within the detector is small. 

To reveal the scaling behavior of the EEC in QCD in small opening angles, the track EEC can also be measured as a function of the variable $z$, which is defined in terms of the angle $\theta_{\rm ij}$ as
\begin{equation}\label{eq:z}
    z = \frac{1 - \cos{\theta_{\rm ij}}}{2}.
\end{equation}
Here, the EEC definition can be analogously written in the form given in Equation \ref{eq:ENC_z}, where all the same notational details as specified above for Equation \ref{eq:ENC} apply.
\begin{equation}\label{eq:ENC_z}
    \text{EEC}(z) = \sum^{\rm n}_{\rm i \; > \; j} \int d\sigma \frac{E_{\rm i}E_{\rm j}}{E^{2}}\delta(z - z_{\rm i j})
\end{equation}
The optimal choice of angular variable in the EEC depends on the physics one aims to study. 
In this analysis, we report both distributions. 

\begin{figure}[ht]
    \centering
    \includegraphics[width=0.75\linewidth]{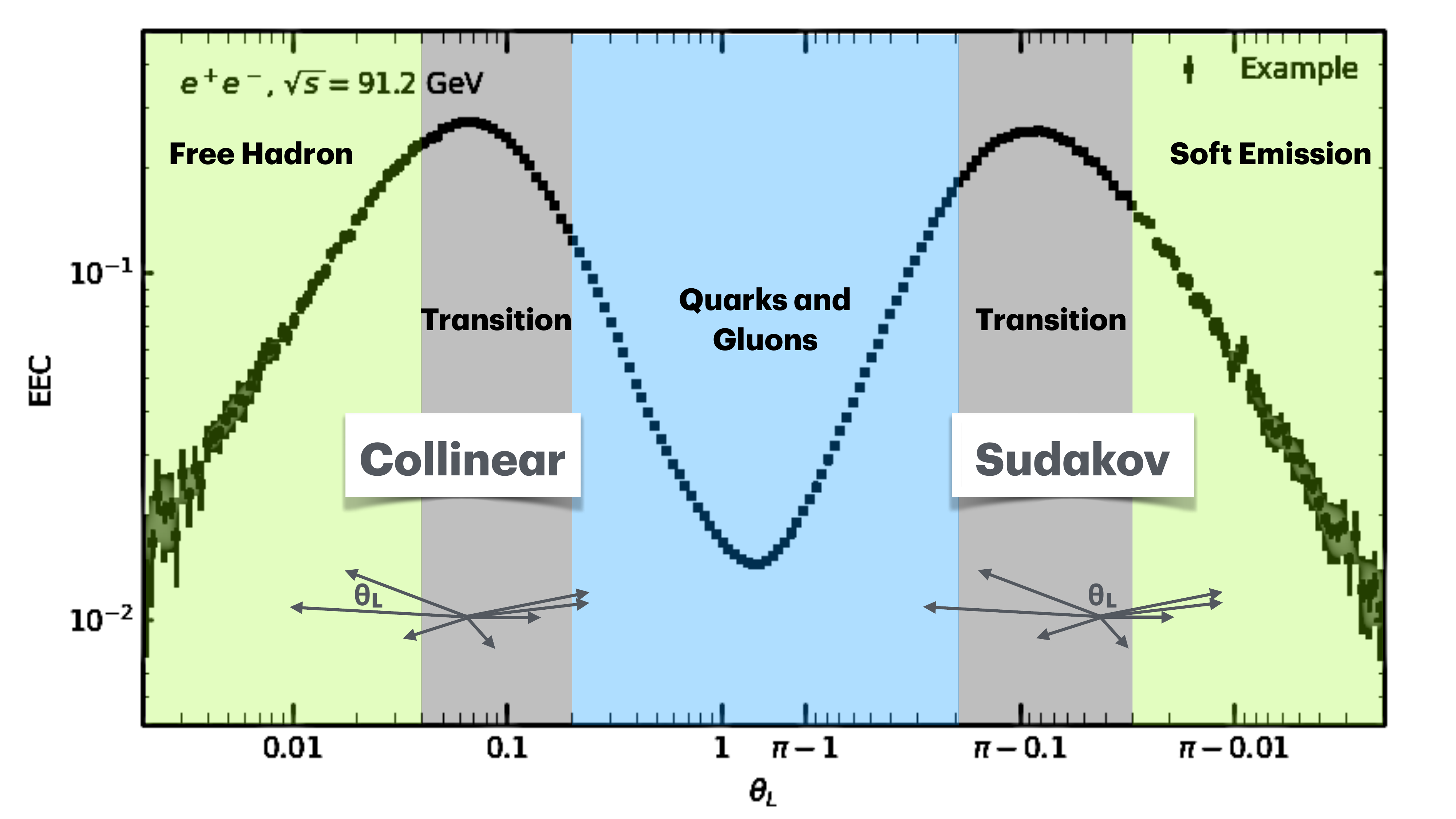}
    \includegraphics[width=0.75\linewidth]{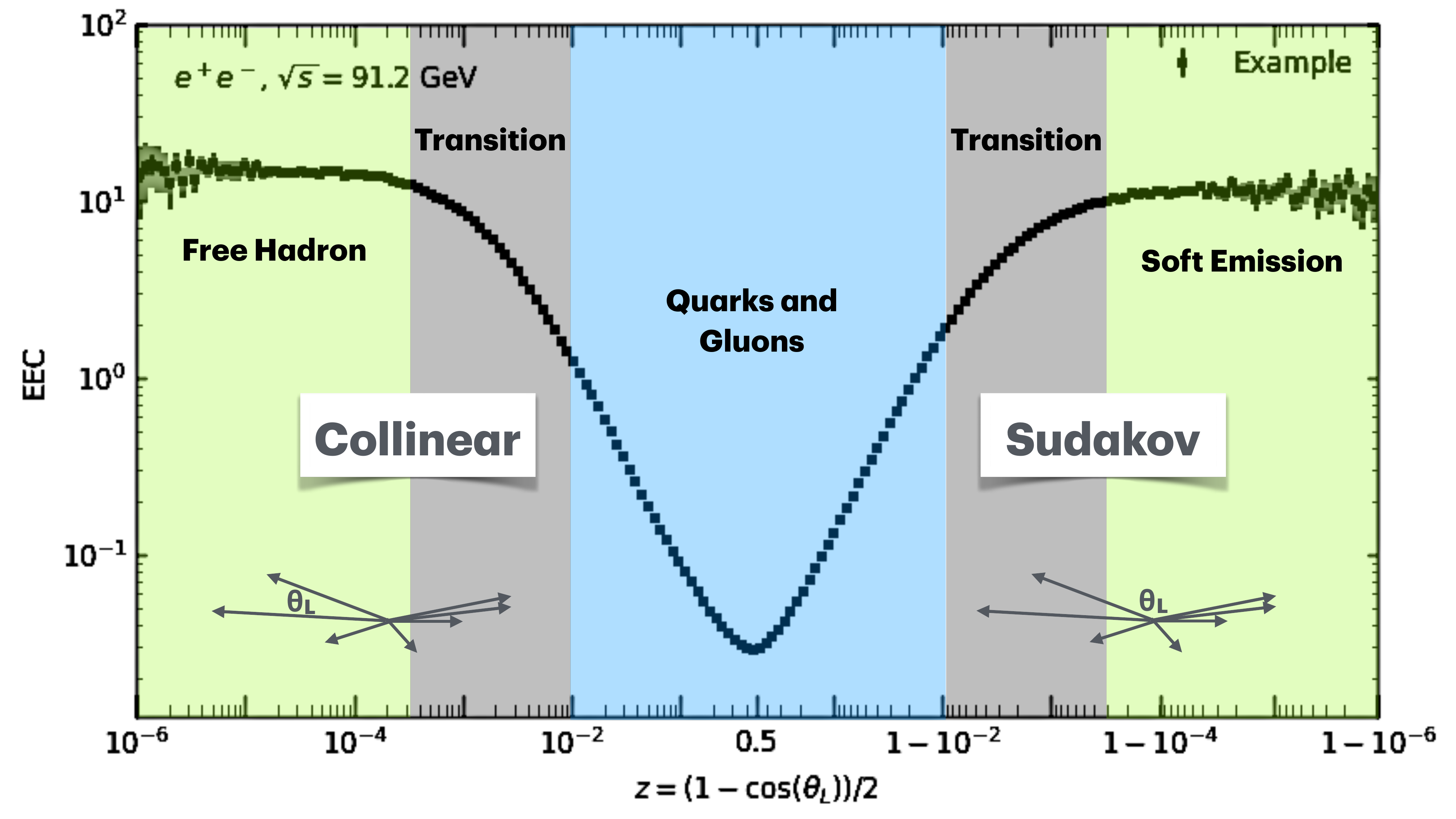}
    \caption{Example distributions of the EEC as a function of $\theta_{\rm L}$ (top) and $z$ (bottom) using a ``flipped double-log style" as described in the main text. The various regions of the distribution are marked for illustration purposes, where the green represents the free hadron (or soft emission) region, the blue represents the quark/gluon region, and the gray represents the transition between these two regions. The collinear region (where $\theta_{\rm L}\sim 0$ and $z \sim 0$) and the back-to-back or Sudakov region (where $\theta_{\rm L}\sim\pi$ and $z \sim 1$) is also marked.}
    \label{fig:ENCcollinearsudakov}
\end{figure}

To illustrate the shape of the EEC definition used here and to highlight the relevant regions, example EEC distributions are shown as functions of $\theta_{\rm L}$ and $z$  in the left and right panels of Figure~\ref{fig:ENCcollinearsudakov}, respectively.
 The distributions here are shown using a so-called ``flipped double-log" style. As the name suggests, there are two log axes, one which increases approaching $\theta_{\rm L} = \pi/2$ ($z = 1/2$) and another ``flipped" log axis decreasing from $\theta_{\rm L} = \pi/2$ ($z = 1/2$) to $\theta_{\rm L} = \pi$ ($z = 1$). This is done in order to highlight three regions of interest. In each distribution, the left hand side, corresponding to smaller angles (and therefore smaller values of $z$) represents the collinear limit. This would be analogous to what is probed by studying the EEC inside of jets, where the EEC exhibits a number of different scaling behaviors cleanly broken up into three regions corresponding to the free hadron region, the quark and gluon region, and the transition between these two regions. Similarly, the right-hand side, corresponding to larger angles (and therefore larger values of $z$), represents the back-to-back. Analogous to the collinear limit, the Sudakov limit also has three characteristic regions, including the quark and gluon region, a region of soft emissions, and the transition between these two regions. 
 However, in the Sudakov limit, the ordering of these distributions is reversed compared to the collinear limit, making the regions appear approximately, though not necessarily perfectly, symmetric when displayed on a double-logarithmic scale

\subsection{Thrust}

Thrust is defined in the center-of-mass frame of an $e^{+}e^{-}$ collision as given in~\cite{PhysRevLett.39.1587} and Equation \ref{eq:thrust} where the sum is over all particles in the event and the maximum is over 3-vectors $\vec{n}$ of unit norm. 
\begin{equation}\label{eq:thrust}
T = \max_{\hat{n}} \frac{\sum_i |\vec{p}_i \cdot \hat{n}|}{\sum_i |\vec{p}_i|}
\end{equation}

The vector $\hat{n}$ that maximizes thrust is known as the thrust axis and the plane which is defined as normal to $\hat{n}$ splits the event into two hemispheres. 
Since the thrust distribution roughly follows a log normal distribution, $P(1 - T) \sim \exp \left[-C \log^2(1 - T)\right]$, where $C$ is a constant, a second variable is defined as $\tau = 1 - T$. 
In this analysis, $\tau$ and $\log \tau$ are used directly in the unfolding. 
A detailed description of the thrust distribution can be found in~\cite{Abbate:2010xh, Benitez:2024nav, Becher:2008cf}. 
The distribution is determined using all charged and neutral particles. 
Similar to EEC, charged particles include electrons, muons, and charged hadrons considered stable within the detector volume (e.g., pions, kaons, protons).
The neutral particles include photons and all long-lived neutral hadrons (e.g., $K^0_L$ and neutrons). 
Short-lived particles like $\pi_0$'s and V0's are not directly used; their stable decay products are. 
Neutrinos contribute to the event's missing energy vector and are also included.
The algorithm implementation is based on the Herwig++ implementation~\cite{Brandt:1978zm, B_hr_2008, rivet_thrust, BUCKLEY20132803}, which is an exact algorithm to calculate thrust rather than one based on heuristics as is done in other implementations~\cite{Belle2Thrust}.

\begin{figure}[ht]
    \centering
    \includegraphics[width=0.95\linewidth]{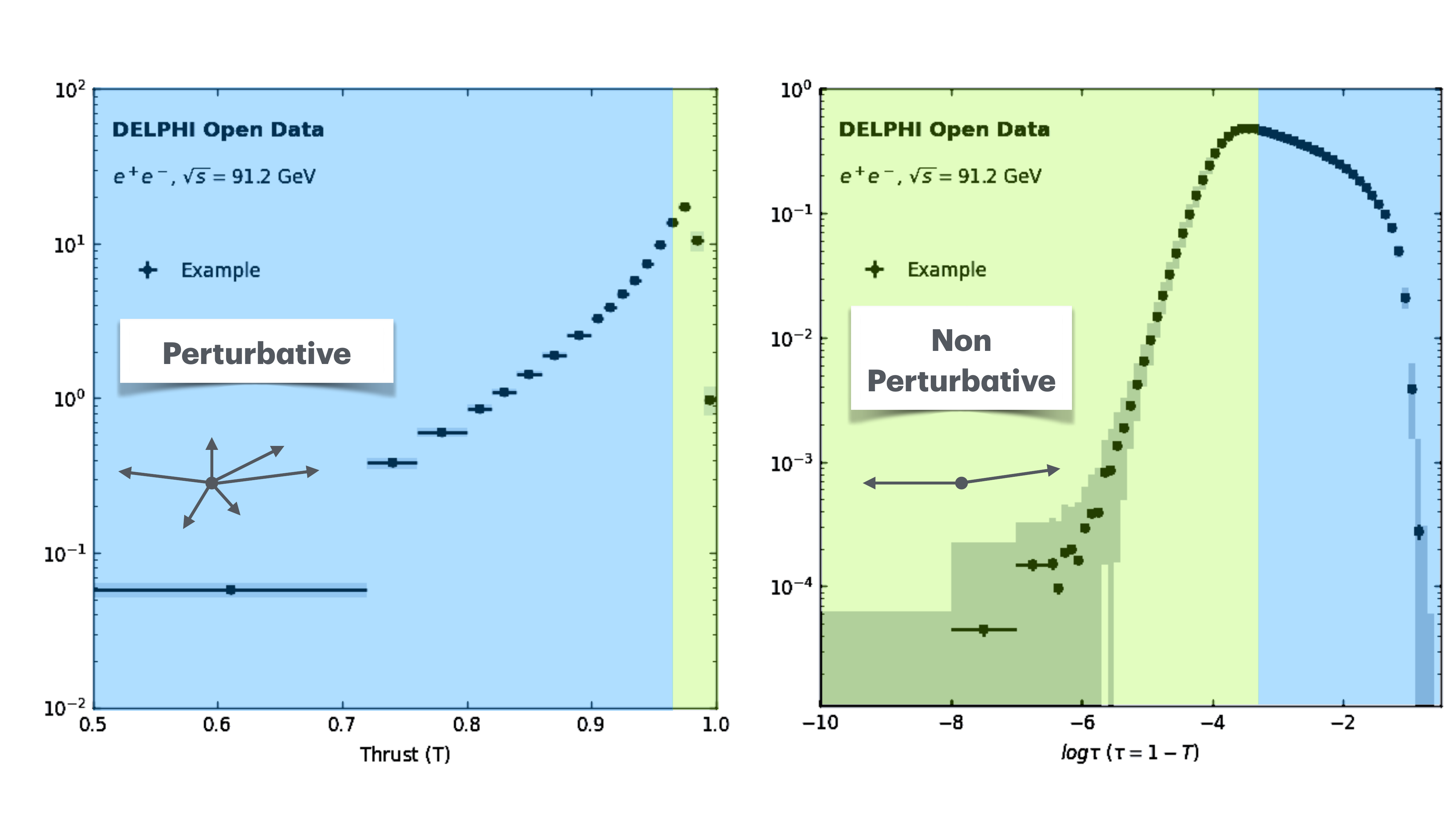}
    \caption{Example distribution of thrust (left) and $\log \tau$ (right). The thrust distribution clearly separates two distinct QCD regimes: the perturbative multi-jet region at lower thrust values (light blue), characterized by a broad distribution from isotropic particle production, and the non-perturbative dijet region (light green) at high thrust values near unity. The sharp dijet peak located just below the thrust maximum results from Sudakov form factor suppression, which represents the probability of no additional parton emission. This suppression makes perfect two-jet configurations less probable, creating the characteristic peaked structure. The $\log \tau$ parameterization provides an enhanced view of the non-perturbative region by expanding the logarithmic scale, offering detailed resolution of the peak structure that complements the linear thrust binning.} 
    \label{fig:thrustLogTau}
\end{figure}

Example distributions are shown in Figure~\ref{fig:thrustLogTau} to illustrate the characteristic features of thrust using two different parameterizations to highlight different physical regimes. 
The left panel shows the distribution as a function of T.
At low thrust values (light blue), the multi-jet or "perturbative" region is present, where events arise from hard gluon radiation, leading to a more spherical or isotropic event topology.
In contrast, the distribution rises sharply into a peak as T approaches 1. 
This is the dijet or "non-perturbative" region (light green), where events are dominated by two back-to-back jets. The characteristic peak structure is a classic feature of QCD, arising from Sudakov suppression: the probability of radiating no additional soft or collinear gluons is suppressed, preventing the distribution from diverging at the perfect dijet limit of T=1.
The right panel, plotted against $\log \tau$, provides a ``zoomed-in'' view of the non-perturbative region. This logarithmic variable effectively "stretches" the peak, revealing its detailed shape with much greater resolution. This representation is the most natural way to plot the P(1-T) log normal distribution; it is essential for precision studies of non-perturbative QCD phenomena.
In this analysis, both the linear thrust and the $\log \tau$ distributions are reported, allowing a comprehensive comparison between data and theory in all regions. 
The linear distribution is highly sensitive to $\alpha_s$, while the $\log \tau$ provides powerful constraints on universal non-perturbative parameters, such as $\rm \Omega_{1}$, through the calculation of its logarithmic moments. 

\section{The DELPHI detector}
\label{sec:delphi}
DELPHI is one of four large general-purpose detectors stationed around LEP. 
The detector collected data from 1989 to 2000, covering center-of-mass energies from the Z boson pole up to 209 GeV, before being dismantled to make way for the construction of the Large Hadron Collider in the LEP tunnel.

\begin{figure}[ht]
    \centering
    \includegraphics[width=0.9\linewidth]{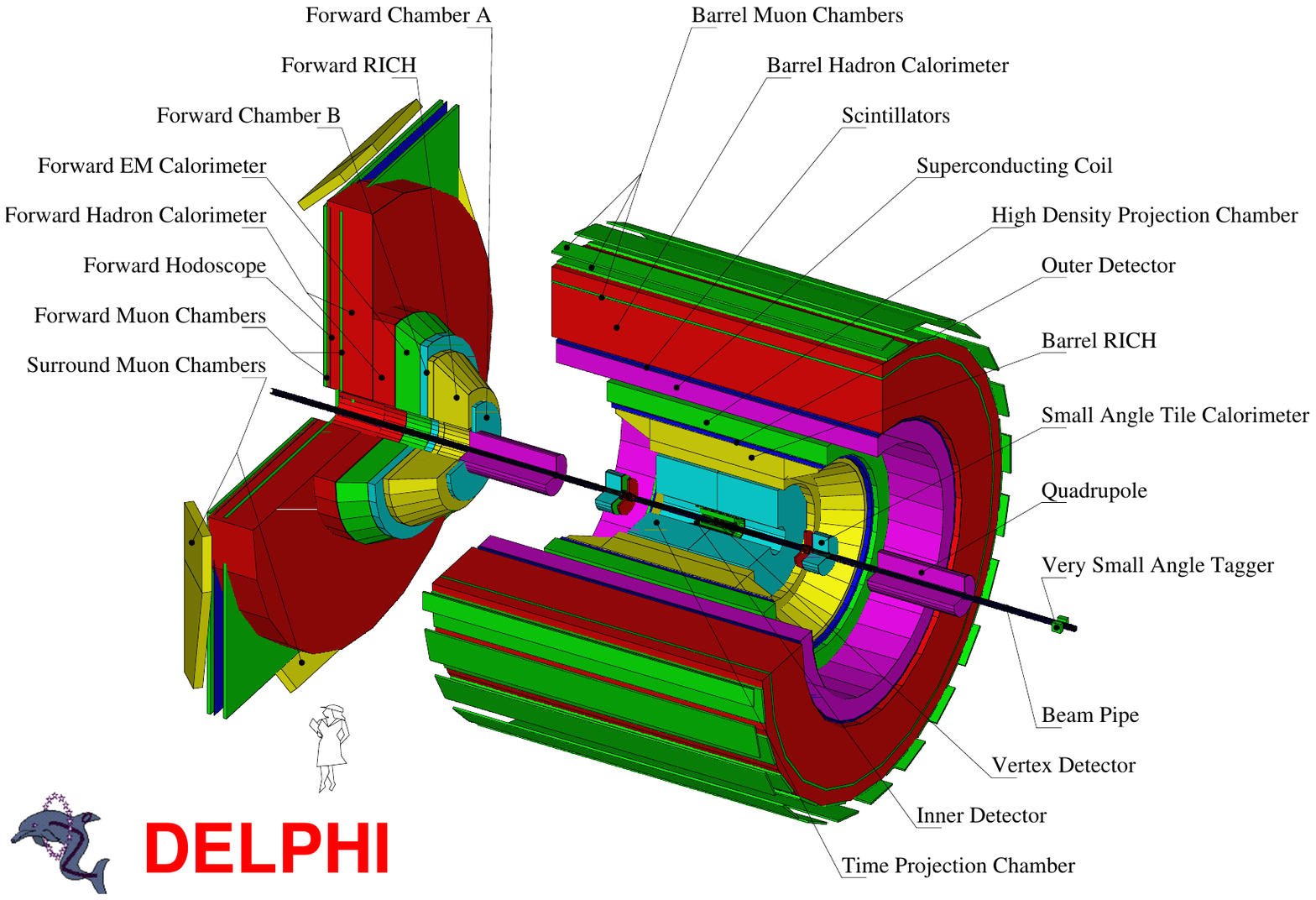}
    \caption{Schematic of the DELPHI detector.}
    \label{fig:delphi}
\end{figure}

As illustrated in the schematic in Figure~\ref{fig:delphi}, DELPHI is a complex apparatus consisting of over 20 sub-detectors arranged in a cylindrical geometry around the interaction point. While a comprehensive description of the full detector and its performance can be found in Refs.~\cite{DELPHI:1990cdc,DELPHI:1995dsm}, the following sections will briefly describe the components most critical for the reconstruction of particles and the measurement of event shapes in this analysis.

In the barrel region, the charged particle tracks are measured by a set of cylindrical tracking detectors whose axes are parallel to the 1.2~T solenoidal magnetic field and to the beam direction. The innermost one is the Microvertex Detector (VD), which is located between the LEP beam pipe and the Inner Detector (ID). The DELPHI Microvertex Detector used from 1991 to 1993 was composed of 3 layers of single-sided silicon microstrip detectors at radii of 6.3, 9, and 11 cm from the beam line, respectively called the closer, inner, and outer layers. To improve the performance of the detector in tracking and especially in the identification of b-hadrons, in 1994, it was upgraded using double-sided silicon detectors, allowing three-dimensional impact parameter reconstruction.

The Time Projection Chamber (TPC) is the main tracking device and is a cylinder of length 3~m, inner radius 30~cm, and outer radius 122~cm. Between polar angles of $39^{\circ}$ and $141^{\circ}$, tracks are reconstructed using up to 16 space points. Outside this region ($21^{\circ}$ to $39^{\circ}$ and $141^{\circ}$ to $159^{\circ}$), tracks can be reconstructed using at least four space points. The Inner Detector (ID) is a cylindrical drift chamber having an inner radius of 12~cm and an outer radius of 28~cm and covers polar angles between $29^{\circ}$ and $151^{\circ}$. It contains a jet chamber section providing 24 $R\phi$ coordinates, surrounded by five layers of proportional chambers giving both $R\phi$ and $z$ coordinates. The Outer Detector (OD) has five layers of drift cells, with radii ranging from 198 to 206 cm, and covers polar angles from $42^{\circ}$ to $138^{\circ}$.

The barrel electromagnetic calorimeter (HPC) covers polar angles between $42^{\circ}$ and $138^{\circ}$. It is a gas-sampling device that provides complete three-dimensional charge information in the same way as a time projection chamber. The excellent granularity allows good separation between close particles in three dimensions and hence good electron identification even inside jets. In the forward region, the tracking is complemented by two sets of planar drift chambers, Forward Chamber A (FCA) and Forward Chamber B (FCB), at distances of $z = \pm 165$~cm and $z = \pm 275$~cm from the interaction point. A lead glass calorimeter (EMF) is used to reconstruct electromagnetic energy in the forward region.

The iron return yoke of the magnet is instrumented with limited streamer mode detectors to create a sampling gas calorimeter, the Hadron Calorimeter (HAC). A new readout system was implemented to capture signals from both the HAC tubes and pads. This system provides a more detailed picture of hadronic showers, which allows for better distinction between showers caused by neutral and charged hadrons and improves muon identification. The readout was installed in the barrel part of the detector in early 1995 and later in the endcaps.

The Ring Imaging CHerenkov (RICH) detectors provide charged particle identification in both the barrel (BRICH) and forward (FRICH) regions. They combine liquid and gas radiators to identify charged particles over most of the momentum range at LEP-1. Though the main structures were installed before startup in 1989, the radiators, fluid systems, chambers, and electronics were installed and brought into operation in stages from 1990 to 1993. The BRICH became fully operational during 1992, and the FRICH at the beginning of 1994.

Muon identification in the barrel region is based on a set of muon chambers (MUB), covering polar angles between $53^\circ$ and $127^\circ$. In the forward region, the muon identification is provided using two sets of planar drift chambers (MUF) covering the angular region between $11^\circ$ and $45^\circ$.


The luminosity determination at DELPHI relies on two forward detectors: the Small angle TIle Calorimeter (STIC) and the Very Small Angle Tagger (VSAT). The primary measurement of the absolute luminosity was derived from the rate of low-angle Bhabha scattering events recorded in the STIC detector. This was complemented by the VSAT, which is used to measure the relative luminosities at different energies. 

The DELPHI trigger system is a four-level, hierarchical apparatus designed to handle high event rates while efficiently selecting physics events. The system comprised two hardware levels (T1, T2) and two software filters (T3, T4). The first level, T1, acts as a loose pre-trigger using simple patterns in individual sub-detectors. The main hardware decision is made at T2, which requires correlations between signals from different sub-detectors to suppress backgrounds from noise and beam-gas interactions. 
The subsequent software levels, T3 and T4, are used to increase the purity of the data sample written to tape. They are designed for high efficiency and redundancy. For processes like Z decays, the efficiency is 100\%~\cite{DELPHITriggerGroup:1994vta}.

\clearpage
\section{Data and simulated samples}
\label{sec:sample}
\subsection{The DELPHI open data}
In August 2024, the DELPHI Collaboration released its entire data collection to the physics community for public access, following FAIR (Findable, Accessible, Interoperable, and Reusable) principles. 
The data spans the entire operational period of DELPHI from 1989 to 2000, covering both LEP-1 operations with collision energies corresponding to the Z boson resonance and LEP-2 high-energy running.
The data are available in various formats, including short DST and extended short DST files optimized for analysis, as well as raw data for specialized studies.
The DELPHI data re-use policy can be found in Ref.~\cite{DELPHI:2024policy}. 
Comprehensive documentation, quick-start guides, and example analyses can be found in Ref.~\cite{DELPHI:2024opendata}. 

Beyond the raw experimental data, the DELPHI Collaboration has also released the complete original software stack, including simulation and reconstruction tools. This unprecedented level of openness enables researchers to perform end-to-end analyses---from passing Monte Carlo generator output through the full DELPHI detector simulation to comparing with actual measurements---a capability not possible for the archived ALEPH dataset, whose simulation software was not preserved. The software release includes the DELPHI DST (Data Summary Tape) analysis framework, Monte Carlo production tools, event reconstruction capabilities, and the graphical event display system. The data and software are distributed through CERN's established systems, including CVMFS (CERN Virtual Machine File System) for software distribution and EOS storage for data access. The availability of both data and software allows for rigorous validation of new theoretical predictions, reanalysis of historical measurements using modern techniques, and the development of novel analysis methods.

This analysis uses data collected by the DELPHI detector at a center-of-mass energy of $\sqrt{s} = 91.2$~GeV during 1994 and 1995.
The 1994 data correspond to an integrated luminosity of 46~pb$^{-1}$. 
In 1995, LEP finished data taking at the $Z^0$ resonance with a final energy scan: $15.1$~pb$^{-1}$ at $\sqrt{s} = 91.2$~GeV, $9.1$~pb$^{-1}$ at $\sqrt{s} = 89.4$~GeV, and $9.4$~pb$^{-1}$ at $\sqrt{s} = 93.0$~GeV.
Only the data taken at $\sqrt{s} = 91.2$~GeV are considered in this analysis.
The total integrated luminosity used is then 61~pb$^{-1}$ at the $Z^0$ resonance peak, providing high-statistics samples of hadronic $Z^0$ decays with negligible impact on initial state radiation.
Data collected in earlier years of LEP-1 operation are not included because the vertex detector (VD) layout was upgraded in 1994, significantly improving its performance.

The DELPHI detector and offline software, denoted by the tags $94\_\text{c}$ and $95\_\text{d}$ in the DELSIM program, differ slightly between 1994 and 1995.
To ensure a consistent treatment of detector effects, the analysis is therefore performed independently on the two datasets. The fully corrected, unfolded distributions from each year are then statistically combined in the final step to produce the results that are presented in this note.

Several sets of Monte Carlo (MC) simulated $e^{+}e^{-} \to q\bar{q}$ samples are used to correct the detector effects. For the 1994 data, two of these samples are sourced from the DELPHI open data portal. They are generated using the KK2f event generator~\cite{Jadach:1999vf,Jadach:2000ir} and fragmented with two different hadronization models: PYTHIA5.7/JETSET7.4~\cite{Sjostrand:1995vx} and ARIADNE~\cite{Lonnblad:1992tz}. The hadronization models in these official samples are tuned, and the final-state hadrons are processed through the full DELPHI detector simulation using DELSIM~\cite{DELSIM} to accurately model the detector response. For the 1995 data, only the PYTHIA5.7/JETSET7.4 sample is used, as the ARIADNE sample is not available from the data portal. A summary of the data and MC samples, including their DELPHI open data portal Digital Object Identifiers (DOI), is provided in Table~\ref{tab:SampleSummary}.

\begin{table}[ht]
\centering
\small
\begin{tabular}{l|l|l}
\hline\hline
\textbf{Type} & \textbf{Sample} & \textbf{DOI} \\
\hline
Data 1994 & \texttt{short94\_c2} & \href{https://doi.org/10.7483/OPENDATA.DELPHI.0XNE.G96F}{\texttt{10.7483/OPENDATA.DELPHI.0XNE.G96F}} \\
Data 1995 & \texttt{short95\_d2} & \href{https://doi.org/10.7483/OPENDATA.DELPHI.K4LR.4PQ4}{\texttt{10.7483/OPENDATA.DELPHI.K4LR.4PQ4}} \\
\hline
PYTHIA 5 1994 & \texttt{sh\_kk2f4146qqpy\_e91.25\_c94\_2l\_c2} & \href{https://doi.org/10.7483/OPENDATA.DELPHI.P4BN.GFID}{\texttt{10.7483/OPENDATA.DELPHI.P4BN.GFID}} \\
PYTHIA 5 1995 & \texttt{sh\_kk2f4146qqpy\_e91.25\_c95\_1l\_d2} & \href{https://doi.org/10.7483/OPENDATA.DELPHI.OJDR.XYNG}{\texttt{10.7483/OPENDATA.DELPHI.OJDR.XYNG}} \\
ARIADNE 1994 & \texttt{sh\_kk2f4146qqardcy\_e91.25\_r94\_2l\_c2} & \href{https://doi.org/10.7483/OPENDATA.DELPHI.0VKG.S2RX}{\texttt{10.7483/OPENDATA.DELPHI.0VKG.S2RX}} \\
\hline\hline
\end{tabular}
\caption{Summary table for data and MC samples used in this note.}
\label{tab:SampleSummary}
\end{table}

Another two sets of simulated events are created using PYTHIA~8.3~\cite{Bierlich:2022pfr} with the default and Dire~\cite{Hoche:2015sya} parton shower algorithms for each of the $94\_\text{c}$ and $95\_\text{d}$ configurations for systematic uncertainty estimations. 
The parton showers are hadronized by PYTHIA~8.312 with the Monash Tune~\cite{Skands:2014pea}, and then processed using the DELSIM software container provided in the DELPHI open data portal. 
The Monash tune is used as it is the default in PYTHIA 8; a comprehensive study of alternative tunes is beyond the scope of this analysis.
The key component of the software pipeline for the simulation is to convert PYTHIA~8 output to the FORTRAN binary compatible with the DELSIM container (see Appendix~\ref{app:analysiscode}). 
The reconstruction efficiency and fake rate of these samples are provided in Appendix~\ref{app:pythia8perf}. 
The results are in good agreement with those from the DELPHI official PYTHIA 5 sample (which is presented in Section~\ref{sec:matching}), validating the PYTHIA-to-DELSIM pipeline. 

\subsection{Reconstruction and selections}
\label{sec:selections}
In DELPHI, the energy flow in each event is determined through an approach that combines all available information from both the tracking detectors and the calorimetric systems. 

Due to the exceptional precision of the DELPHI tracking system (see Section~\ref{sec:matching}), the momentum and energy of charged particles are determined using tracking information. For charged particles identified by the tracking system, particle masses are assigned based on the comprehensive DELPHI particle identification framework, which combines information from the RICH detectors, specific energy loss measurements in the TPC, and other detector subsystems.

The electromagnetic calorimeters primarily provide energy measurements for photons, and the hadron calorimeters are specifically utilized to detect and measure the energy of long-lived neutral hadrons such as neutrons, $\rm K_L^0$\footnote{An important detail of this work is the inclusion of V0 decay products in the generator-level definition, which is necessary for consistent comparisons with modern convention. Note that this may differ from the treatment in some previous measurements; see Appendix~\ref{app:Ks} for further discussion.}, and other neutral particles that do not leave tracks in the detector but deposit energy through hadronic interactions in the calorimetric material.

The thrust is measured using both charged and neutral particles, while the EEC is measured using only charged particles, reconstructed and identified by the energy flow algorithm. 
Neutrinos are treated as the total missing momentum ($p_{\rm miss}$) and included in the thrust calculation for data and detector-level MC. 
The $p_{\rm miss}$ is defined as the negative of the vector sum of all the charged and neutral particles' 3-momentum. 
In the gen-level MC, neutrinos are included as individual particles. 
The selection criteria for charged and neutral particles are summarized in Table~\ref{tab:SelectionSummary}. These criteria, optimized by the DELPHI collaboration, are taken from Ref.~\cite{DELPHI:2003yqh}.

Charged particles are required to fall within the detector acceptance, such that the polar angles ($\theta$) are between 20 and 160 degrees and transverse momenta $p_{\rm T}$ are greater than 0.4~GeV, following the common DELPHI selection. This requirement is applied to both generator-level and detector-level MC in the unfolding procedure described in Section~\ref{sec:unfolding}\footnote{The generator-level track selection will be corrected to full phase-space after unfolding for the final fully-corrected results (see Section~\ref{sec:acccorr})}. Track candidates are also required to be of so-called ``high-quality". To meet this requirement, they must have a measured length greater than 30~cm and a relative error from track fitting less than 1$\sigma$. An additional requirement is placed on the impact parameter to have a radial displacement from the interaction point of $d_{0} <4$ cm and a longitudinal displacement of $z_{0}<10$ cm. The requirements on impact parameter and track quality are applied to data and detector-level MC only. 

For neutral particles, the acceptance requirements are 20 and 160 degrees and 0.5~GeV for $\theta$ and energy ($E$), respectively. 
\begin{table}[ht]
\centering
\begin{tabularx}{\textwidth}{l|l}
\hline\hline
\multicolumn{2}{l}{Charged particles}  \\
\hline
Acceptance              & $20^\circ\le\theta\le160^\circ$ \\
                        & $0.4~\text{GeV} \le p_{\rm T} \le 100~\text{GeV}$ \\
High quality tracks     & measured track length $\ge 30~\text{cm}$ \\
                        & $\Delta p/p \le1.0 $ \\
Impact parameter        & $d_0\le4$~cm, $z_0\le10$~cm \\
\hline\hline
\multicolumn{2}{l}{Neutral particles}  \\
\hline
Acceptance              & $20^\circ\le\theta\le160^\circ$ \\
                        & $0.5~\text{GeV} \le E $ \\
\hline\hline
\multicolumn{2}{l}{Event selection}  \\
\hline
Hadronic events         & $30^\circ \le \theta_{\rm thrust} \le 150^\circ$\\
                        & at least 7 good tracks \\
                        & $E_{\rm tot} \ge 0.5E_{\rm cm}$ \\
\hline\hline
\end{tabularx} 
\caption{Summary table for particle and event selections.}
\label{tab:SelectionSummary}
\end{table}

In addition to the track and neutral particle selections, a standard hadronic event selection is applied to suppress background processes such as two-photon interactions, beam-gas and beam-wall interactions, and leptonic final states including $e^{+}e^{-}\to \tau^{+}\tau^{-}$ process, following Ref.~\cite{DELPHI:2003yqh}. 
These selections require events to have at least 7 selected charged tracks, total reconstructed energy, including both neutral and charged particles, more than half of the center of mass energy, and the polar angle of the thrust axis to be between 30$^\circ$ and 150$^\circ$. 
According to Ref.~\cite {DELPHI:2003yqh}, the purity of these selections is above 99\%. 
These selections are applied to both data and detector-level MC samples.
After these selections, approximately 1.2 million and 400 thousand events remain for the 1994 and 1995 data samples, respectively.
The event selection efficiency, together with the track and neutral particle acceptance, will be corrected using the procedure described in Section~\ref{sec:corrections}.

\subsection{Data and Monte Carlo comparisons}\label{sec:DataMCComp}
\begin{figure}[ht!]
    \centering
    \includegraphics[width = 0.32\textwidth]{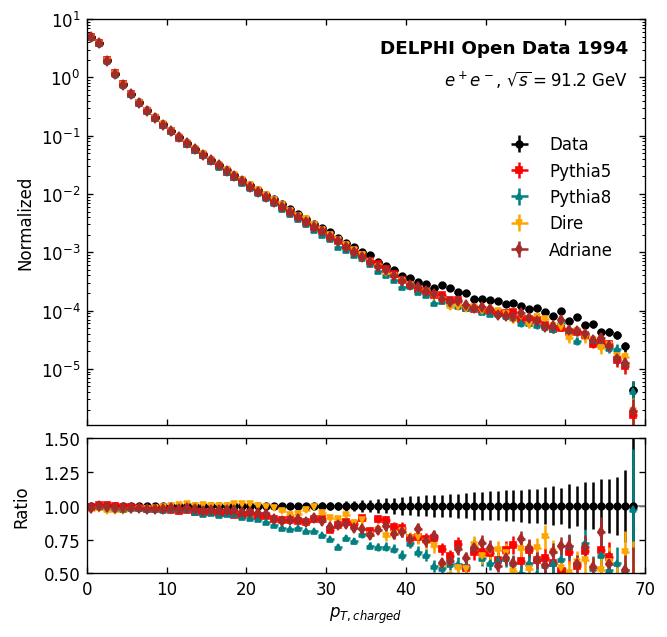}
    \includegraphics[width = 0.32\textwidth]{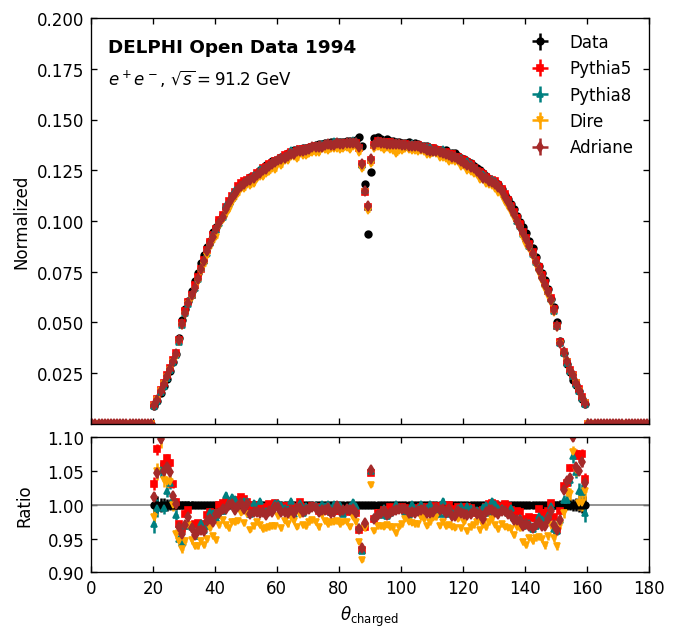}
    \includegraphics[width = 0.32\textwidth]{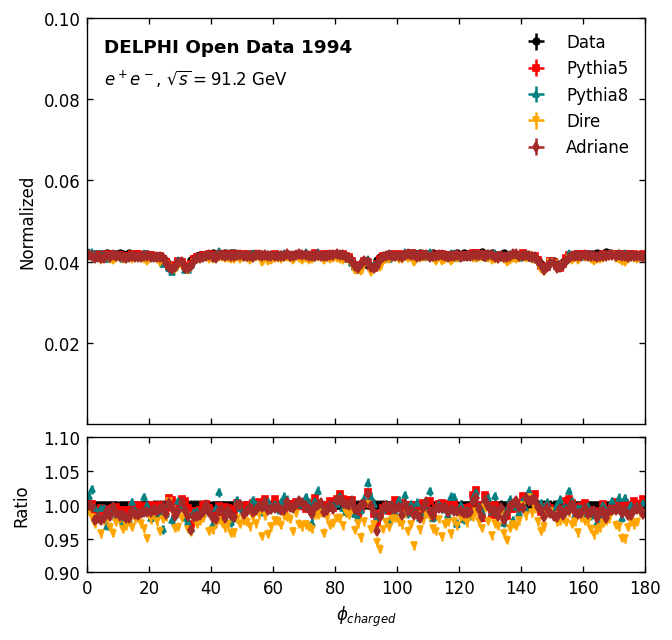}
    \caption{Comparisons of track $p_{\rm T}$, $\theta$, and $\phi$ in 1994 data (black), reconstructed PYTHIA~5.7/JETSET~7.4 (red), ARIADNE (green), PYTHIA 8.3 (orange), and PYTHIA 8.3 Dire (brown) samples. Note that the corresponding number of selected events of each sample normalizes the distributions.}
    \label{fig:MCdatacomp1}
\end{figure}

\begin{figure}[ht!]
    \centering
    \includegraphics[width = 0.32\textwidth]{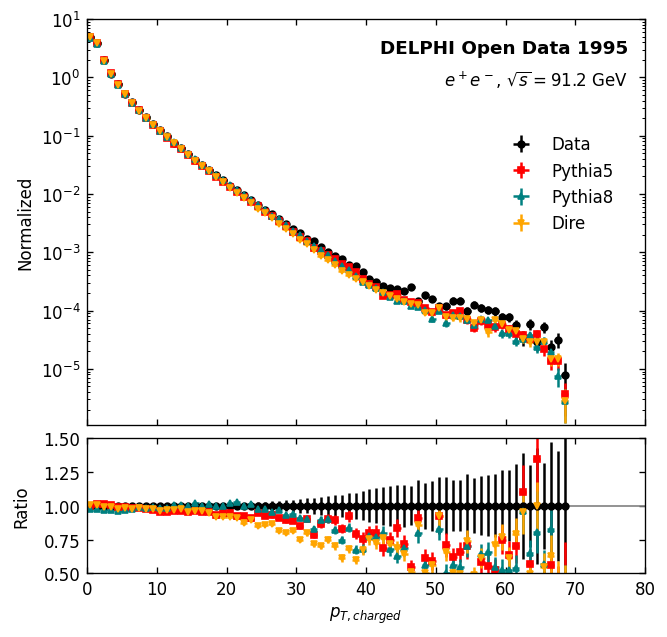}
    \includegraphics[width = 0.32\textwidth]{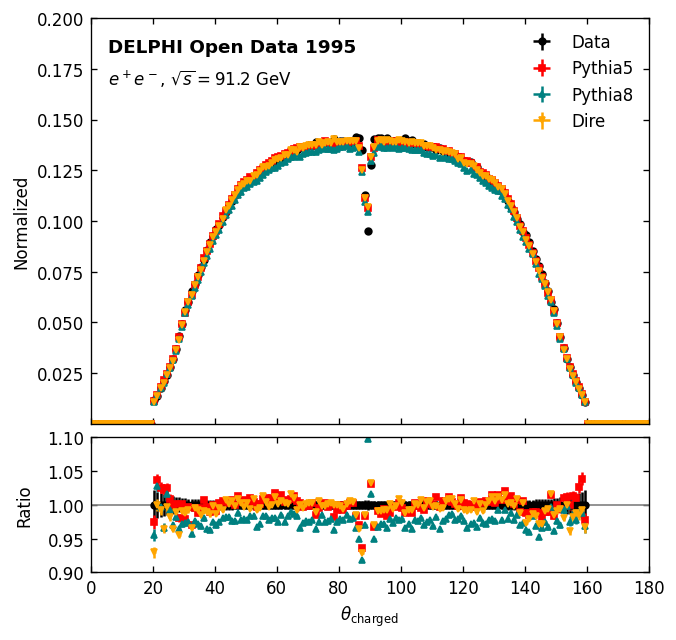}
    \includegraphics[width = 0.32\textwidth]{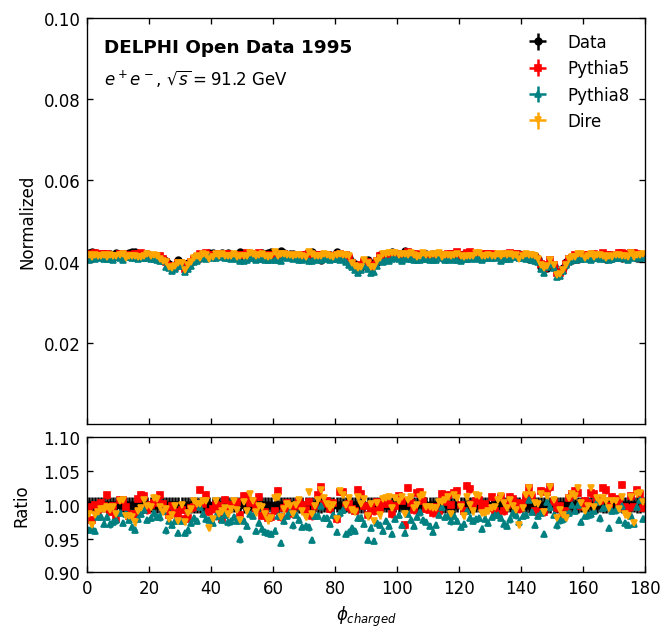}
    \caption{Comparisons of track $p_{\rm T}$, $\theta$, and $\phi$ in 1995 data (black), reconstructed PYTHIA~5.7/JETSET~7.4 (red), ARIADNE (green), PYTHIA 8.3 (orange), and PYTHIA 8.3 Dire (brown) samples. Note that the corresponding number of selected events of each sample normalizes the distributions.}
    \label{fig:MCdatacomp195}
\end{figure}

To assess the quality of the detector simulation and the underlying physics models, and understand the potential biases that will be corrected by the unfolding procedure, the 1994 and 1995 data are compared to several MC samples at the detector level. These comparisons are performed after the event and particle selections are applied. It is important to note that any observed differences are a convolution of both the generator-level modeling and the detector simulation.

\begin{figure}[ht!]
    \centering
    \includegraphics[width = 0.32\textwidth]{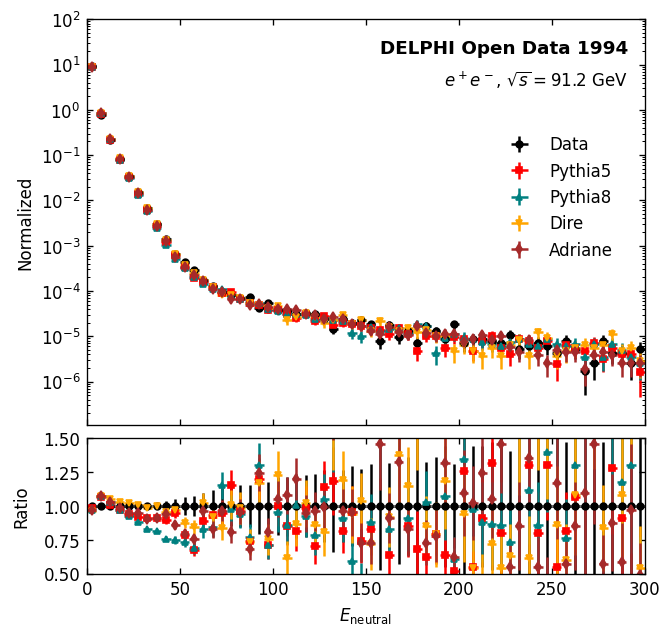}
    \includegraphics[width = 0.32\textwidth]{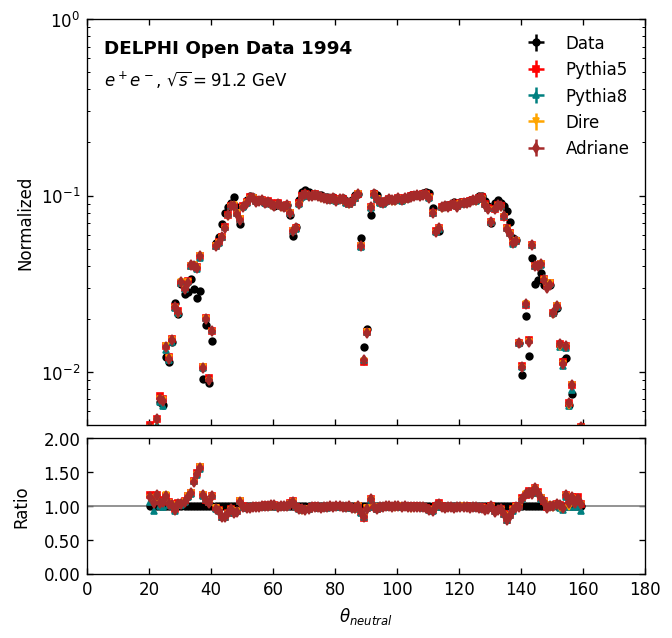}
    \includegraphics[width = 0.32\textwidth]{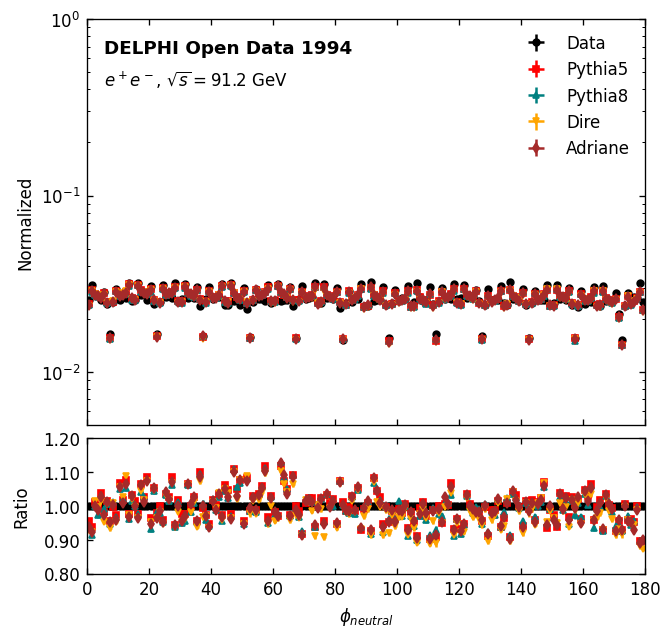}
    \caption{Comparisons of neutral particle $E$, $\theta$, and $\phi$ in 1994 data (black), reconstructed PYTHIA~5.7/JETSET~7.4 (red), ARIADNE (green), PYTHIA 8.3 (orange), and PYTHIA 8.3 Dire (brown) samples. Note that the corresponding number of selected events of each sample normalizes the distributions.}
    \label{fig:MCdatacomp2}
\end{figure}

\begin{figure}[ht!]
    \centering
    \includegraphics[width = 0.32\textwidth]{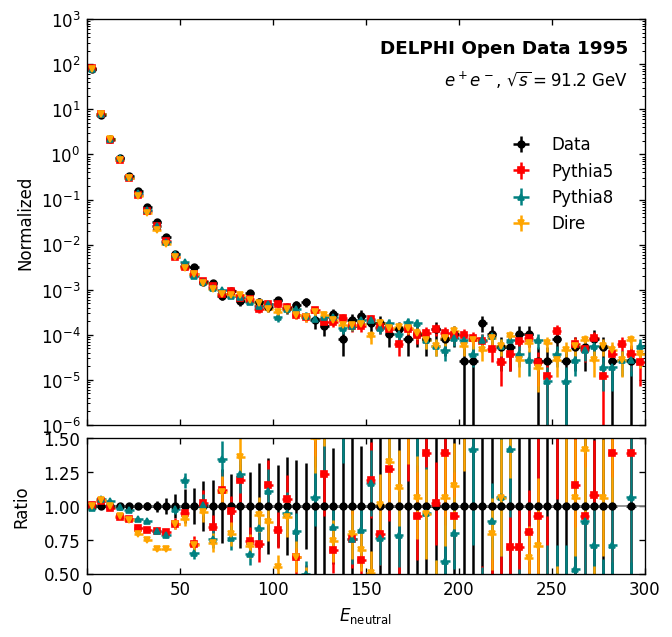}
    \includegraphics[width = 0.32\textwidth]{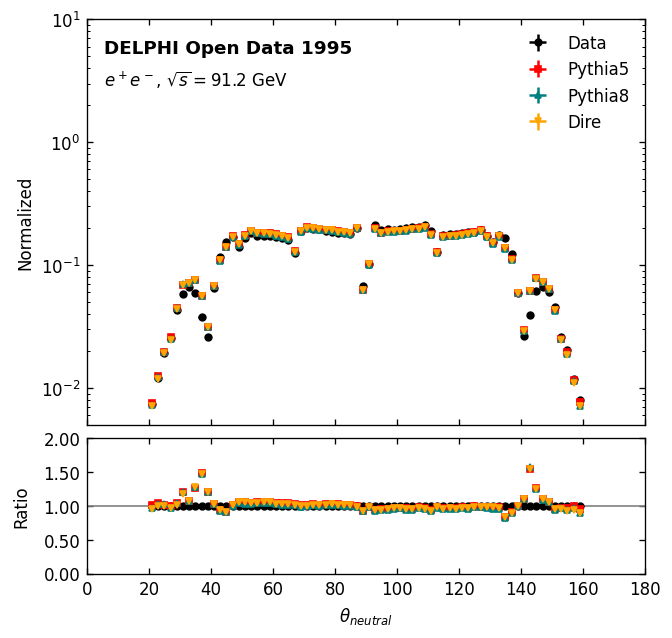}
    \includegraphics[width = 0.32\textwidth]{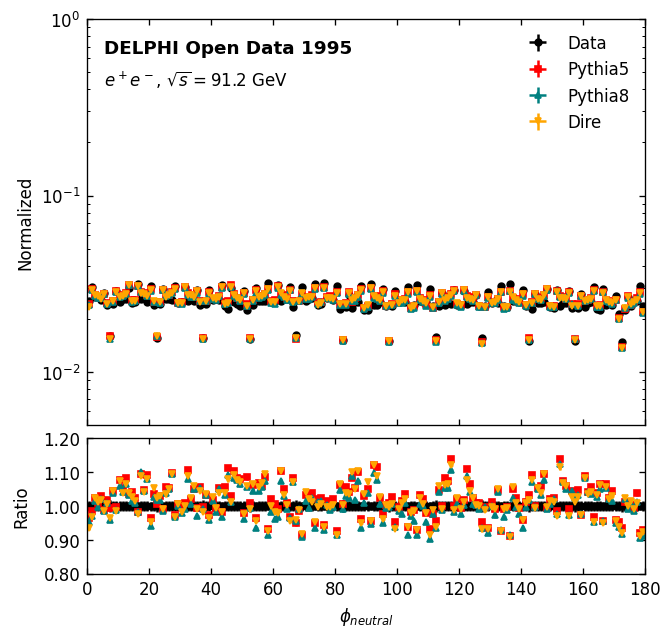}
    \caption{Comparisons of neutral particle $E$, $\theta$, and $\phi$ in 1995 data (black), reconstructed PYTHIA~5.7/JETSET~7.4 (red), ARIADNE (green), PYTHIA 8.3 (orange), and PYTHIA 8.3 Dire (brown) samples. Note that the corresponding number of selected events of each sample normalizes the distributions.}
    \label{fig:MCdatacomp295}
\end{figure}

The comparisons of charged track kinematics ($p_{\rm T}$, $\theta$, and azimuthal angle $\phi$) for 1994 and 1995 are shown in Figs.~\ref{fig:MCdatacomp1} and~\ref{fig:MCdatacomp195}, respectively. The corresponding distributions for neutral particles are shown in Figs.~\ref{fig:MCdatacomp2} and~\ref{fig:MCdatacomp295}. The dips in the angular distributions correspond to known geometric features of the detector, such as cracks, which are generally well-reproduced by the simulations.

\begin{figure}[ht!]
    \centering
    \begin{minipage}{0.45\textwidth}
        \includegraphics[width=\textwidth]{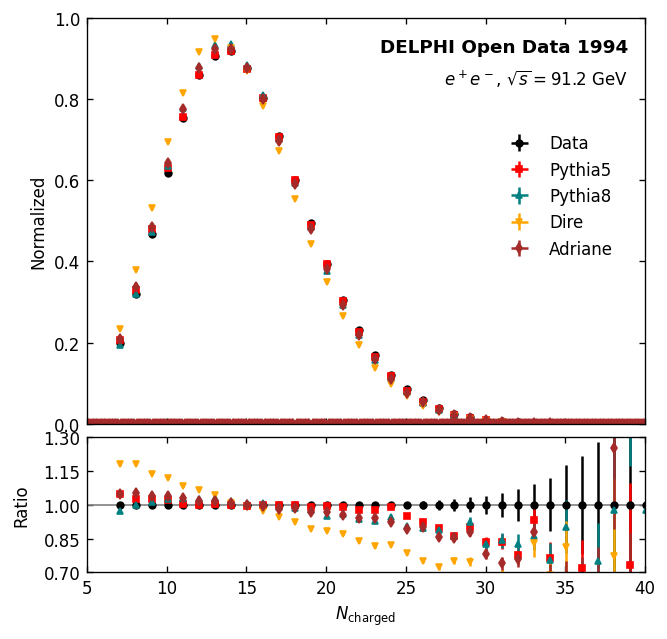}
        \centering
        \text{(a)}\\
    \end{minipage}
    \begin{minipage}{0.45\textwidth}
        \includegraphics[width=\textwidth]{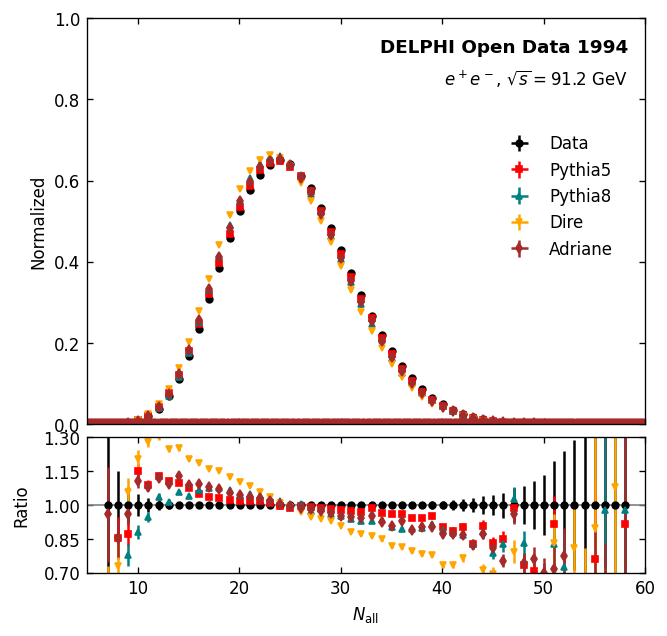}
        \centering
        \text{(b)}\\
    \end{minipage}\\[0.5em]
    \begin{minipage}{0.45\textwidth}
        \includegraphics[width=\textwidth]{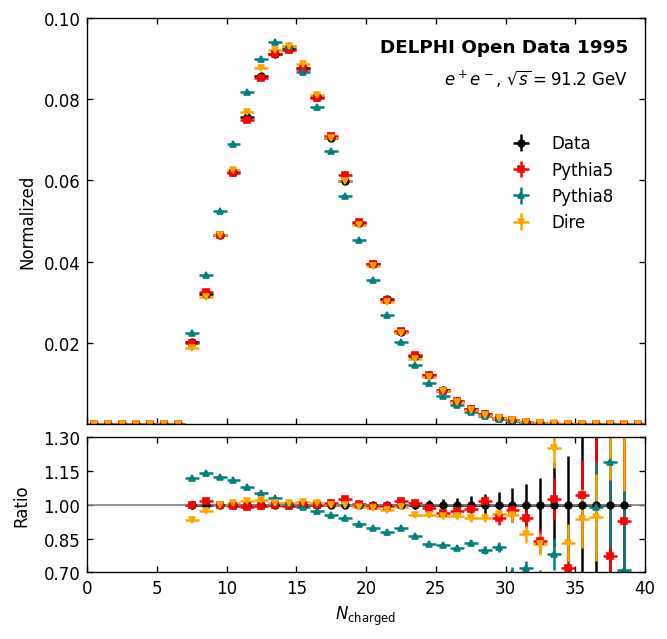}
        \centering
        \text{(c)}\\
    \end{minipage}
    \begin{minipage}{0.4\textwidth}
        \includegraphics[width=\textwidth]{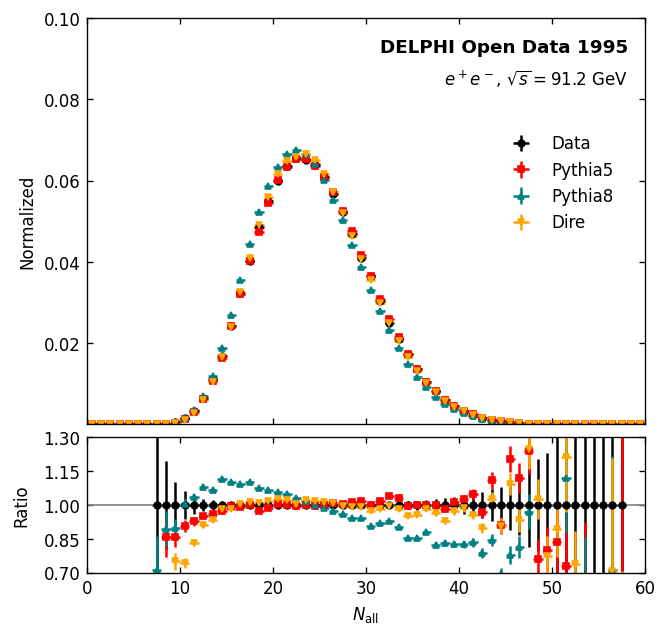}
        \centering
        \text{(d)}\\
    \end{minipage}\\[0.5em]
    \caption{Comparisons of the (a) 1994 charged track multiplicity, (b) 1994 all particle multiplicity, (c) 1995 charged track multiplicity, and (b) 1995 all particle multiplicity data (black), reconstructed PYTHIA~5.7/JETSET~7.4 (red), ARIADNE (green), PYTHIA 8.3 (orange), and PYTHIA 8.3 Dire (brown) samples. Note that the corresponding number of selected events of each sample normalizes the distributions.}
    \label{fig:MCdatacomp3}
\end{figure}

\begin{figure}[ht!]
    \centering
    \includegraphics[width = 0.45\textwidth]{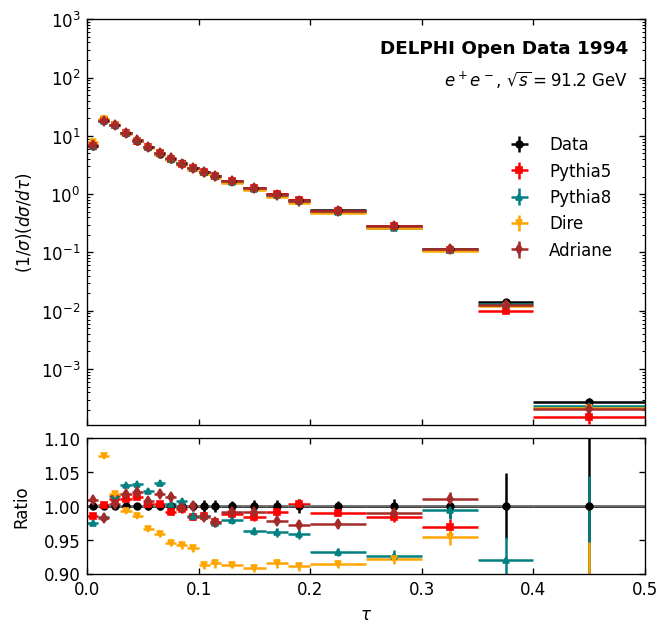}
    \includegraphics[width = 0.45\textwidth]{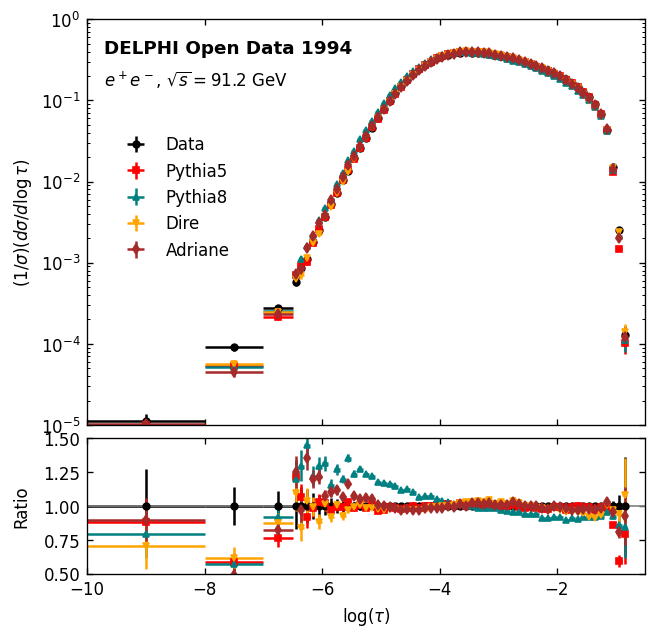}
    \caption{Comparisons of the thrust variables $\tau$ (left) and $\log\tau$ (right) in 1994 data (black), reconstructed PYTHIA~5.7/JETSET~7.4 (red), ARIADNE (green), PYTHIA 8.3 (orange), and PYTHIA 8.3 Dire (brown) samples. }
    \label{fig:MCdatacomp4}
\end{figure}

\begin{figure}[ht!]
    \centering
    \includegraphics[width = 0.45\textwidth]{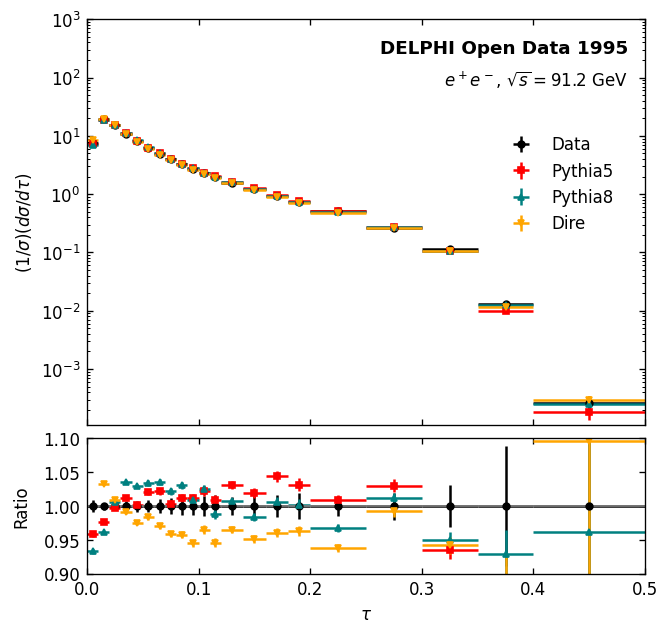}
    \includegraphics[width = 0.45\textwidth]{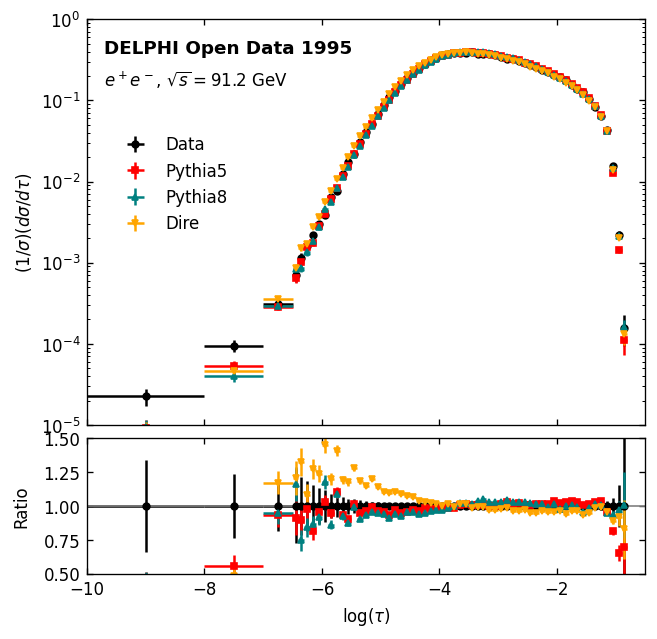}
    \caption{Comparisons of the thrust variables $\tau$ (left) and $\log\tau$ (right) in 1995 data (black), reconstructed PYTHIA~5.7/JETSET~7.4 (red), ARIADNE (green), PYTHIA 8.3 (orange), and PYTHIA 8.3 Dire (brown) samples. }
    \label{fig:MCdatacomp495}
\end{figure}

\begin{figure}[ht!]
    \centering
    \includegraphics[width = 0.6\textwidth]{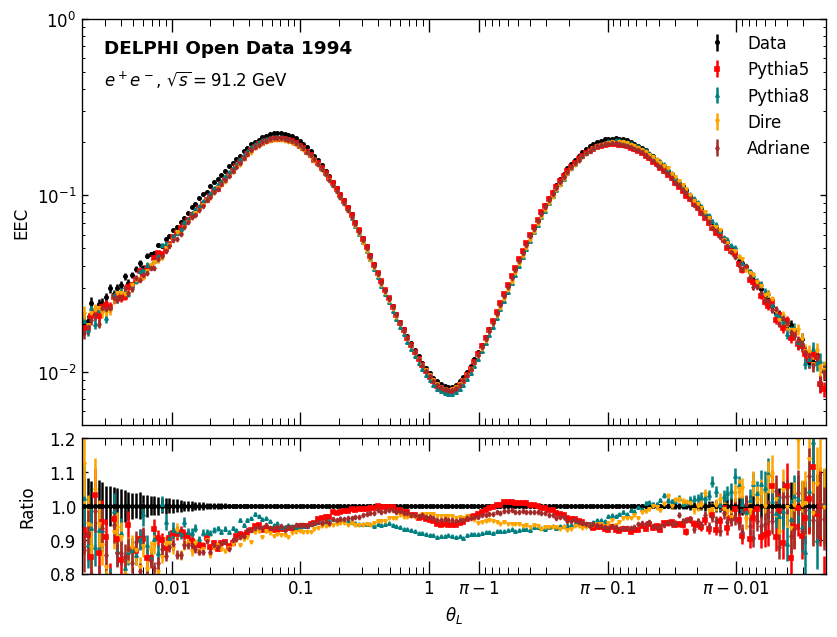}
    \includegraphics[width = 0.6\textwidth]{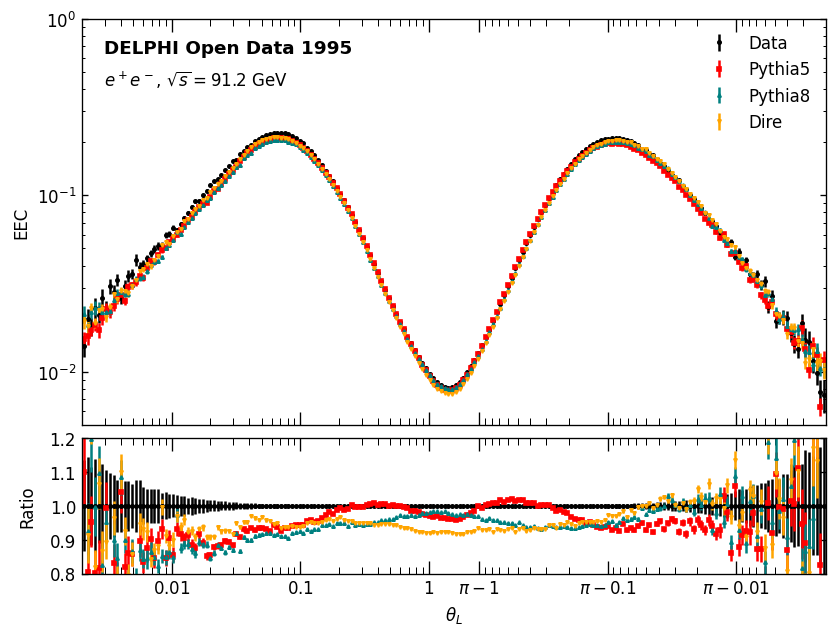}
    \caption{Comparisons of the EEC in data (black), reconstructed PYTHIA~5.7/JETSET~7.4 (red), ARIADNE (green), PYTHIA 8.3 (orange), and PYTHIA 8.3 Dire (brown) samples for 1994 (left) and 1995 (right). }
    \label{fig:MCdatacomp5}
\end{figure}

The event multiplicity and shape variables are also compared. The charged and total particle multiplicities for both years are shown in Figure~\ref {fig:MCdatacomp3}. Comparisons of the thrust variables, $\tau (= 1-\rm T)$ and $\log \tau$, are shown for 1994 and 1995 in Figs.\ref{fig:MCdatacomp4} and\ref{fig:MCdatacomp495}. Finally, the EEC distributions are shown in Figure~\ref{fig:MCdatacomp5}. For all distributions, each sample is normalized to its own number of selected events.

For the charged track $p_T$ distribution, PYTHIA~5 and ARIADNE show good agreement with the data at low momentum, while PYTHIA~8 performs better in the intermediate range of $10 < p_T < 30$ GeV. A notable discrepancy is observed in the high-$p_T$ tail ($p_T > 30$ GeV), where all MC models underestimate the data. This could originate from the modeling of intrinsic parton transverse momentum at the generator level or effects like fake tracks in the reconstructed data sample. However, as demonstrated in Section~\ref{sec:trackTail}, the impact of this high-$p_T$ discrepancy on the final EEC and thrust measurements is found to be minimal. 

The neutral particle energy spectrum has two characteristic shapes: a steeply falling distribution from physical processes below 50 GeV and a long tail attributed to calorimeter noise. \textbf{While the DELPHI simulation models this noise tail reasonably well, an additional upper cut of 50 GeV is placed on the neutral energy to explicitly exclude this region from this analysis.} 
The impact of this cut on the final result is demonstrated to be negligible (less than 0.1\%). 
At very low energy, the level of agreement varies between generators and data-taking years. These inconsistencies are carefully evaluated as systematic uncertainties in Section~\ref{sec:simUncert}.

For charged track multiplicities, most samples describe the data well, with the exception of PYTHIA8, which underestimates the distribution. For the total particle multiplicity, PYTHIA 5 agrees well with the 1995 data but slightly underestimates the 1994 data. Regarding the event shapes, most generators agree reasonably well with the thrust data, while for EEC, PYTHIA 5 and ARIADNE better describe the perturbative region, and the PYTHIA 8 samples better capture the Sudakov region. The level of agreement for any given observable is an entangled effect of the generator-level physics and the detector-level simulation. Overall, the PYTHIA 5 sample provides the best agreement with data, while the PYTHIA 8 Dire sample consistently fails to describe the data across most variables.

\textbf{Therefore, the PYTHIA5.7 sample will be used to derive the nominal unfolding and correction factors. The ARIADNE, PYTHIA 8, and PYTHIA~8 Dire samples will be used to evaluate systematic uncertainties related to the physics modeling.}

\clearpage
\section{EEC analysis}
\label{sec:analysis}
\begin{figure}[ht!]
    \centering
    \includegraphics[width=\linewidth]{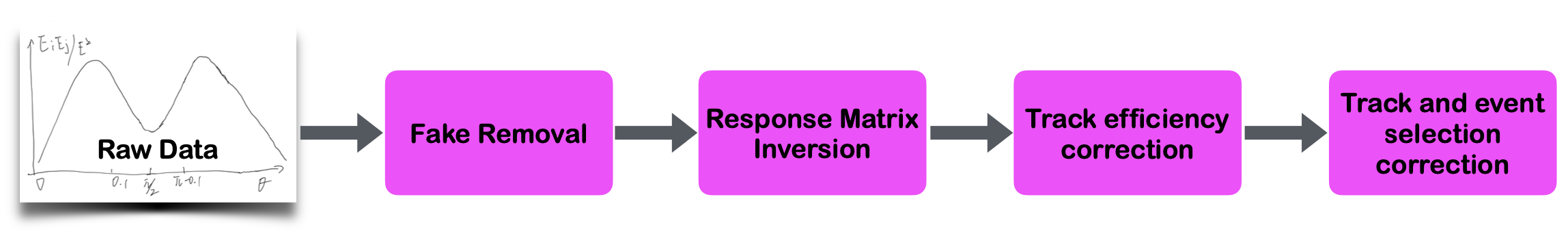}
    \caption{Overview of the various steps of the analysis in their sequential order.}
    \label{fig:analysisOutline}
\end{figure}

An outline of the analysis flow is shown in Figure \ref{fig:analysisOutline}. This outline covers all parts of the procedure following event selection (discussed in Section \ref{sec:sample}). The systematic uncertainty determinations are also discussed in Section \ref{sec:syst}. 

For the EEC analysis, a procedure based on per-track pair-level 2D unfolding is performed as follows. The first step is to remove the contribution to the EEC resulting from reconstructed particle pairs lacking a matched generator-level pair (herein referred to as "fake"). The matching procedure will be described in Section~\ref{sec:matching}. The matched track pairs are then used to build the two-dimensional response matrix in angles (here, $\theta_{\rm L}$ or $z$, depending on the choice of variable) and the energy weights ($\rm E_{\rm 1}E_{\rm 2}/E^{2}$). The response matrix is described in Section~\ref{sec:response}. The EEC distribution is then unfolded using the response matrix. The matching procedure also gives generator pairs that lack a matched reconstructed-level pair (herein referred to as "inefficiency"). A correction of this effect is applied to the distribution unfolded with the response matrix. The whole unfolding procedure, including fake subtraction, unfolding with response matrix, and efficiency correction, is described in Section~\ref{sec:unfoldingresults}. To properly handle the strong bin-to-bin correlations inherent to the EEC, a full covariance matrix is constructed in Section~\ref{sec:cov}. The output after unfolding is a two-dimensional histogram in angle and energy, which must then be projected onto the angular axis in order to construct the final observable. This process is discussed in Section \ref{sec:projection}. Finally, a correction on the resulting distribution will be performed in order to correct the observable to the so-called ``full phase-space" so that it can be compared to MC and analytic predictions. These corrections will be discussed in Section \ref{sec:acccorr}.

For the sake of clarity and conciseness in the main text, the figures presented in this section are primarily based on the 1994 dataset. The 1995 analysis uses the same procedure, and the comprehensive set of corresponding plots for the 1995 analysis is provided in Appendix~\ref{app:1995EEC}.

\subsection{Unfolding}
\label{sec:unfolding}
Experimental measurements are inherently subject to detector effects, which cause the reconstructed ``rec'') observables to deviate from their truth (``gen'') values. This smearing, driven by the finite resolution of the detector (as illustrated in Figs.~\ref{fig:theta_res}, \ref{fig:phi_res}, and \ref{fig:phi_pt}), must be corrected for using a procedure known as unfolding. In this analysis, two quantities can be smeared: the opening angle $\theta_{\rm L}$ or $z$ and the energy weighting of the particles $E_1 E_2 / E^2$. Therefore, a 2-dimensional unfolding procedure is used to correct the measured EEC distribution. Note that one large advantage of measuring the EEC distributions in $e^{+}e^{-}$ collisions is that the exact momentum transfer is known, due to the colliding objects being fundamental particles. In the case of measuring the EEC in hadronic collision systems, one must also independently correct for the smearing in the total momentum transfer, increasing the difficulty of the unfolding problem. The remainder of this section is organized as follows: In Section~\ref{sec:response}, the binning and procedure used to form the response matrix are described. In Section~\ref{sec:unfoldingresults}, the results of the unfolded procedure are presented.

The binning utilized for the angular axis (either $z$ or $\theta_{\rm L}$) uses a ``double log'' style where 200 variable bins are chosen to be evenly spaced on a log scale: 100 log bins ranging from $\theta_{\rm L} = 0.002$ to $\theta_{\rm L} = \pi/2$ and 100 ``flipped'' log-style bins ranging from $\theta_{\rm L} = \pi - 0.002$ to $\theta_{\rm L} = \pi/2$. The binning for the \(z\) parameterization is defined by a direct analytical conversion of these \(\theta_L\) bin boundaries using Equation~\ref{eq:z}. As a result, the relative uncertainties are identical for both parameterizations. \textbf{Although the unfolding is performed independently for the \(\theta_L\) and \(z\) distributions, for conciseness, only the \(\theta_L\) results will be shown in the following sub-sections.}

\subsubsection{Matching and track reconstruction performance}
\label{sec:matching}
In the matching procedure, a correspondence is built between the reconstructed and generator-level information, which will then be used to correct the measured distribution from data. 

The single track matching is performed via the Hungarian Method~\cite{hungarianMatching}, which is a standard algorithm for solving assignment problems in polynomial time. Hungarian maximization is most easily explained using a matrix formulation where the $i$th row and the $j$th column represent the cost of assigning $j$ to $i$, and the algorithm uses this structure to make a unique assignment that minimizes the total cost. Therefore, a central component of the Hungarian method is the metric for determining the cost of a given entry, herein referred to as the matching metric. There are a variety of different choices of metrics that may be applied. In this case, the angle between two tracks is used as the default matching metric, and a cutoff of 0.05 is imposed so that no pair separated by $\theta > 0.05$ can be matched. The dependence of the results on this particular choice of matching scheme and cutoff value is accounted for as a systematic uncertainty. See Section~\ref{sec:matchingUncert} for more details. 

\begin{figure}[ht!]
    \centering
    \includegraphics[width=0.4\linewidth]{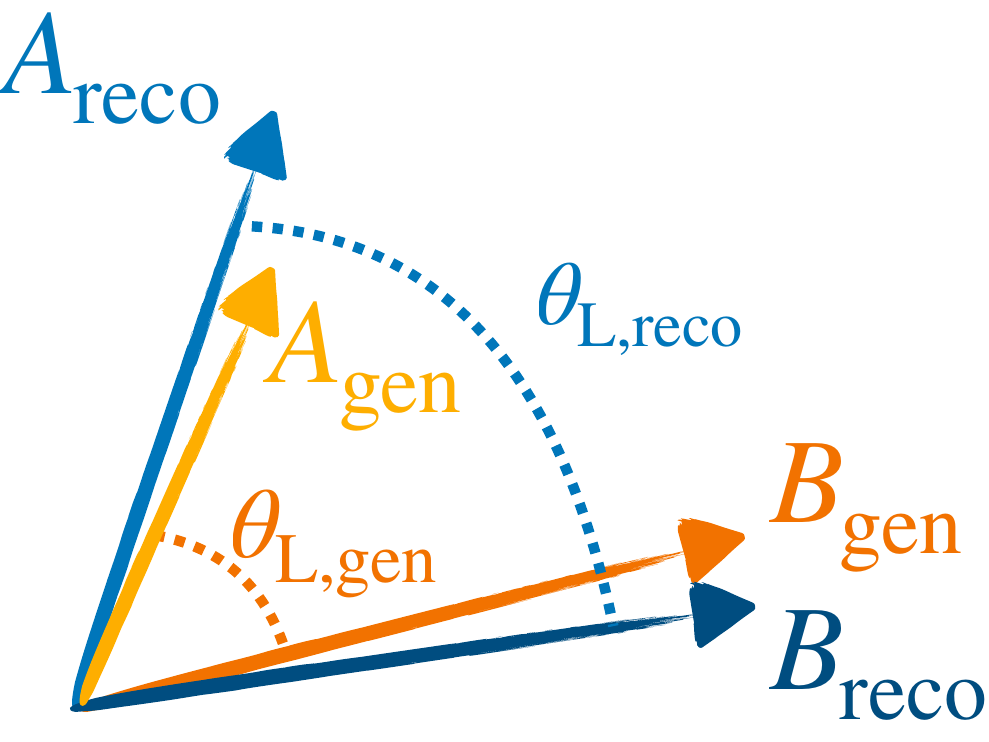}
    \caption{Cartoon visualization of single track and pairwise matching. Tracks shown in blue are considered to be at the reconstructed (reco) level, and tracks shown in orange are considered to be at the generator (gen) level.}
    \label{fig:matchedPairCartoon}
\end{figure}

The result of the single-track matching is used to build the match at the pair level. The conceptual distinction between single track and pair matching can be found in Figure~\ref{fig:matchedPairCartoon}. A track pair is matched when both tracks in the pair are matched. If at least one reco-level track is unmatched, the reco-level pair is considered pair-wise ``fake''. If at least one gen-level track is unmatched, the gen-level pair is included in pair-wise ``inefficiency''. 

\begin{figure}[ht!]
    \centering
    \includegraphics[width=0.45\linewidth]{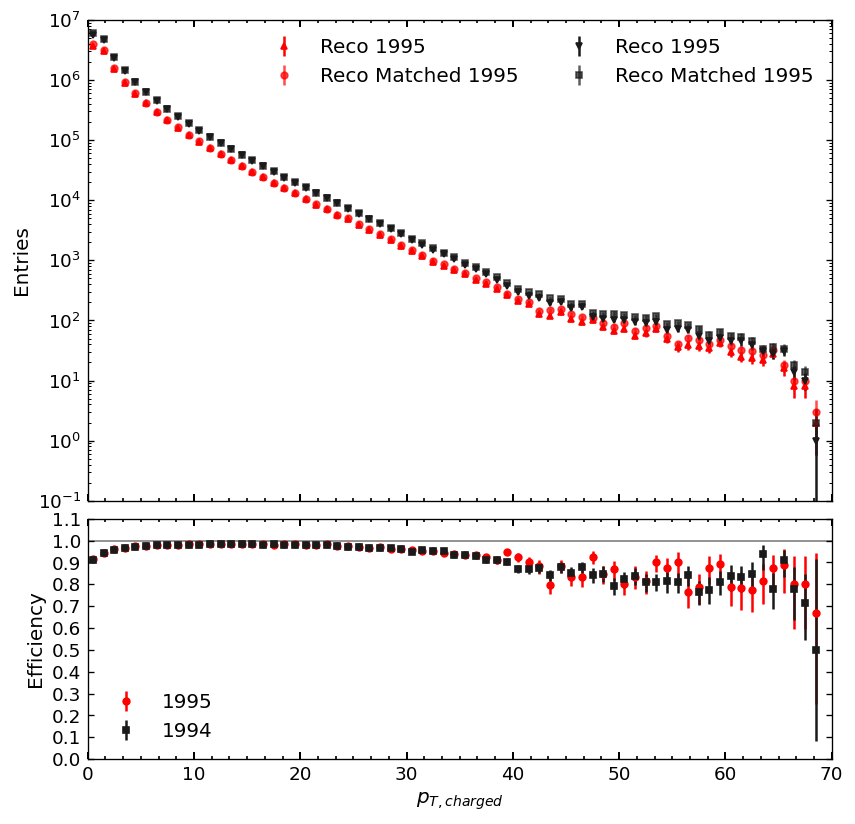}
    \includegraphics[width=0.45\linewidth]{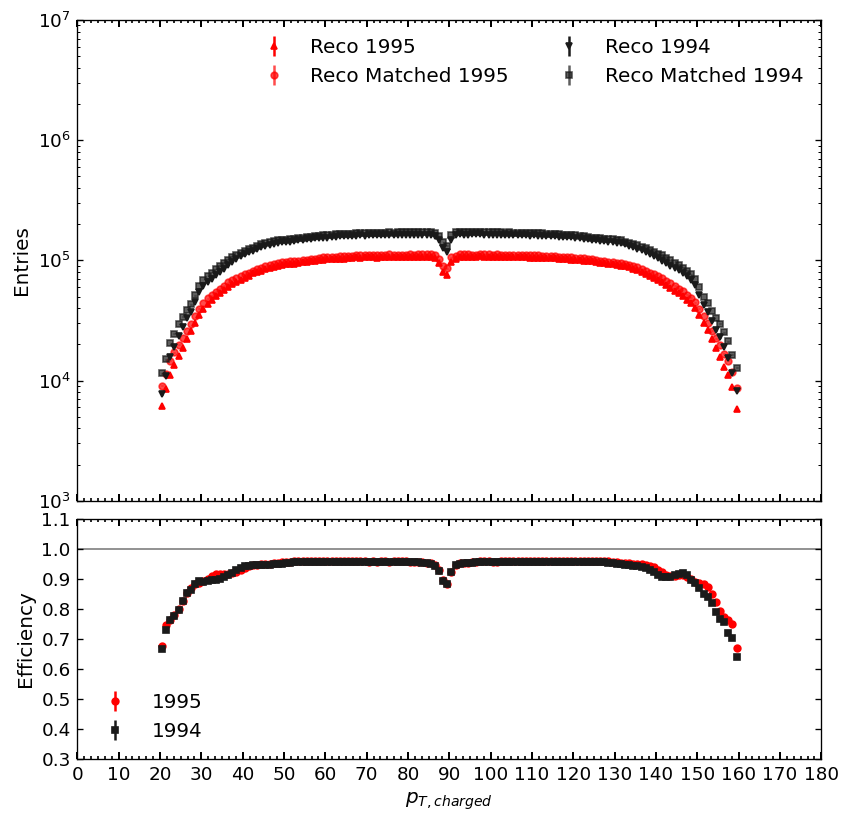}
    \caption{Single track fake rate from PYTHIA 5 MC as a function of reconstructed $p_{\rm T}$ (left) and $\theta$ (right) for 1994 (black) and 1995 (red) samples, respectively. The upper panels show the distributions of matched (triangle markers) and all (circular and square markers) detector-level tracks, while the lower panels show the fake rate ratio.}
    \label{fig:trk_fake}
\end{figure}

Single-track matching performance is shown below. In Figure~\ref{fig:trk_eff}, matched generator-level track $p_{\rm T}$ (left) and $\theta$ (right) are compared to all generator-level tracks. The ratio, shown in the lower panel, indicates the matching efficiency. In this figure, results from the 1994 and 1995 PYTHIA 5 samples are shown in red and black, respectively. In the upper panel, triangle markers are for generator-level matched tracks, and circular and square markers are for all generator-level tracks. 
Similarly, in Figure~\ref{fig:trk_fake}, matched reconstruction-level track $p_{\rm T}$ (left) and $\theta$ (right) are compared to all reconstruction-level tracks for 1994 (red) and 1995 (black) samples. The ratio, shown in the lower panel, indicates the fake rate. 
In the upper panel, triangle markers are for detector-level matched tracks, and circular and square markers are for all detector-level tracks. 

As shown, the effects of the fake tracks and inefficiencies are very similar in the 1994 and 1995 samples, and are concentrated in the region near $\theta=90^\circ$. This degraded performance originates from the cathode membrane structure of the TPC and is accounted for in the unfolding procedure through fake subtraction and efficiency correction.

\begin{figure}[ht!]
    \centering
    \includegraphics[width=0.45\linewidth]{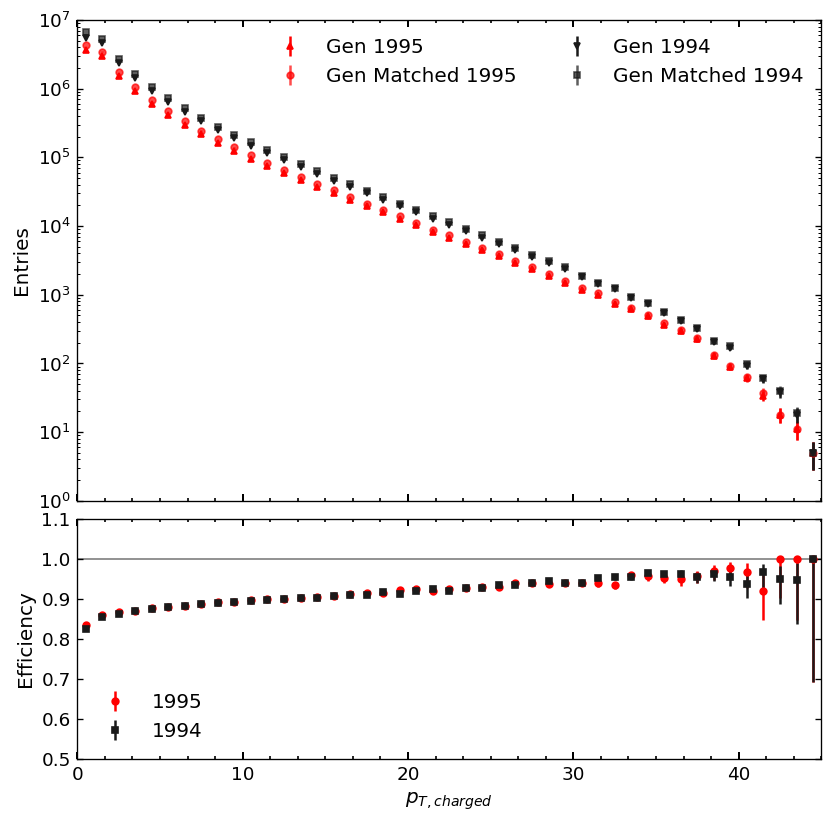}
    \includegraphics[width=0.45\linewidth]{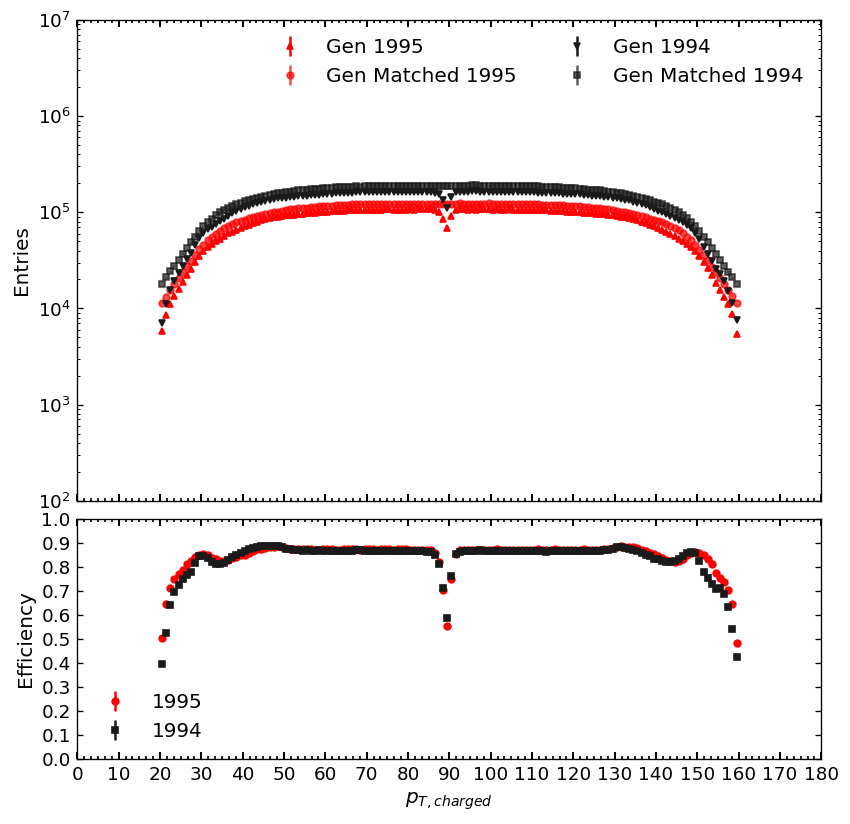}
    \caption{Single track matching efficiency from PYTHIA 5 MC as a function of generator-level $p_{\rm T}$ (left) and $\theta$ (right) for 1994 (black) and 1995 (red) samples, respectively. The upper panels show the distributions of matched (triangle markers) and all (circular and square markers) generator-level tracks, while the lower panels show the efficiency ratio.}
    \label{fig:trk_eff}
\end{figure}

\begin{figure}[ht!]
    \centering
    \includegraphics[width=\linewidth]{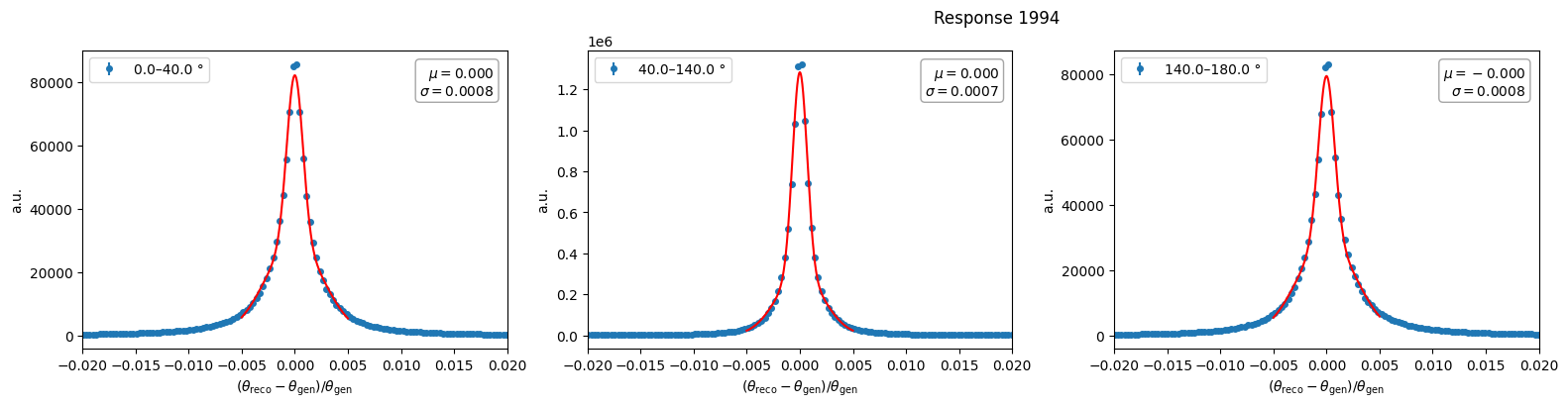}
    \caption{Track $\theta$ resolution distributions from PYTHIA 5 MC in different detector regions: forward endcap (left), barrel (middle), and backward endcap (right). The $\theta$ response is calculated as $(\theta_{\rm reco} - \theta_{\rm gen}) / \theta_{\rm gen}$. Double Gaussian fits are overlaid to extract the resolution parameters. The figures here are for the 1994 sample only; the track angular resolution from the 1995 sample, shown in Appendix~\ref{app:1995}, has a similar level of performance. }
    \label{fig:theta_res}
\end{figure}

\begin{figure}[ht!]
    \centering
    \includegraphics[width=\linewidth]{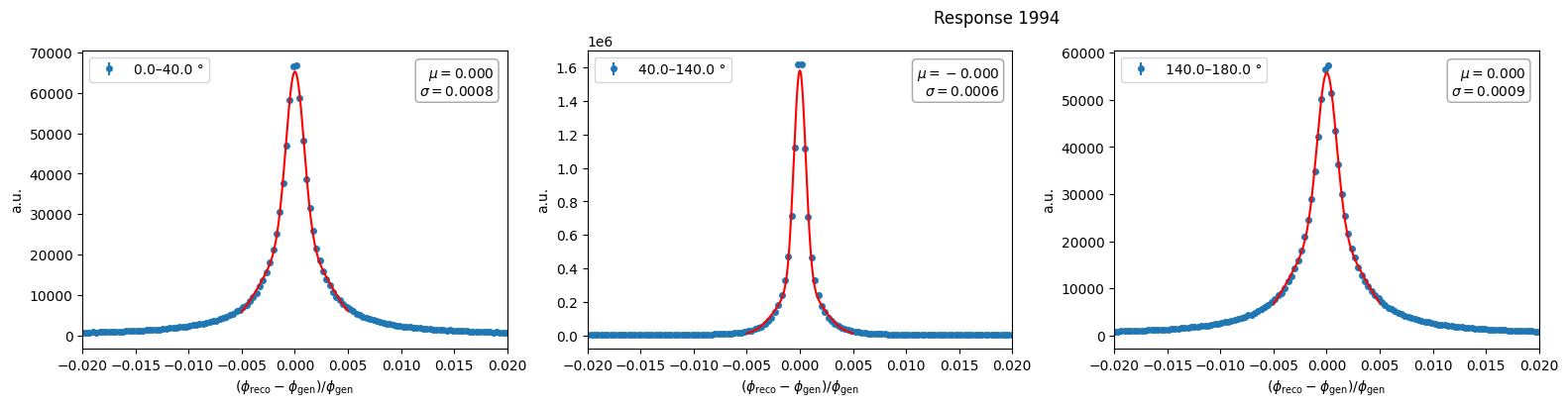}
    \caption{Track $\phi$ resolution distributions from PYTHIA 5 MC in different detector regions: forward endcap (left), barrel (middle), and backward endcap (right). The $\phi$ response is calculated as $(\phi_{\rm reco} - \phi_{\rm gen}) / \phi_{\rm gen}$. Double Gaussian fits are overlaid to extract the resolution parameters. The figures here are for the 1994 sample only; the track angular resolution from the 1995 sample, shown in Appendix~\ref{app:1995}, has a similar level of performance.}
    \label{fig:phi_res}
\end{figure}

\begin{figure}[ht!]
    \centering
    \includegraphics[width=\linewidth]{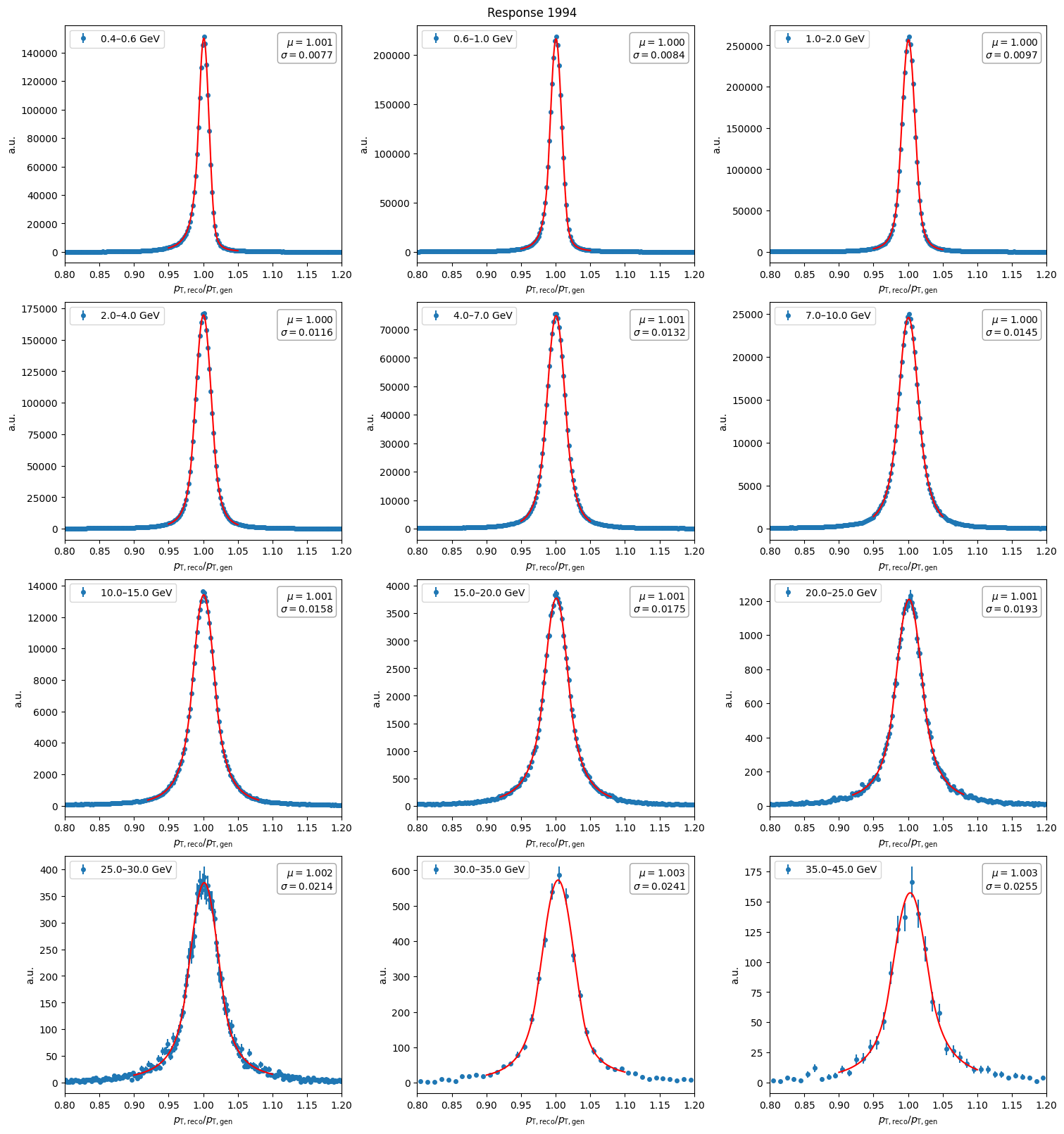}
    \caption{Track $p_{\rm T}$ resolution distributions from PYTHIA 5 MC in different $p_{\rm T}$ ranges. The $p_{\rm T}$ response is calculated as $p_{\rm T, reco} / p_{\rm T, gen}$. Double-sided crystal ball fits are overlaid to extract the response and resolution parameters. The figures here are for the 1994 sample only; the track momentum resolution from the 1995 sample, shown in Appendix~\ref{app:1995}, has a similar level of performance.}
    \label{fig:phi_pt}
\end{figure}

The track momentum and angular resolution are also derived using the matched tracks. Figure~\ref{fig:theta_res} and Figure~\ref{fig:phi_res} show the responses of $\theta$ and $\phi$ in the barrel (middle) and the two endcaps (left and right). The angular response is calculated as $(\theta_{\rm reco} - \theta_{\rm gen}) / \theta_{\rm gen}$ and $(\phi_{\rm reco} - \phi_{\rm gen}) / \phi_{\rm gen}$, respectively. Figure~\ref{fig:phi_pt} shows the track $p_{\rm T}$ response in various $p_{\rm T}$ bins. The $p_{\rm T}$ response is calculated as $p_{\rm T, reco} / p_{\rm T, gen}$. Simple fits with double Gaussian and double-sided crystal ball functions are performed on the distributions to extract the angular and momentum responses and resolutions. The mean of the response distributions, $\mu$, is consistent with zero for the angular variables and one for the momentum variables. This indicates that there is no significant scale shift in the reconstruction of track parameters between the detector level and the generator level. The angular and momentum resolutions ($\sigma$) are in the range of 0.0006 to 0.0008 and 0.005 to 0.02, respectively. These values are in good agreement with the performance benchmarks originally documented by the DELPHI Collaboration in Ref.~\cite{DELPHI:1995dsm}. Note that the figures here are for the 1994 sample only. The track angular and momentum resolutions from the 1995 sample, shown in Appendix~\ref{app:1995}, have a similar level of performance.  

\clearpage

\subsubsection{Response matrix}
\label{sec:response}
The unfolding procedure relies on a correspondence between the measured and truth-level quantities. The response matrix is what captures this correspondence, created by utilizing a mapping between the generator and reconstructed-level MC that is then inverted to correct the measured data. Here, the response matrix is formed with the PYTHIA 6 MC, where the correspondence between the detector- and generator-level distributions is formed using the matching procedure described in Section~\ref{sec:matching}. When filling the response matrix, pairs in a given event are matched, and then the bin that corresponds to the reconstructed and generator-level energy and angle is filled. 

The energy binning utilized in the $\rm E_i E_j / E^2$ axis of the response matrix is provided in Table~\ref{tab:energyBinning}. The choice of binning represents a balance between two competing factors. While a finer binning would, in principle, further reduce the small non-closure effects in the 2D distribution projection (see Section~\ref{sec:projection}), the current binning scheme is found to be optimal.
With the chosen bins, the correction factors and their associated systematic uncertainties are already subdominant, indicating that the binning is sufficiently fine. Adopting a finer binning would offer little improvement in physics precision while substantially increasing computational resource requirements. Therefore, the current binning is used for the final analysis.

\begin{table}[ht!]
    \centering
    \caption{Bin boundaries for the energy weight axis, \(E_i E_j / E_{\text{vis}}^2\), used in the unfolding procedure.}
    \label{tab:energyBinning}
    \begin{tabular}{c}
        \toprule
        Energy Weight Bin Boundaries \\
        \midrule
        \texttt{0.00000, 0.00010, 0.00013, 0.00016, 0.00020, 0.00025} \\
        \texttt{0.00032, 0.00040, 0.00050, 0.00063, 0.00079, 0.00100} \\
        \texttt{0.00126, 0.00158, 0.00200, 0.00251, 0.00316, 0.00398} \\
        \texttt{0.00501, 0.00631, 0.00794, 0.01000, 0.01259, 0.01585} \\
        \texttt{0.02000, 0.02512, 0.03162, 0.03981, 0.05012, 0.06310} \\
        \texttt{0.07943, 0.10000, 1.00000} \\
        \bottomrule
    \end{tabular}
\end{table}

\begin{figure}[ht!]
    \centering
    \includegraphics[width=0.95\linewidth]{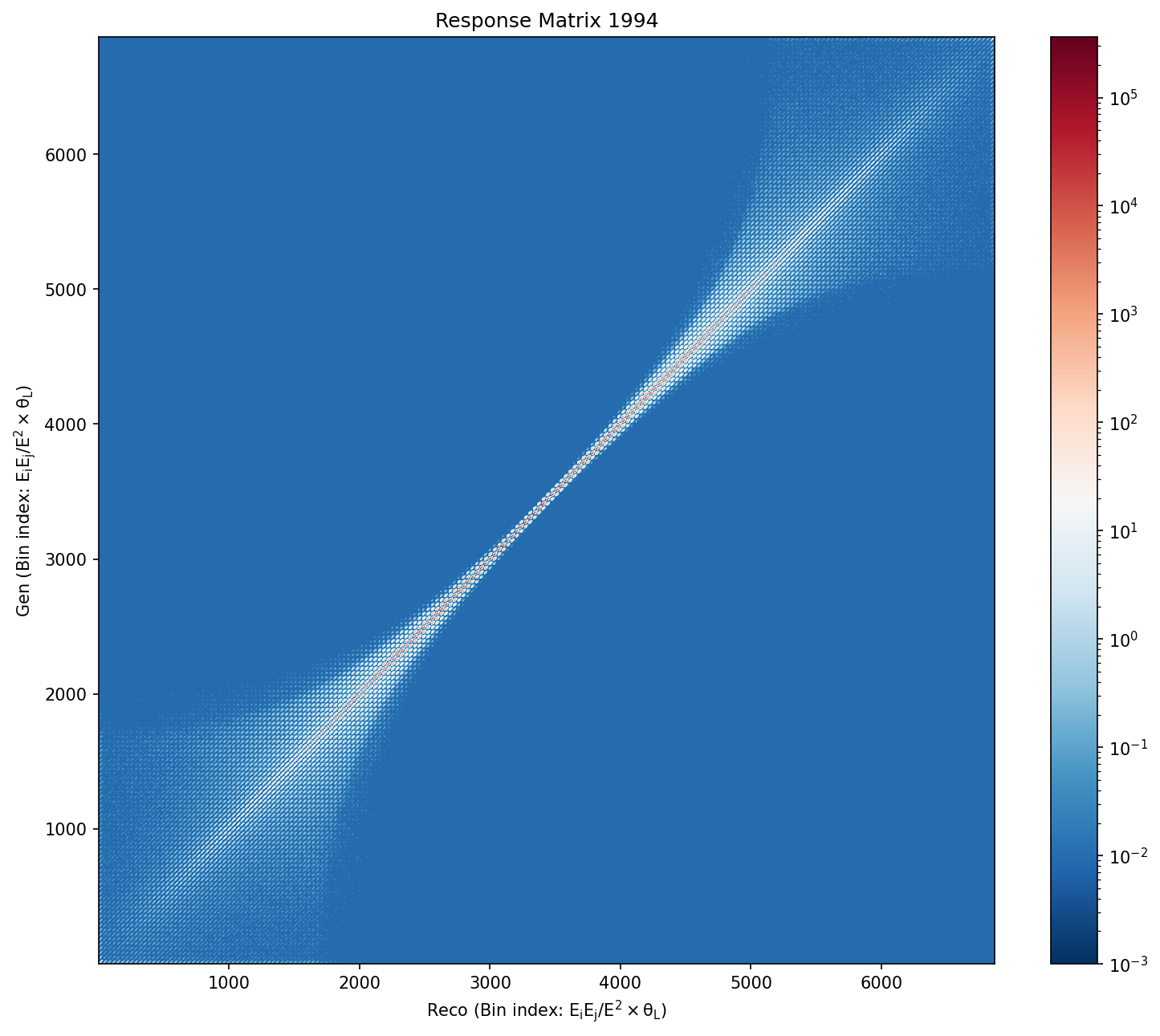}
    \caption{Unnormalized response matrix for EEC unfolding showing the correspondence between reconstructed (x-axis) and generator-level (y-axis) distributions for the $\theta_{\rm L}$ parametrization in 1994. Here, the 2D response matrix is visualized by structuring it so that each $\theta_{\rm L}$ bin forms a block containing all energy dimension correspondences. The strong correlation along the diagonal indicates good detector resolution.}
    \label{fig:response}
\end{figure}

The (unnormalized) response matrix for 1994 is shown in Fig.~\ref{fig:response}. Here, the 2D response matrix (4D matrix) is visualized by flattening (structuring) so that each $\theta_{\rm L}$ bin forms a block containing all energy dimension correspondences. The size of the off-diagonal entries compared to the diagonal entries quantifies the magnitude of the smearing effect. As can be seen, despite some deterioration in the very small and very large $\theta_{\rm L}$ regions, the smearing effect is very small given the excellent detector built by the DELPHI collaboration.

\clearpage
\subsubsection{Unfolded distributions}
\label{sec:unfoldingresults}
As mentioned before, the unmatched detector-level pairs are treated as "fakes". 
\begin{equation}\label{eq:fakefraction}
    f(\theta_{\rm L, reco} \; \text{or} \; z) = \text{EEC}_{\rm all \; pair}(\theta_{\rm L, reco} \; \text{or} \; z) - \text{EEC}_{\rm matched\; pairs}(\theta_{\rm L, reco} \; \text{or} \; z). 
\end{equation}
The unmatched generator-level pairs are treated as inefficiencies.
\begin{equation}\label{eq:matcheff}
    \epsilon(\theta_{\rm L, gen} \; \text{or} \; z) = \frac{\text{EEC}_{\rm matched \; pair}(\theta_{\rm L, gen} \; \text{or} \; z)}{\text{EEC}_{\rm all\; pairs}(\theta_{\rm L, gen} \; \text{or} \; z)}
\end{equation}
In the unfolding implementation below, fakes are first subtracted from the data before the response matrix unfolding, and inefficiencies are accounted for as a multiplicative efficiency correction after the unfolding with the response matrix. 

The unfolding with response matrix is done using the D'Agostini iterative method~\cite{D'Agostini:265717} in the \texttt{RooUnfold}~\cite{Brenner:2019lmf} package. 
The D'Agostini method uses early stopping to regulate the smoothness of the unfolded distributions. 
The regularization parameter, defined by the number of iterations $N_{\rm iter}$, is determined as follows: after each iteration, the $\chi^2/\rm ndf$ between the distributions from the current and previous iterations is computed.
If the $\chi^2$ falls below 0.05, the current iteration is chosen as the optimal number of iterations.
Based on this criterion, $N_{\rm iter}$ is set to 4 in this analysis. 

For 2D unfolding, the unfolded 2-dimensional $E_{i}E_{j}/E^{2}$ and $\theta_{\rm L}$ histogram is projected to one dimension taking the bin centers in each $E_{i}E_{j}/E^{2}$ bin as the approximation for the bin averages. 
The potential bias created by the projection procedure will be corrected as described in Section~\ref{sec:projection}.
A closure test comparing the unfolded and generated distributions using the same $\textsc{pythia}$ 5 MC is performed. 
Close to perfect closure is obtained.

\begin{figure}[ht!]
    \centering
    \includegraphics[width=0.6\linewidth]{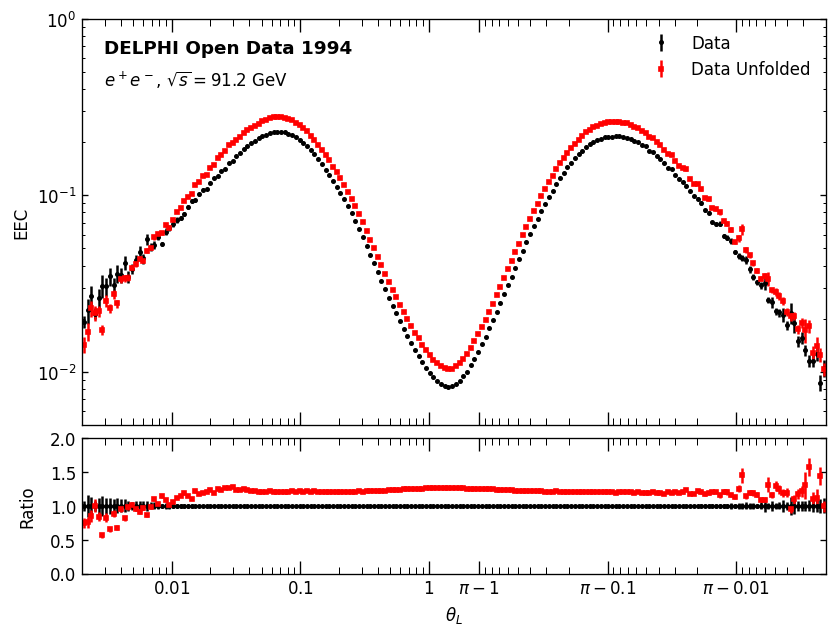}
    \caption{Comparison of the unfolded (red) and raw (black) 1994 data distributions. The unfolded distribution shown here includes fake subtraction and efficiency correction.}
    \label{fig:unfoldEEC}
\end{figure}

The impact of the full correction procedure on the 1994 data is shown in Figure~\ref{fig:unfoldEEC}, which compares the final unfolded distribution to the initial raw data.
The unfolded distribution shown here includes fake subtraction and efficiency correction.
A large shape correction is evident at small opening angles ($\theta_{\rm L} < 0.03$), while a uniform normalization correction of approximately 17\% is applied across the rest of the spectrum.

\begin{figure}[H]
    \centering
    \includegraphics[width=0.5\linewidth]{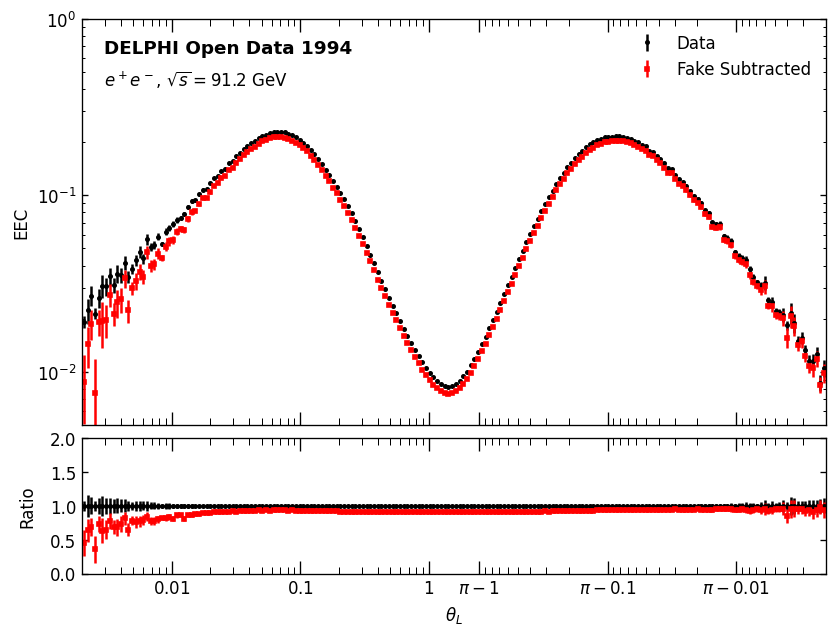}
    \caption{Comparison of the response matrix  unfolded (red) and fake-subtracted (black) 1994 data distributions.}
    \label{fig:fakeCorrUnf}
\end{figure}

\begin{figure}[H]
    \centering
    \includegraphics[width=0.5\linewidth]{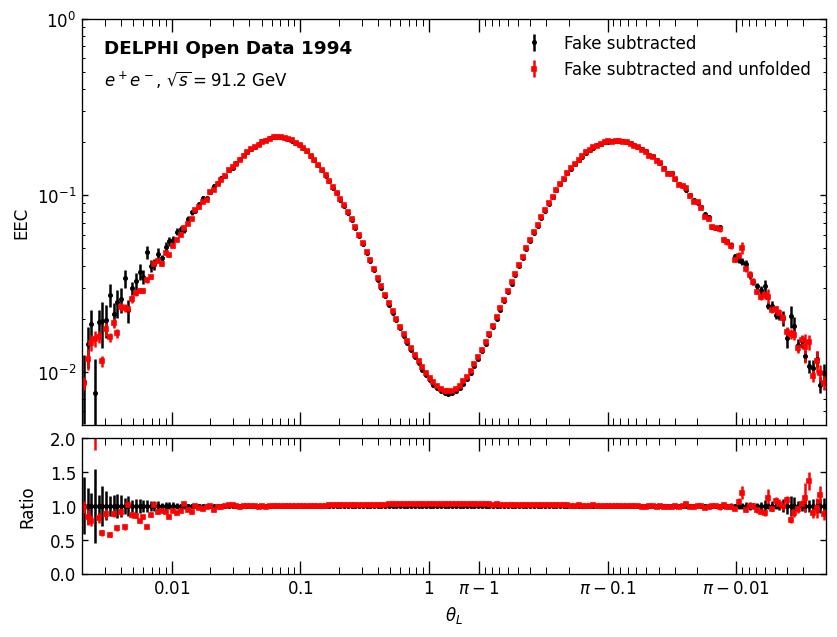}
    \caption{Comparison of the efficiency corrected (red) and response matrix unfolded (black) 1994 data distributions.}
    \label{fig:unfoldCorrUnf}
\end{figure}

\begin{figure}[H]
    \centering
    \includegraphics[width=0.5\linewidth]{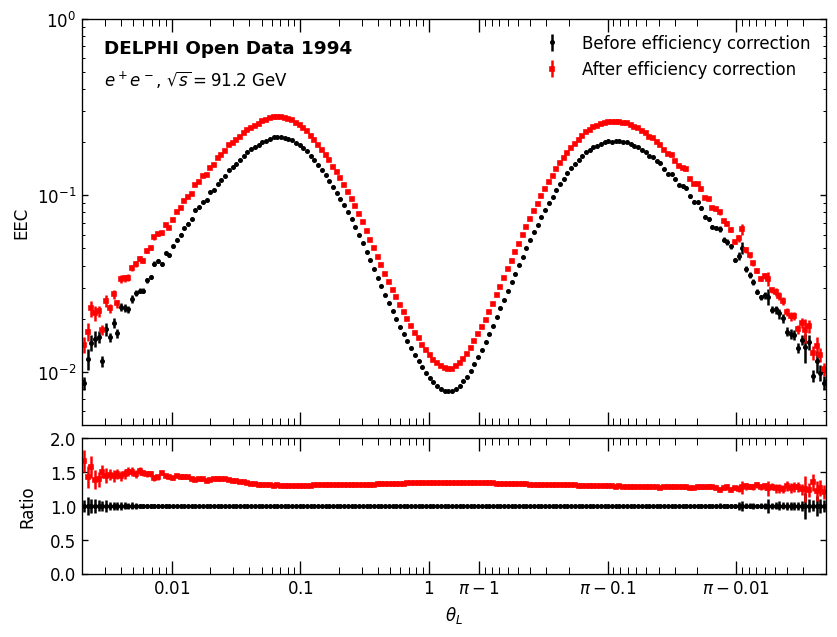}
    \caption{Comparison of the unfolded (red) and raw (black) 1994 data distributions.}
    \label{fig:effCorrUnf}
\end{figure}

To disentangle the contributions from each stage of the correction, the effects are shown sequentially in Figures~\ref{fig:fakeCorrUnf}, \ref{fig:unfoldCorrUnf}, and \ref{fig:effCorrUnf}.
\begin{enumerate}
    \item \textbf{Fake Subtraction (Fig.~\ref{fig:fakeCorrUnf}):} the raw data is compared with the distribution after fake subtraction.

    \item \textbf{Unfolding for Resolution Smearing (Fig.~\ref{fig:unfoldCorrUnf}):} the fake-subtracted data distribution is compared to the distribution after unfolding with the response matrix.

    \item \textbf{Efficiency Correction (Fig.~\ref{fig:effCorrUnf}):} the distribution after unfolding with the response matrix is compared to the distribution after efficiency correction.
\end{enumerate}

As implied by these plots, the large corrections at small $\theta_{\rm L}$ are primarily due to degraded track reconstruction performance when two tracks are very close. This proximity can lead to track merging, resulting in inefficiencies, or split tracks, which generate fakes. A distinct feature in the detector response is visible around $\theta = 90^\circ$ (polar angle), corresponding to a known inefficiency in the TPC at the cathode plane as shown in Figure~\ref{fig:trk_eff}.
The final correction for reconstruction efficiency accounts for the majority of the overall 17\% normalization shift.

It is worth noting that in Figure~\ref{fig:unfoldCorrUnf}, the statistical uncertainties in the small-angle region appear to decrease after unfolding. This can be an artifact of the regularization used in the unfolding algorithm, which can introduce a bias in exchange for stabilizing the result. The potential effect for such a bias is quantified and included as a systematic uncertainty, as detailed in Section~\ref{sec:unfoldingUncert}. Other than the very small opening angle region ($\theta_{\rm L} < 0.02$), statistical uncertainties slightly increase after unfolding. 

\subsection{Covariance matrix}
\label{sec:cov}
The statistical uncertainty and covariance matrix of the EEC distribution are non-trivial and cannot be estimated by simple Poissonian statistics. The EEC distribution is constructed from a weighted sum of particle pairs, and a single event can contribute entries to multiple bins, inducing significant bin-to-bin correlations. To determine these uncertainties robustly, a full covariance matrix was calculated by treating each of the events as an independent statistical sample. The covariance for the EEC distribution is then calculated using the unbiased sample covariance formula:
\begin{equation}\label{eq:ENC}
V_y = \frac{1}{N(N-1)}(\sum_{i=1}^{N_{evt}} \vec{y}_{evt}^{(i)} \left(\vec{y}_{evt}^{(i)}\right)^T-\frac{1}{N}\sum_{i=1}^{N_{evt}}\vec{y}_{evt}^{(i)}(\sum_{i=1}^{N_{evt}}\vec{y}_{evt}^{(i)})^T)
\end{equation}
in which $\vec{y}_{evt}^{(i)}$ represents the single event EEC and $\sum_{i=1}^{N_{evt}}\vec{y}_{evt}^{(i)}$ represents unnormalized, event-summed EEC. The derivation of the equation can be seen in Appendix~\ref{app:eq5.1}. 

\begin{figure}[ht!]
    \centering
    \includegraphics[width=0.95\linewidth]{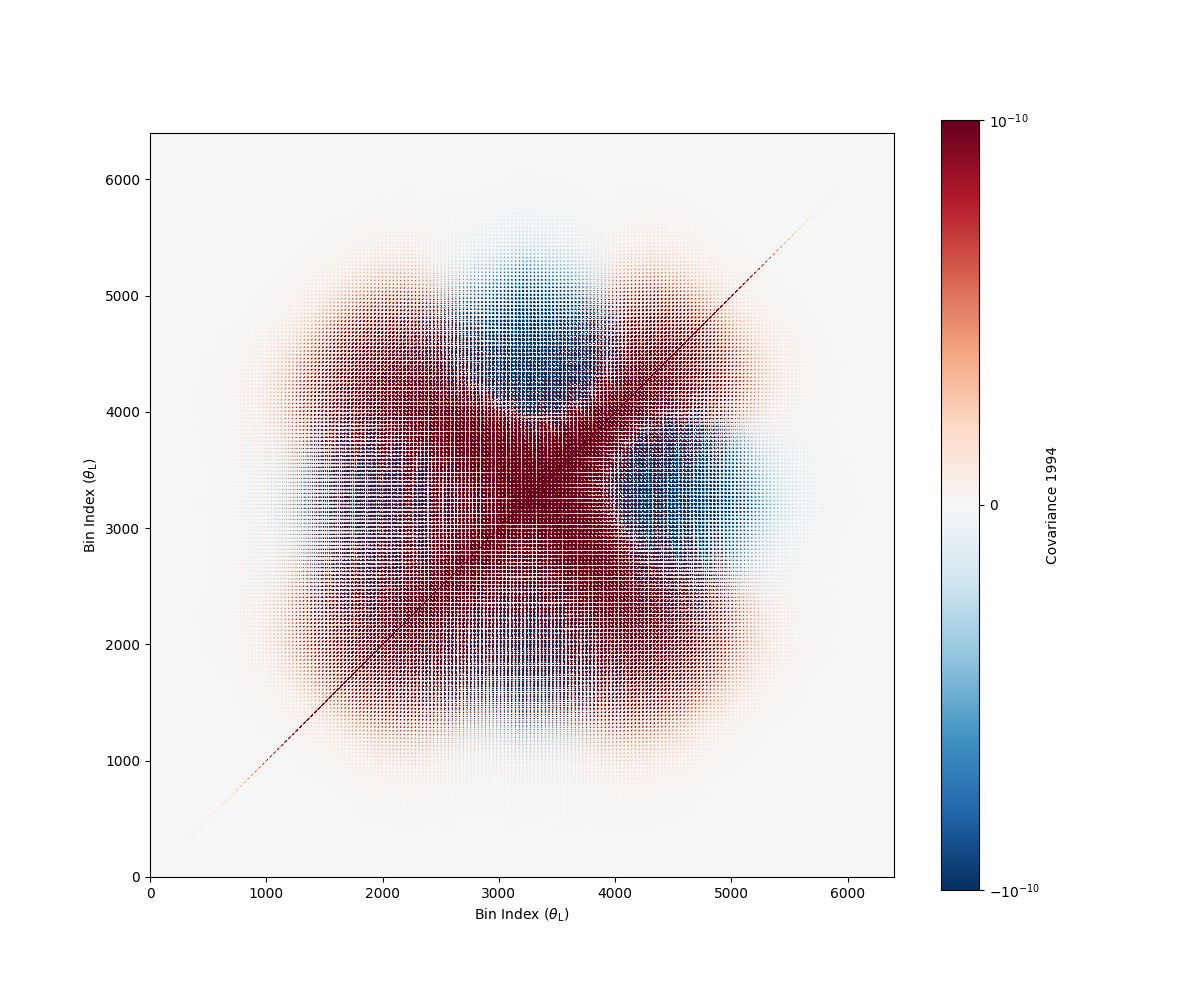}
    \caption{Covariance matrix for 1994 data. Here, the 2D covariance matrix is visualized by structuring it so that each $\theta_{\rm L}$ bin forms a block containing all energy dimension correspondences.}
    \label{fig:covariance}
\end{figure}

The covariance matrix for EEC unfolding from 1994 data is shown in Fig.~\ref{fig:covariance}. Similar to the response matrix, the 2D response matrix (4D matrix) is visualized by flattening (structuring) so that each $\theta_{\rm L}$ bin forms a block structure, where each block corresponds to a single $\theta_{\rm L}$ bin and contains the $32\times32$ sub-matrix of correlations along the energy axis.

\begin{figure}[ht!]
    \centering
    \includegraphics[width=0.45\linewidth]{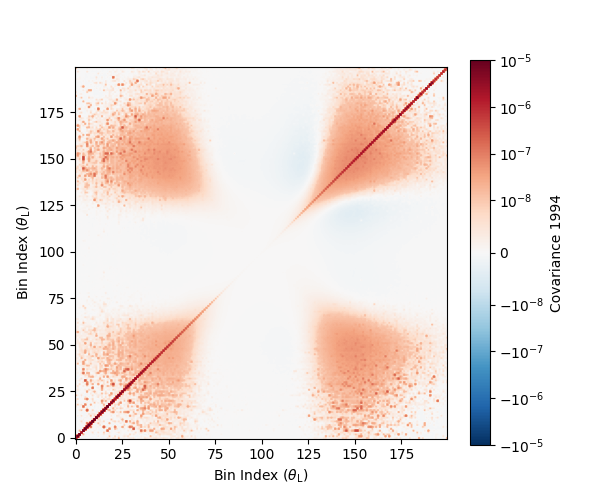}
    \includegraphics[width=0.45\linewidth]{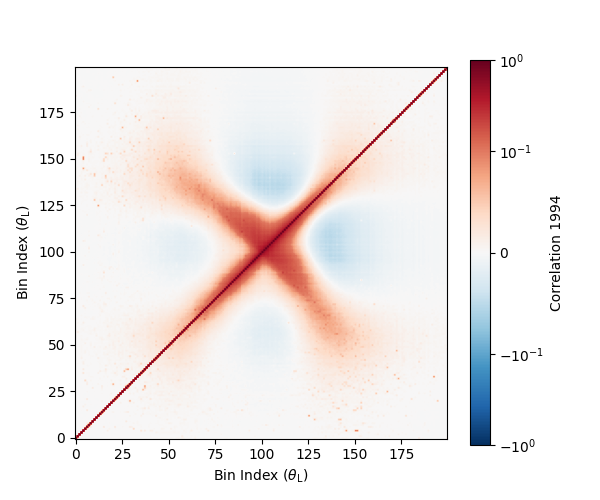}
    \caption{Covariance (left) and correlation (right) matrices of the one-dimensional EEC distribution as a function of $\theta_{\rm L}$ from 1994 data.}
    \label{fig:correlation}
\end{figure}

Subsequently, the covariance of the one-dimensional EEC distribution as a function of r is obtained via a weighted projection over the EEC axis, where the content of each EEC bin is weighted by its corresponding bin center value. The covariance and correlation matrices of the one-dimensional EEC are shown in Figure~.\ref{fig:correlation} left and right, respectively. 

\begin{figure}[ht!]
    \centering
    \includegraphics[width=0.55\linewidth]{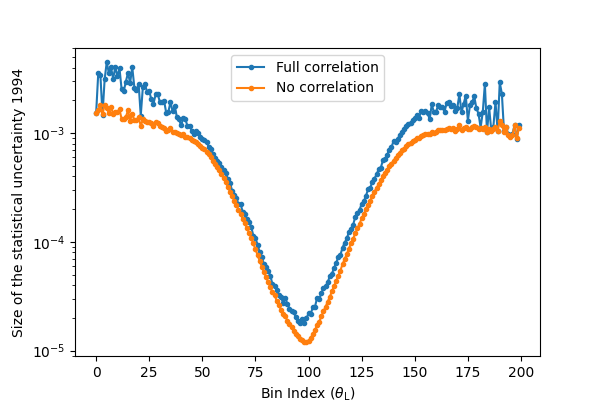}
    \caption{Statistical error from simple Poissonian statistics (orange) and full covariance matrix (blue) from 1994 data. }
    \label{fig:stat}
\end{figure}

The statistical uncertainty for the final EEC distribution is taken as the square root of the diagonal elements of this propagated covariance matrix. A comparison between this error estimation using the full covariance matrix and a naive Poissonian assumption is shown in Figure~\ref{fig:stat}. As demonstrated, neglecting the significant bin-to-bin correlations inherent in the EEC observable would lead to a substantial underestimation of the statistical uncertainty. For this analysis, the full covariance matrix was determined for the measured data and propagated through all subsequent correction and unfolding steps.

\begin{figure}[ht!]
    \centering
    \includegraphics[width=0.45\linewidth]{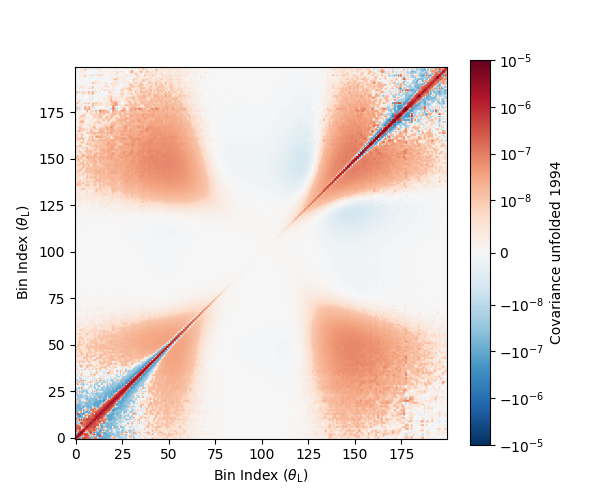}
    \includegraphics[width=0.45\linewidth]{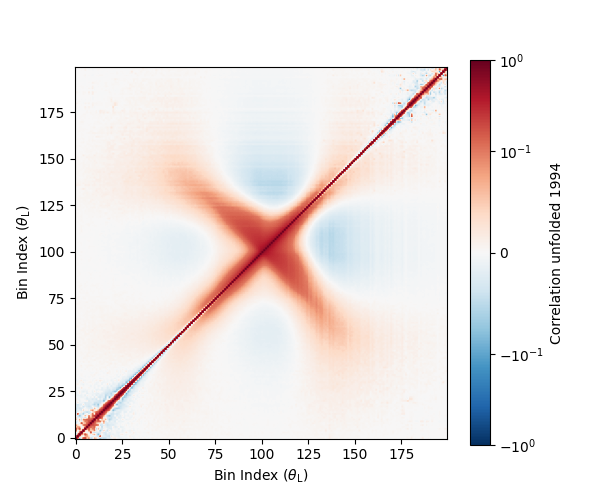}
    \caption{Covariance (left) and correlation (right) matrices of the one-dimensional EEC distribution as a function of $\theta_{\rm L}$ after unfolding for 1994.}
    \label{fig:covUnfold}
\end{figure}

The covariance matrix shown in Figure~\ref{fig:covariance} is propagated through the unfolding procedure through linear approximation, i.e., in each iteration, the covariance matrix is propagated by 
\begin{equation}\label{eq:IBUerror}
\rm Cov = J \ Cov_{input} \ J^T 
\end{equation}
where $\rm J$ is an effective Jacobian matrix that linearizes the non-linear unfolding algorithm, characterizing the response of the final unfolded distribution to variations in the measured input.
This propagation is handled by the internal \texttt{SetMeasuredCov} function in the \texttt{RooUnfold} package. 
While this method is an approximation, its accuracy is expected to be high. In this analysis, the response matrix is strongly diagonal, indicating minimal bin-to-bin migration. This near-linearity ensures that the linear error propagation is a reliable and accurate estimate of the true statistical uncertainty. More computationally intensive methods, such as bootstrapping via pseudo-experiments, are not expected to give significantly different results and are thus considered beyond the scope of this analysis. The final covariance and correlation matrices of the unfolded distribution are shown in Figure~\ref{fig:covUnfold}.
The covariance and correlation matrices after unfolding for 1994 are shown in Figure~\ref{fig:covUnfold}. As shown, the unfolding introduces some extra correlation in the small-angle and back-to-back regions. 

\subsection{Additional corrections}
\label{sec:corrections}
In this Section, additional corrections in this analysis are presented. Several closure checks were performed to make sure that the corrections were applied correctly. These closure checks are presented in Appendix \ref{app:CorrectionClosure}.
As mentioned before, the binning used for $\theta_{\rm L}$ and $z$ results are exactly the same, where the boundaries used are simply converted between $\theta_{\rm L}$ and $z$ by Equation~\ref{eq:z}. Therefore, the correction factors for the $\theta_{\rm L}$ and $z$ parametrizations are the same, and for all the plots shown below, only $\theta_{\rm L}$ will be used.
\subsubsection{2D projection correction}
\label{sec:projection}

\begin{figure}[ht!]
    \centering
    \includegraphics[width=0.6\linewidth]{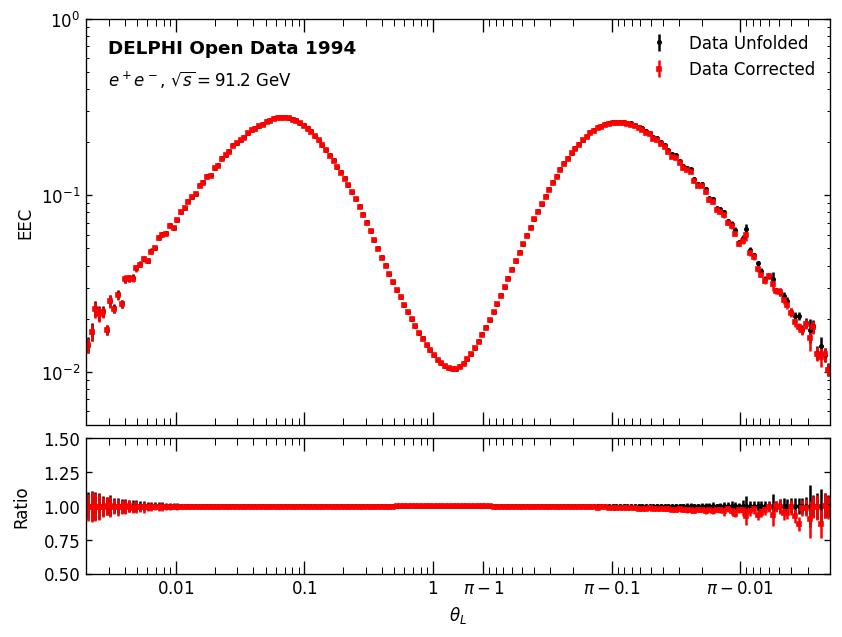}
    \caption{Comparison of the unfolded 1994 data before (black) and after (red) binning projection correction.}
    \label{fig:binningcorr}
\end{figure}

The result of the 2D unfolding procedure described in Section \ref{sec:unfolding} is 2D histogram with an angular dimension in bins of $z$ or $\theta_{\rm L}$ and an energy dimension in bins of $E_{\rm i}E_{\rm j}/E^{2}$. However, the final result presented will be of a one-dimensional histogram of the form given in Equation \ref{eq:ENC}. To convert the two-dimensional histogram in terms of angle and energy weight into the correct form, a projection is performed weighted by the bin center in the $E_{\rm i}E_{\rm j}/E^{2}$ axis to account for the energy weights. Therefore, this will create a difference in the projection and the form given in the Equation \ref{eq:ENC} as the bin center in the $E_{\rm i}E_{\rm j}/E^{2}$ would normally be exact for each pair, but is now approximated by the bin width of the pair that this corresponds to. This difference results in a non-closure. The size of the non-closure, estimated to be less than 5\% across the entire range, is then used as a correction factor for this effect. This correction results in a corresponding systematic uncertainty described in Section \ref{sec:projectionUncert}. 
Note that having finer bins along the energy axis would reduce this systematic uncertainty, but would increase computing resource usage. For this analysis, the energy bins were chosen as reported in Section \ref{sec:response}. 
The size of the correction is shown in Fig.~\ref{fig:binningcorr}. 

\subsubsection{Acceptance and event selection corrections}
\label{sec:acccorr}
The event and track selection efficiency correction is performed by comparing the EEC distributions before and after the event and charged-particle selections as a function of $z$ and/or $\theta_{\rm L}$. Note that this includes the charged particle selection requirements so that the fully-corrected distribution can be directly compared to theoretical calculations. This is written in Equation \ref{eq:evtselcorr} where $\text{EEC}^{\rm gen}_{\rm Selected}$ refers to the EEC distribution after the standard event selections written in Table \ref{tab:SelectionSummary} and $\text{EEC}^{\rm gen}_{\rm ALL}$ refers to the EEC distribution before these standard selections. To apply this correction, the EEC distribution is divided by this correction factor such that it properly reflects the distribution prior to event selection that can be compared to MC simulations.

\begin{equation}\label{eq:evtselcorr}
    \text{Event Sel.} = \frac{\text{EEC}^{\rm gen}_{\rm Selected}}{\text{EEC}^{\rm gen}_{\rm All}}
\end{equation}

\begin{figure}[ht!]
    \centering
    \includegraphics[width=0.6\linewidth]{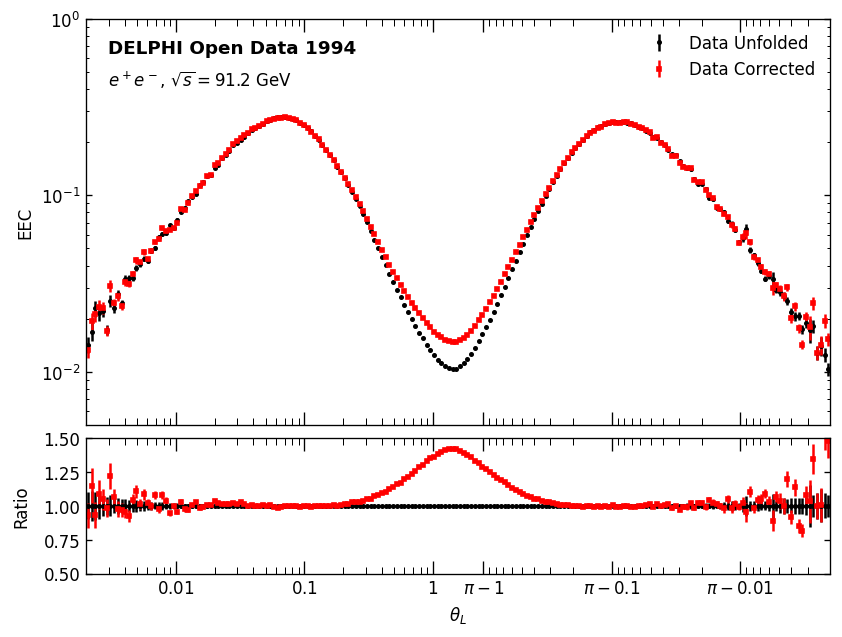}
    \caption{Event selection efficiency corrections as a function of $\theta_{\rm L}$ for 1994. The ratio is shown in the bottom panel, which represents the inverse of the event selection efficiency correction.}
    \label{fig:evtSelCorr}
\end{figure}

Figure~\ref{fig:evtSelCorr} shows the correction factor for event and track selection efficiencies as a function of $\theta_{\rm L}$. A prominent feature of this correction is a significant inefficiency (corresponding to a large correction factor) in the central region, with the minimum efficiency occurring around $\theta_{\rm L} = \pi/2$ ($z = 1/2$).

This inefficiency is primarily driven by the topology of dijet events relative to the detector geometry. The effect is largest for events where the primary dijet axis is perpendicular to the beam line (i.e., in the ``midrapidity'' region). For such events, any soft radiation emitted perpendicularly to the two main jets is preferentially directed along the beam pipe, where the detector has no acceptance. Since these lost particles would have formed pairs with angles of approximately $\pi/2$ relative to the particles in the main jets, their loss creates the pronounced dip in efficiency observed in the central angular region.

\subsection{Systematic uncertainties}
\label{sec:syst}
Seven major systematic uncertainty groups are considered in this work.  
They are discussed in the following sub-sections. 

\subsubsection{Track efficiency and momentum scale}
\label{sec:trkUncert}
Uncertainties related to the reconstruction and modeling of charged particles are evaluated. These include the tracking efficiency and the momentum scale calibration. 
The DELPHI collaboration performed detailed calibrations, tuning the hit efficiency in simulation and using $V^0$
 decays to correct the momentum scale~\cite{Elsing:2000fv, Osterberg:1998xxx, DELPHI:1995dsm}. To evaluate the residual uncertainties, we follow a prescription similar to that used in previous DELPHI and CMS analyses~\cite{DELPHI:2000uri, CMS:2024mlf}.

\begin{figure}[ht!]
    \centering
    \includegraphics[width=0.6\linewidth]{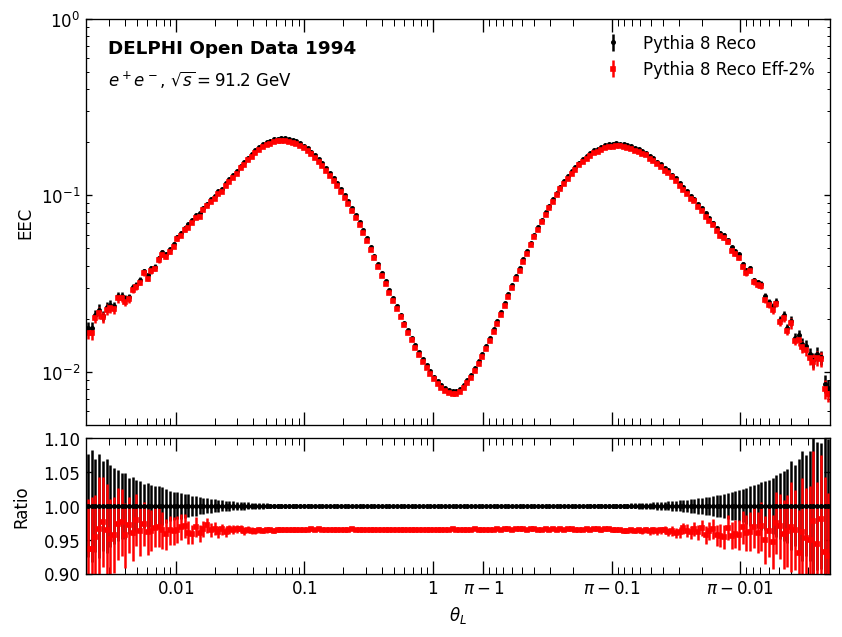}
    \caption{The relative uncertainty on the reconstructed track EEC distribution from the track efficiency variation. The effect is shown as the ratio of the varied result (with 2\% of tracks removed) to the nominal result.}
    \label{fig:trackEffUncert}
\end{figure}

\noindent\textbf{Track efficiency}
The uncertainty from tracking efficiency is estimated by randomly removing 2\% of the reconstructed tracks from each event in the Monte Carlo sample. As shown in Figure~\ref{fig:trackEffUncert}, this variation results in an approximately 4\% change in the overall normalization, with negligible shape-dependent effects.

\begin{figure}[ht!]
    \centering
    \includegraphics[width=0.6\linewidth]{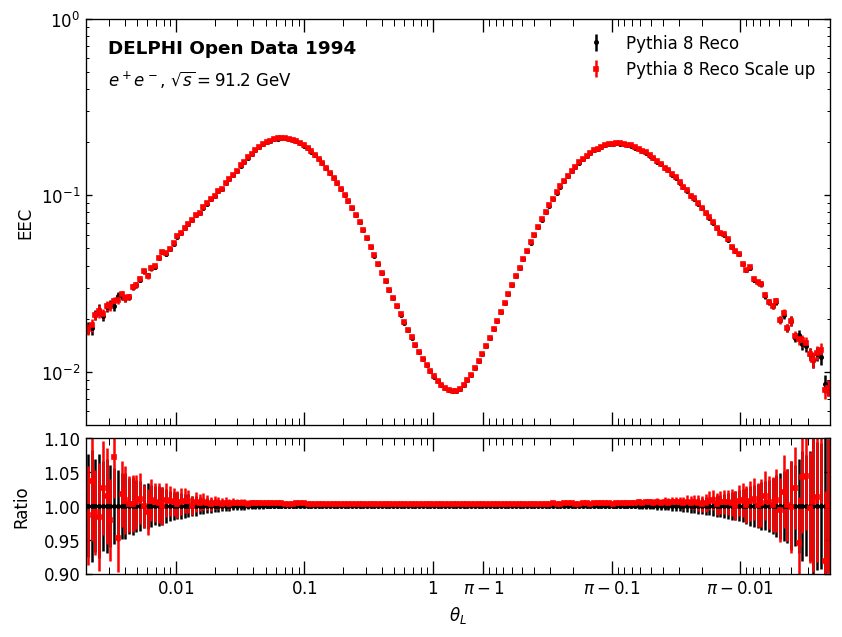}
    \caption{The relative uncertainty on the reconstructed EEC distribution from the track momentum scale variation.}
    \label{fig:trackScaleUncert}
\end{figure}

\noindent\textbf{Momentum scale}
To account for potential mis-calibration, the momentum of each charged track is shifted by a $p_{\rm T}$-dependent factor, ranging from 0.1\% at low $p_{\rm T}$ to 2.5\% at high $p_{\rm T}$. The effect of this kinematic shift is less 1\% across all $\theta_{\rm L }$ range and are shown in the Figure~\ref{fig:trackScaleUncert}.

\subsubsection{High $p_{\rm T }$ tracks}
\label{sec:trackTail}
\begin{figure}[ht!]
    \centering
    \includegraphics[width=0.6\linewidth]{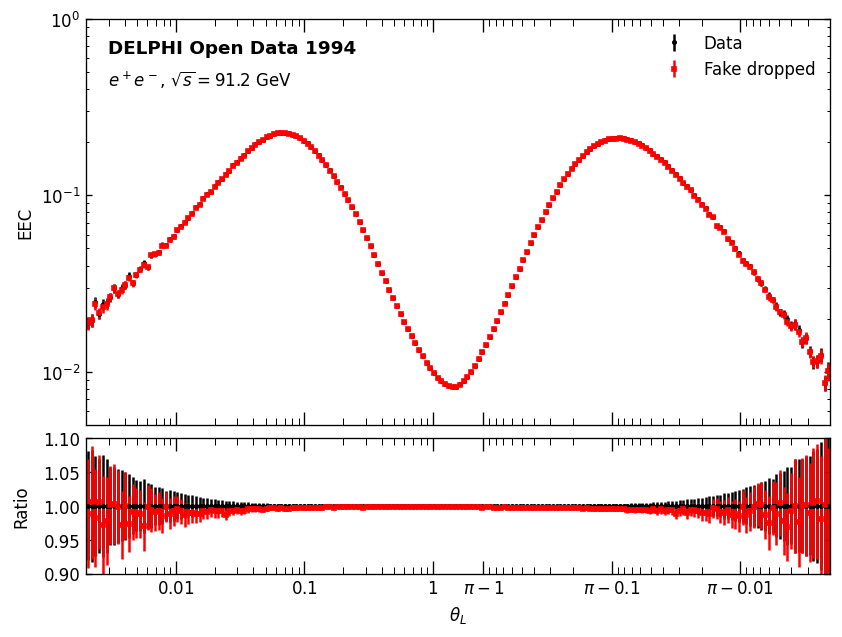}
    \caption{The relative uncertainty on the unfolded EEC distribution from the high-$p_T$
 track contamination study.}
    \label{fig:trackFakeUncert}
\end{figure}

As shown in Section~\ref{sec:DataMCComp}, a significant discrepancy exists between data and simulation in the high-$p_T$ track spectrum. This systematic addresses the possibility that the data contains a source of high-$p_T$ background tracks not modeled in the simulation. The potential impact is estimated by removing a fraction of these tracks from the data. Specifically, a $p_T$-dependent fraction, ranging from 10\% to 40\% of charged tracks with $p_T > 30$ GeV, is randomly removed from the data sample before unfolding. The full difference between the nominal unfolded result and the result from this modified data sample is taken as a conservative estimate of this uncertainty, as shown in Fig.~\ref{fig:trackFakeUncert}.

\subsubsection{Track matching}
\label{sec:matchingUncert}
The matching procedure used to select corresponding pairs between data and the MC has an impact on the pairs that fill the response matrix, therefore impacting the final result. The nominal matching procedure used is described in Section \ref{sec:matching}. 

\begin{figure}[ht!]
    \centering
    \includegraphics[width=0.6\linewidth]{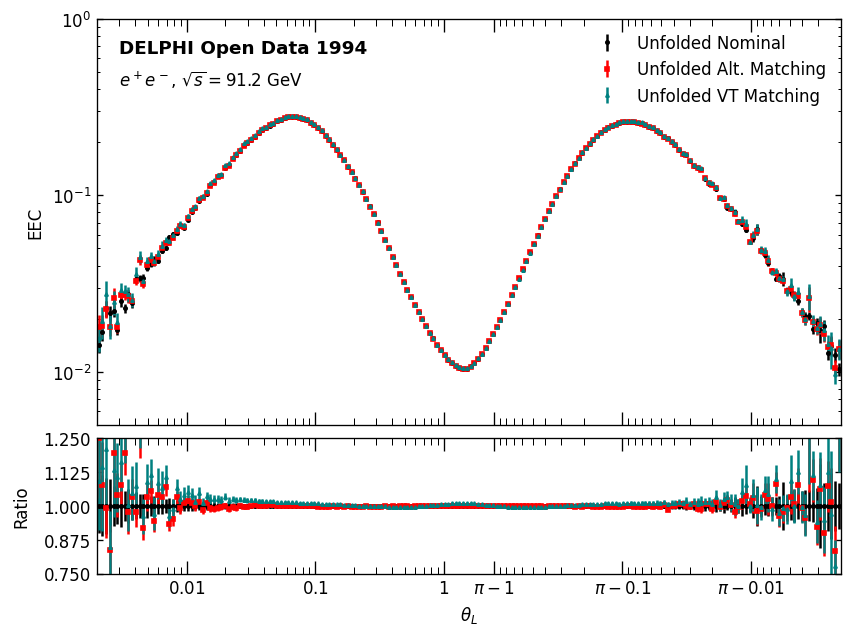}
    \caption{Nominal unfolded result (black) compared to results from an alternative matching scheme (red) and an alternative cut-off (teal).}
    \label{fig:matchinguncert}
\end{figure}

\noindent\textbf{Matching scheme}
This systematic uncertainty is evaluated by taking the difference between the nominal matching scheme (as described in Section \ref{sec:matching}) and an alternative scheme. 
The alternative scheme is taken as the one from the official DELSIM code, in which the reconstructed objects are matched to the GEANT simulation histories to create a correspondence of simulated and reconstructed objects. 
The unfolded procedure is then repeated using the alternative scheme and compared to the nominal result. 
The difference is less than 1\% as shown in Fig.~\ref{fig:matchinguncert}. 

\noindent\textbf{Cut-off value}
In the nominal scheme, a cut-off value of is applied to determine the maximum $\theta_{\rm L}$ between 2 tracks can be called a match. 
This value will therefore implicitly affect the matching efficiency and fake rate. 
A cross-check using a very tight (VT) cut-off value of 0.005 is performed to study the biases from this choice of the cut-off value.
The difference between the nominal choice and the VT choice, shown in Fig.~\ref{fig:matchinguncert}, is also included as a systematic uncertainty. 

\subsubsection{Unfolding uncertainties}
\label{sec:unfoldingUncert}
Uncertainties inherent to the unfolding procedure itself are evaluated, including the choice of physics model for the response matrix and the regularization strength.

\noindent\textbf{Model Dependence}
The unfolding result can be biased by mismodeling in the MC generator used to derive the response matrix. To quantify this uncertainty, the entire unfolding procedure is repeated using three alternative samples: ARIADNE, PYTHIA 8, and PYTHIA 8 Dire. For each, a new detector response matrix is constructed and used to unfold the data.
Figure~\ref{fig:prior} presents a comparison of the final unfolded results. The resulting differences vary systematically across the kinematic regions, with deviations of approximately 1\% in the central region, 5\% at the non-perturbative to perturbative transition, and up to 15\% in the far non-perturbative tail. The maximum envelope of the differences between the nominal PYTHIA 5.7 result and the variations from PYTHIA 8, ARIADNE, and PYTHIA 8 Dire is taken as the systematic uncertainty.

\begin{figure}[ht!]
\centering
\includegraphics[width=0.6\linewidth]{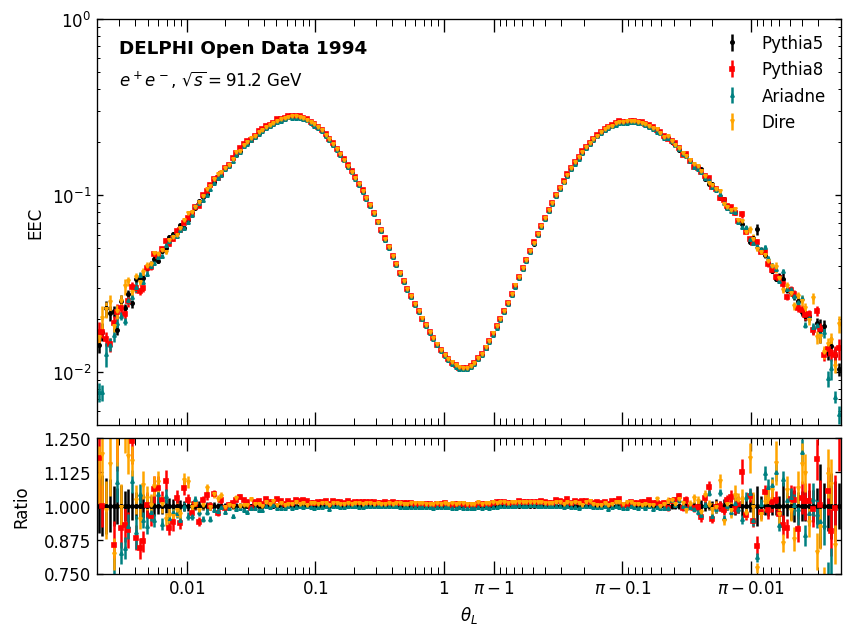}
\caption{Comparison of the unfolded EEC distributions obtained using response matrices derived from four different generators: PYTHIA 5.7 (black), ARIADNE (teal), PYTHIA 8 (red), and PYTHIA 8 Dire (orange).}
\label{fig:prior}
\end{figure}

\begin{figure}[ht!]
\centering
\includegraphics[width=0.6\linewidth]{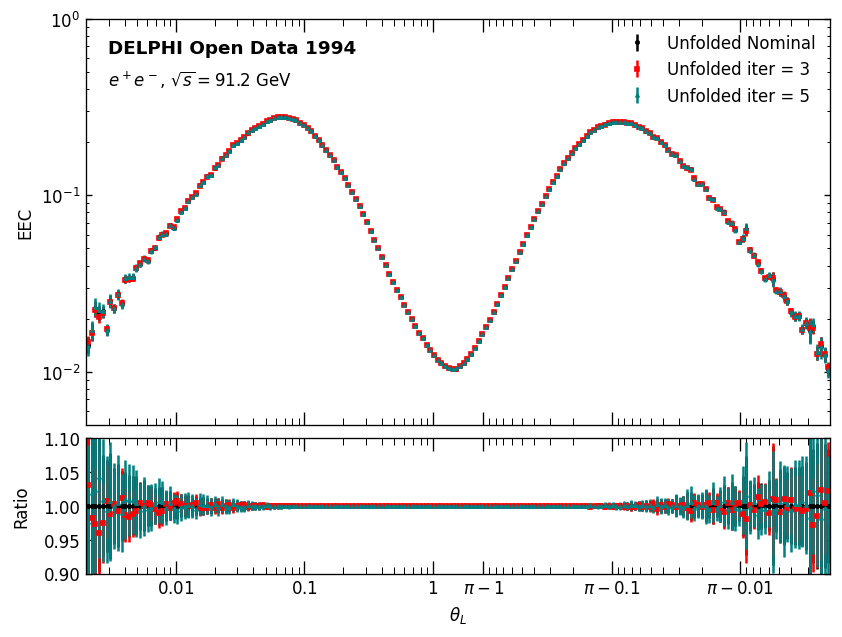}
\caption{Comparison of the unfolded EEC distributions obtained using 3 (red), 4 (black, nominal), and 5 (teal) iterations in the D'Agostini method.}
\label{fig:iter}
\end{figure}

\noindent\textbf{Regularization}
Another potential source of uncertainty is the choice of regularization strength in the unfolding algorithm. For the iterative D'Agostini method, this is controlled by the number of iterations. The effect is evaluated by comparing the nominal result (obtained with 4 iterations) to the results obtained using 3 and 5 iterations. As shown in Fig.~\ref{fig:iter}, the differences are negligible in the bulk of the distribution and increase to approximately 2\% at very small and large angles. The maximum observed differences are taken as the uncertainty.

The dependence of the result on the choice of the \(E_i E_j / E^2\) binning has also been studied, and its effect was found to be negligible.

\subsubsection{2D binning projection correction}
\label{sec:projectionUncert}
\begin{figure}[ht!]
\centering
\includegraphics[width=0.6\linewidth]{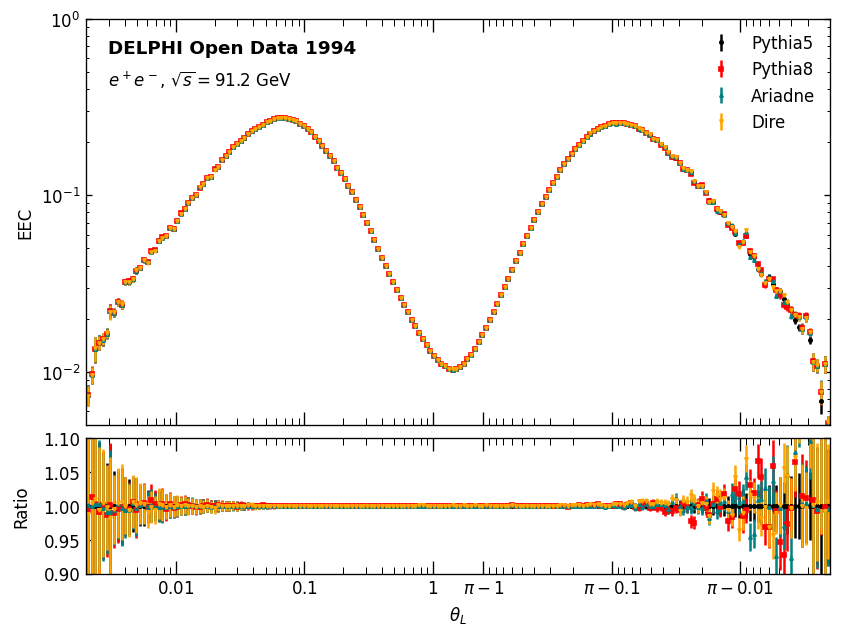}
\caption{Unfolded data distributions corrected by binning projection correction factors derived from PYTHIA 5.7 (black), ARIADNE (teal), and PYTHIA 8 (red), and PYTHIA 8 Dire (orange). }
\label{fig:binningCorrSys}
\end{figure}

As mentioned in Section \ref{sec:projection}, the procedure to construct the final observable introduces a non-closure that is corrected for in a corresponding correction.
The systematic uncertainty associated with the correction is estimated by comparing the correction factor derived from  PYTHIA 5.7, ARIADNE, PYTHIA 8, and PYTHIA 8 Dire MC samples. The comparison is shown in Fig.~\ref{fig:binningCorrSys}. The maximum envelope of the differences between the nominal PYTHIA 5.7 result and the variations from PYTHIA 8, ARIADNE, and PYTHIA 8 Dire is taken as the systematic uncertainty. 

\subsubsection{Track and event selection correction}
\label{sec:selectionUncert}
Similar to the 2D projection correction, the track and event selection correction is model-dependent. 
Therefore, the systematic uncertainties are estimated similarly: comparing results using corrections derived from the nominal and alternative MC samples. 
Fig.~\ref{fig:acceptcorr} shows the comparison. 
Same as before, the maximum envelope of the differences between the nominal PYTHIA 5.7 result and the variations from PYTHIA 8, ARIADNE, and PYTHIA 8 Dire is taken as the systematic uncertainty. 

\begin{figure}[ht!]
\centering
\includegraphics[width=0.6\linewidth]{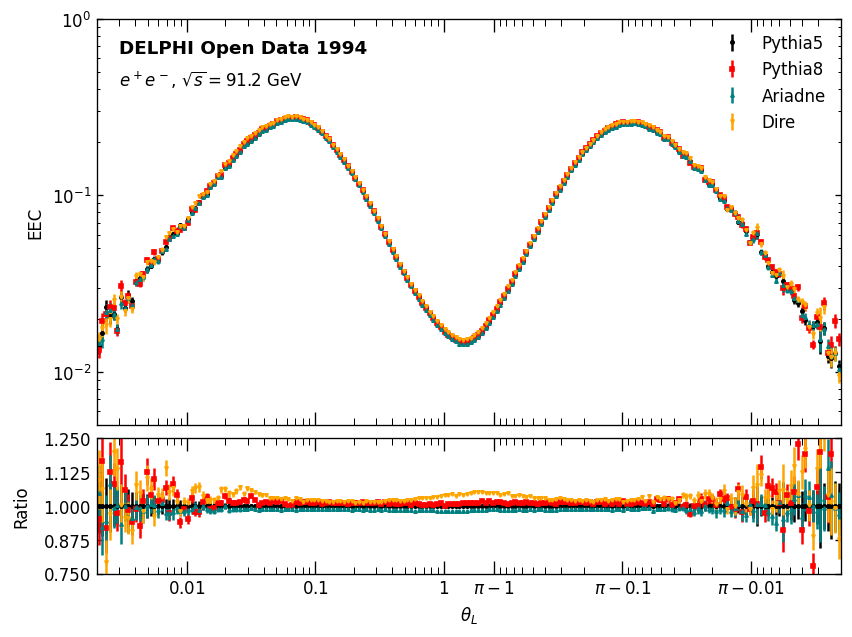}
\caption{Unfolded data distributions corrected by track and event selection correction factors derived from PYTHIA 5.7 (black), ARIADNE (teal), PYTHIA 8 (red), and PYTHIA 8 Dire (orange).}
\label{fig:acceptcorr}
\end{figure}

\subsubsection{Summary of systematics}
The various sources of systematic uncertainty are summarized in Fig.~\ref{fig:systematicSummary}. The total systematic uncertainty is calculated by adding each source in quadrature.

The dominant contributions vary across the angular range. In the central region, the largest uncertainty comes from the charged track efficiency. In the tails of the distribution, at very small and very large opening angles, the uncertainty is dominated by the model dependence of the unfolding procedure and the acceptance correction.
The relative systematic uncertainties for the \(z\) parameterization are identical to those for \(\theta_L\), as the binning scheme for \(z\) is a direct conversion from the \(\theta_L\) binning via Equation~\ref{eq:z}.
A sharp increase in the total uncertainty is observed in the extreme forward and backward regions, where \(\theta_L < 0.006\) (corresponding to an angular separation smaller than \(1^\circ\)), and \(\theta_L > \pi-0.006\). This kinematic boundary coincides with the region of degraded angular resolution for the DELPHI detector, as shown in Figs.\ref{fig:phi_res} and\ref{fig:theta_res}.
A detailed breakdown of the shape of each uncertainty component will be made available on HEPData after final paper publication. 

\begin{figure}[ht!]
    \centering
    \includegraphics[width=0.9\linewidth]{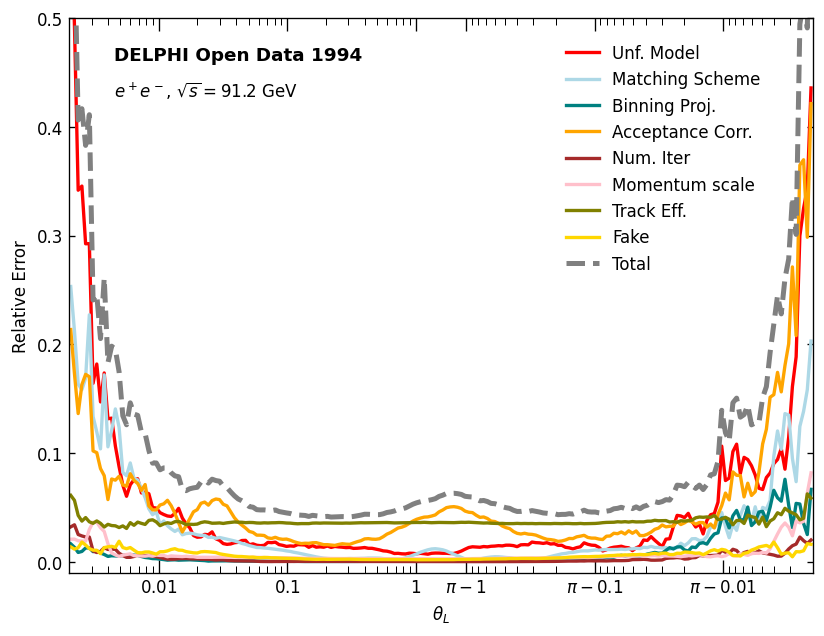}
    \caption{Summary of the systematic uncertainties. The total uncertainty (gray dashed) is calculated as the quadratic sum of the individual uncertainty sources, including event and track selection corrections~\ref{sec:selectionUncert} (orange), unfolding model dependence~\ref{sec:unfoldingUncert} (red), matching scheme~\ref{sec:matchingUncert} (blue), 2D binning projection~\ref{sec:projectionUncert} (teal), unfolding regularization strength~\ref{sec:unfoldingUncert} (brown), track efficiency (olive) and momentum scale (pink)~\ref{sec:trkUncert}, and fake~\ref{sec:trackTail} (gold). }
    \label{fig:systematicSummary}
\end{figure}

\clearpage
\section{Thrust analysis}\label{sec:thrust}
For thrust, a ``simple'' 1D unfolding is performed to correct the detector-level distribution back to the truth-level one. 
The output of the unfolding corresponds to the truth-level distribution for particles within the detector's acceptance. 
An additional correction on particle and event selections is then applied to the unfolded thrust distributions to obtain the final result. 

As mentioned earlier, for the sake of clarity and conciseness in the main text, the figures presented in this section are based on the 1994 dataset. The 1995 analysis uses the same procedure, and the comprehensive set of corresponding plots for the 1995 analysis is provided in Appendix~\ref{app:1995Thrust}.

\subsection{Unfolding and event selection corrections}
\begin{figure}[ht!]
    \centering
    \includegraphics[width = 0.45\textwidth]{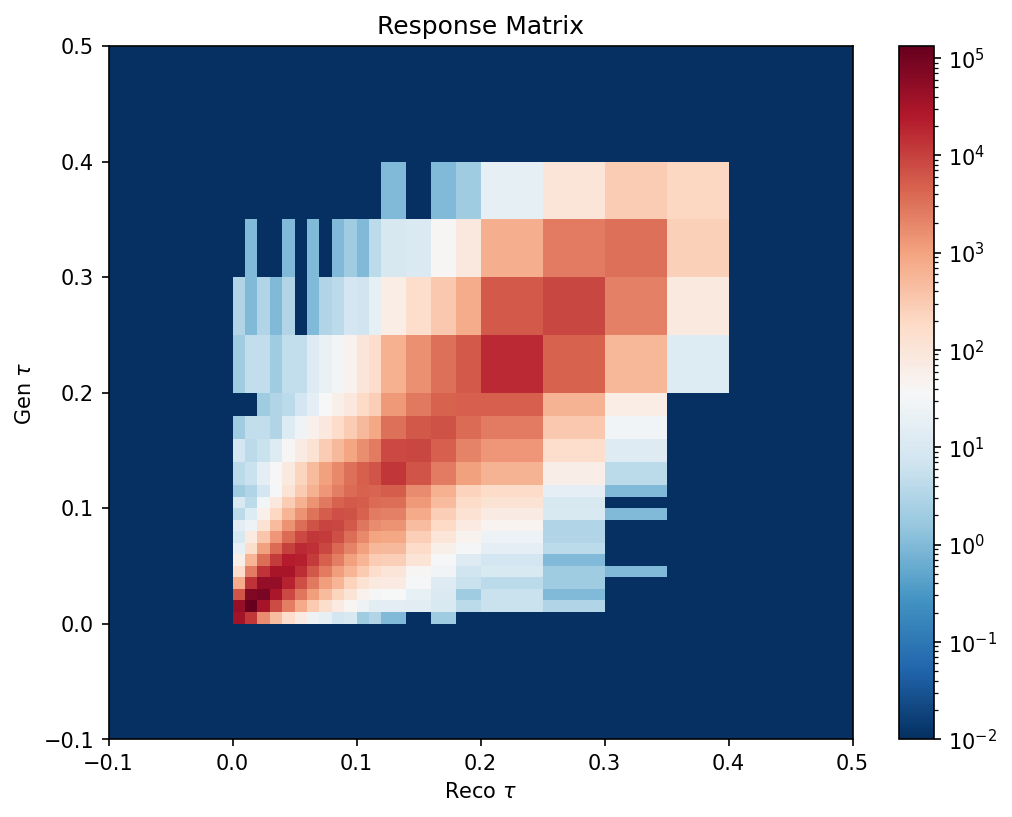}
    \includegraphics[width = 0.45\textwidth]{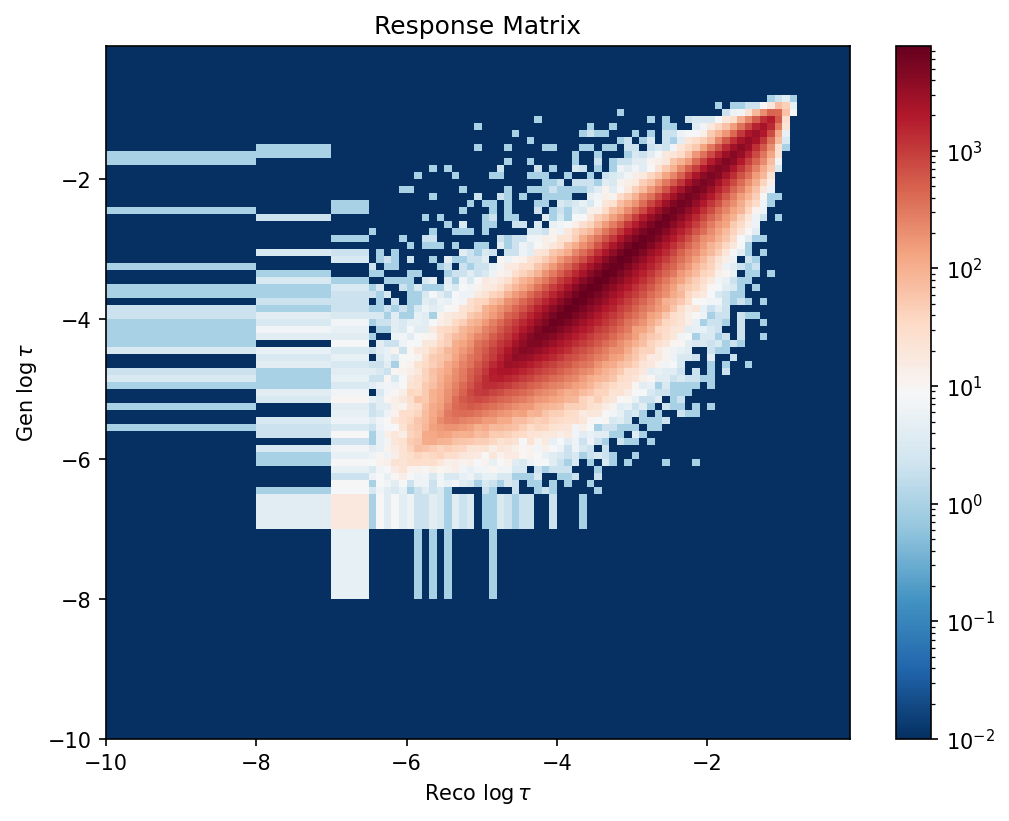}
    \caption{Response matrices for $\tau$ (left) and $\log\tau$ (right).}
    \label{fig:responseThrust}
\end{figure}

Two different binning are used for $\tau (= \rm 1-T)$ and $\log\tau$. Specifically, the binning for $\tau$ is kept the same as the original DELPHI publication~\cite{DELPHI:2003yqh, DELPHI:2000uri} to facilitate a direct comparison of the results. The response matrices are constructed by correlating the reconstructed-level and generator-level distributions, as shown in Fig.~\ref{fig:responseThrust}. Fig.~\ref{fig:responseThrust}.  

\noindent\textbf{Unfolding}
Similar to EEC, the unfolding is done using the D'Agostini iterative method~\cite{D'Agostini:265717} in the \texttt{RooUnfold}~\cite{Brenner:2019lmf} package.  
The number of iterations is chosen to be five, at which point the $\chi^2/\rm ndf$ between the distributions from the current and previous iterations drops below 0.05, indicating convergence.

\begin{figure}[ht!]
    \centering
    \includegraphics[width = 0.45\textwidth]{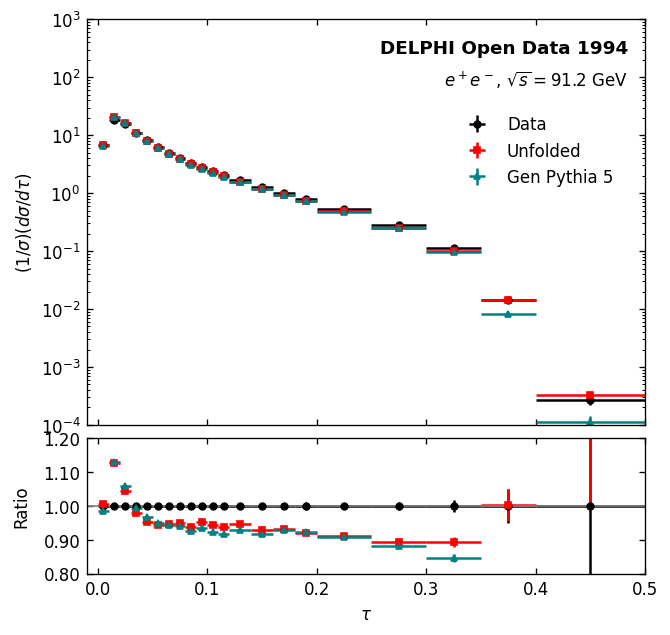}
    \includegraphics[width = 0.45\textwidth]{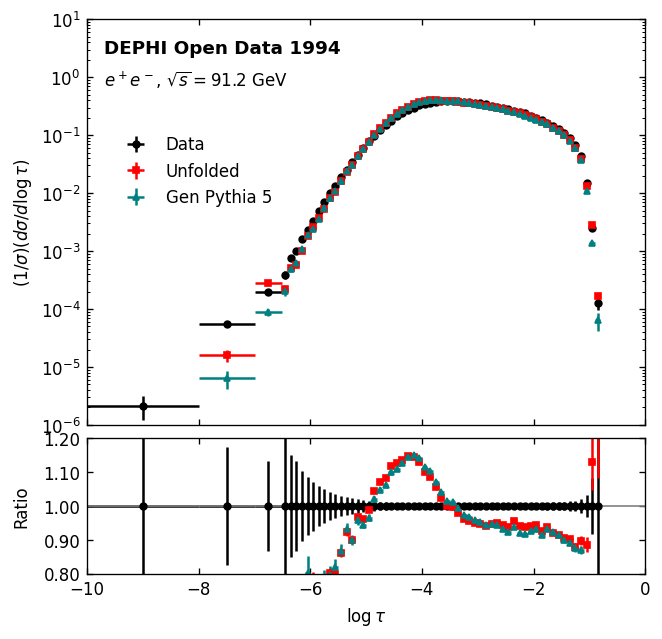}
    \caption{Raw (black) and unfolded (red) data distributions for $\tau$ (left) and $\log\tau$ (right). Also shown are the generator-level distributions from PYTHIA 5 (teal).}
    \label{fig:unfoldThrust}
\end{figure}

The unfolded data is shown in Fig.~\ref{fig:unfoldThrust}. As expected, the unfolding procedure shifts the distribution towards smaller values of $\tau$ (higher Thrust). This is because detector effects, such as finite resolution and reconstruction inefficiencies, tend to make the measured events appear more spherical (less jet-like) than they truly are. The unfolding corrects for this "smearing" effect, restoring the distribution to its sharper, more pencil-like particle-level shape.

\noindent\textbf{Event and track selection corrections}
The unfolded distribution corresponds to the particle-level distribution within the defined fiducial phase space. 
To compare with theoretical predictions, a final correction factor is applied. This factor extrapolates the result from the fiducial region to the full phase space. This correction is derived from the PYTHIA 5 sample by taking the ratio of the particle-level distribution in the full phase space to that in the fiducial phase space.
Before applying the procedure to the data, a closure test on full analysis chain was performed using only MC events, and perfect closure was observed (See Appendix~\ref{app:CorrectionClosure}.

\begin{figure}[ht!]
    \centering
    \includegraphics[width = 0.45\textwidth]{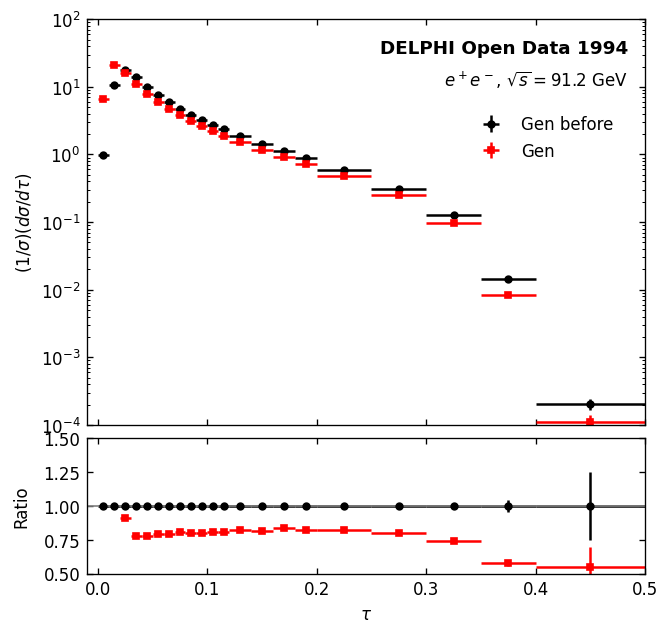}
    \includegraphics[width = 0.45\textwidth]{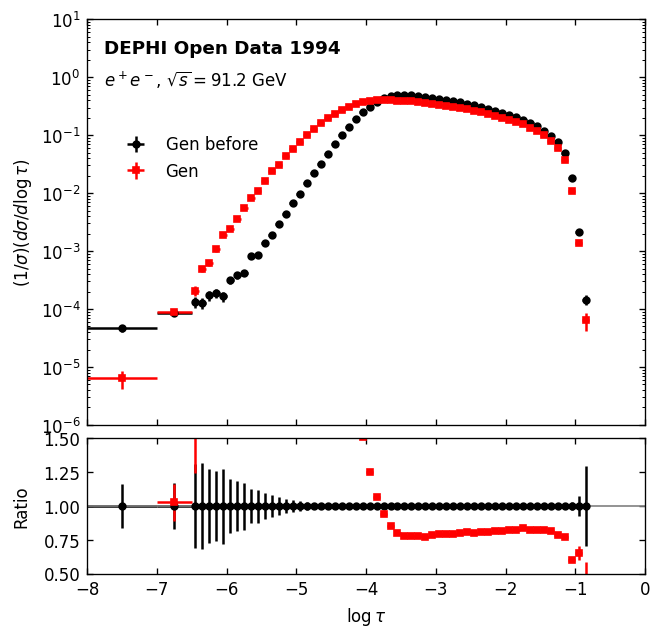}
    \caption{Unfolded data distributions before (red) and after (black) correction to the full phase space for $\tau$ (left) and $\log\tau$ (right).}
    \label{fig:correctedThrust}
\end{figure}

The data distribution before and after this acceptance correction is shown in Fig.~\ref{fig:correctedThrust}. As expected, including particles in the forward and backward regions makes the events appear more spherical. This is because many of the additional particles at low polar angles do not lie perfectly along the primary jet axis, thus increasing the value of $\tau$. 

\subsection{Systematics}
\subsubsection{Experimental systematics}
The following sources of experimental systematic uncertainty were considered: 1) charged particle tracking efficiency, 2) charged particle momentum scale, 3) modeling of high-$p_{\rm T}$ tracks, 4) neutral particle detection efficiency, and 5) neutral particle energy scale. The method used to estimate each uncertainty is briefly illustrated below. For each variation, the observable is re-calculated, and the difference with respect to the nominal MC/data result is taken as the systematic uncertainty.

\begin{figure}[ht!]
    \centering
    \includegraphics[width = 0.45\textwidth]{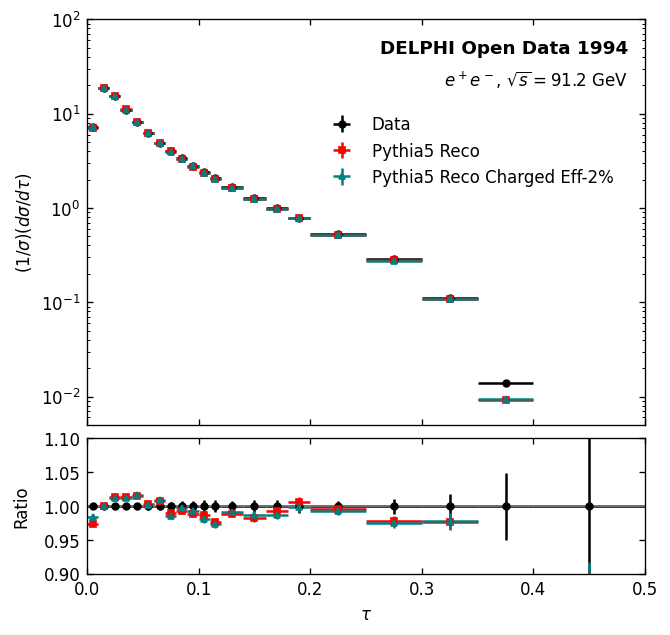}
    \includegraphics[width = 0.45\textwidth]{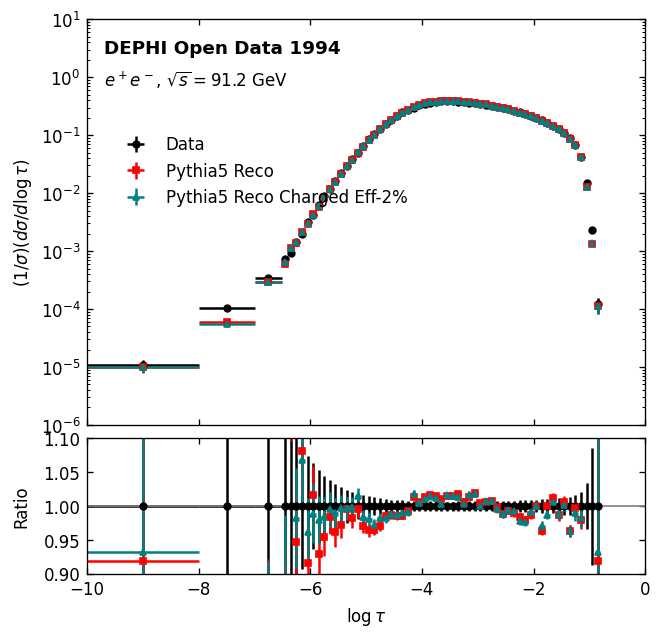}
    \caption{Effect of the charged particle efficiency uncertainty on the reconstructed $\tau$ (left) and $\log\tau$ (right) distributions. The nominal MC is shown in red, the varied MC (with 2\% of tracks removed) is in teal, and the data is shown in black for reference.}
    \label{fig:effChargedSysThrust}
\end{figure}

\begin{figure}[ht!]
    \centering
    \includegraphics[width = 0.45\textwidth]{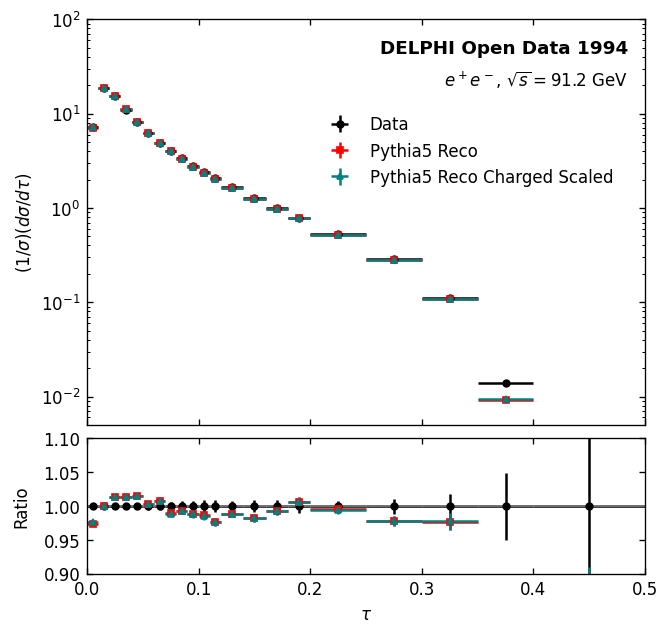}
    \includegraphics[width = 0.45\textwidth]{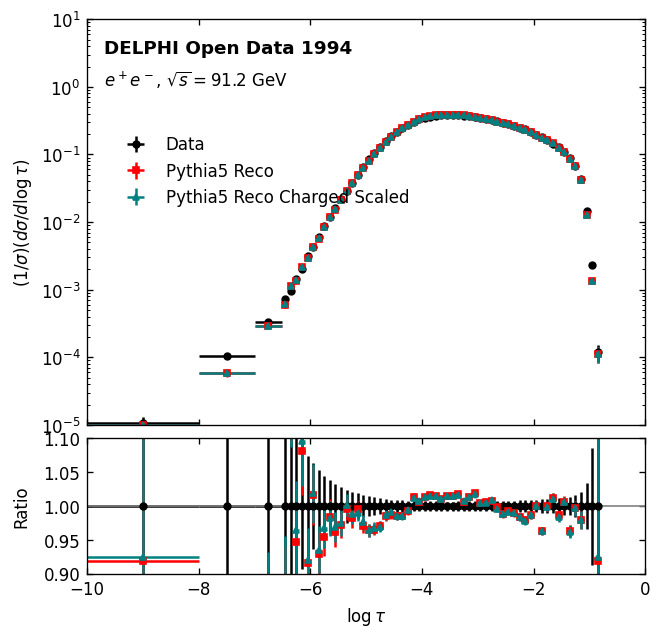}
    \caption{Effect of the charged particle momentum scale uncertainty on the reconstructed $\tau$ (left) and $\log\tau$ (right) distributions. The nominal MC (red) is compared to the varied MC (teal) where track momenta have been shifted, and the data is shown in black for reference.}
    \label{fig:scaleChargedSysThrust}
\end{figure}

\begin{figure}[ht!]
    \centering
    \includegraphics[width = 0.45\textwidth]{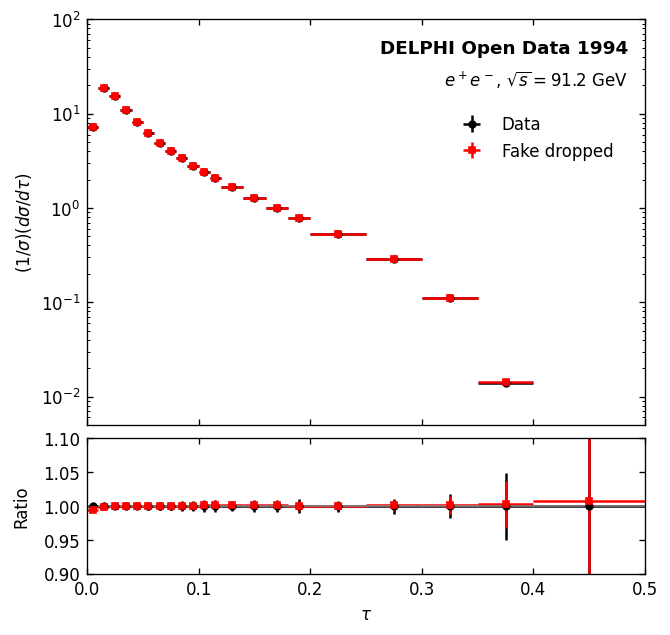}
    \includegraphics[width = 0.45\textwidth]{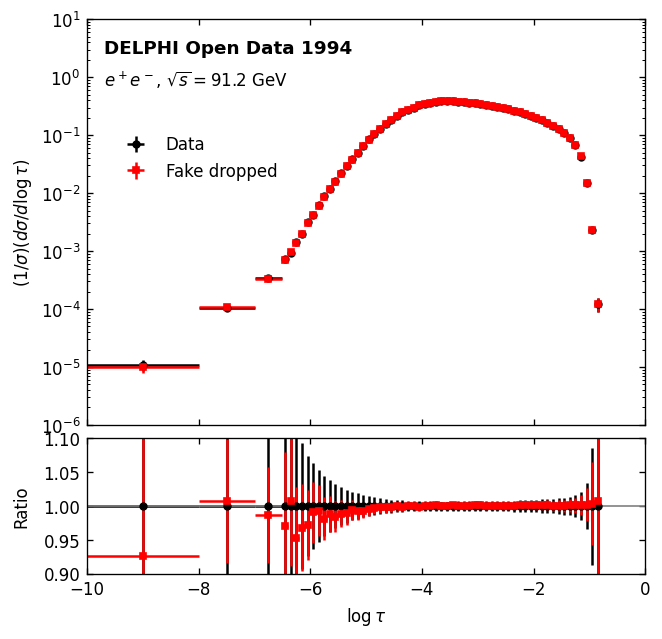}
    \caption{Estimation of the uncertainty from potential high-$p_{\rm T}$ fake track contamination. The nominal data distribution (black) is compared to the varied data distribution (red), where a fraction of tracks with $p_{\rm T} > 30$ GeV has been randomly removed.}
    \label{fig:tailChargedSysThrust}
\end{figure}

\begin{figure}[ht!]
    \centering
    \includegraphics[width = 0.45\textwidth]{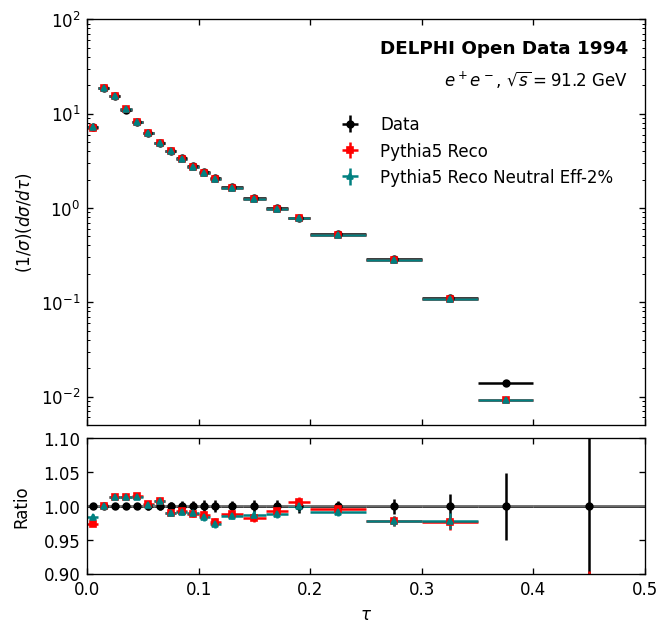}
    \includegraphics[width = 0.45\textwidth]{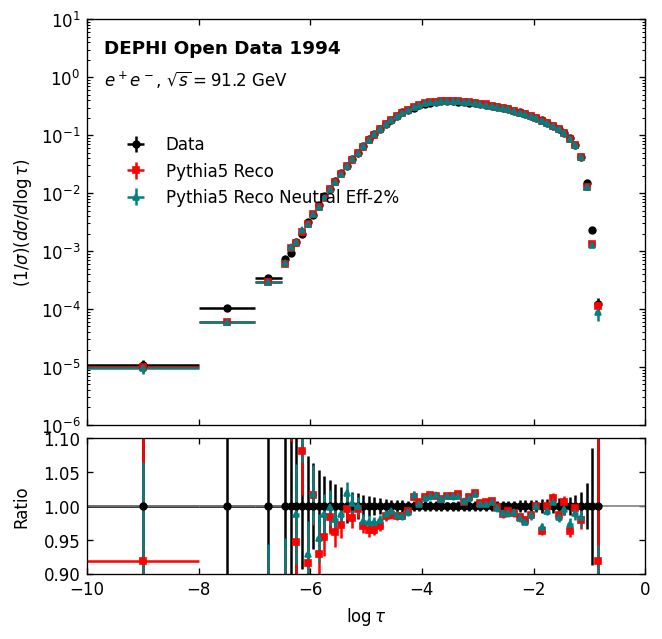}
    \caption{Effect of the neutral particle efficiency uncertainty on the reconstructed $\tau$ (left) and $\log\tau$ (right) distributions. The nominal MC (red) is compared to the varied MC (teal) with 2\% of neutral particles removed. Data is shown in black for reference.}
    \label{fig:effNeutralSysThrust}
\end{figure}

\begin{figure}[ht!]
    \centering
    \includegraphics[width = 0.45\textwidth]{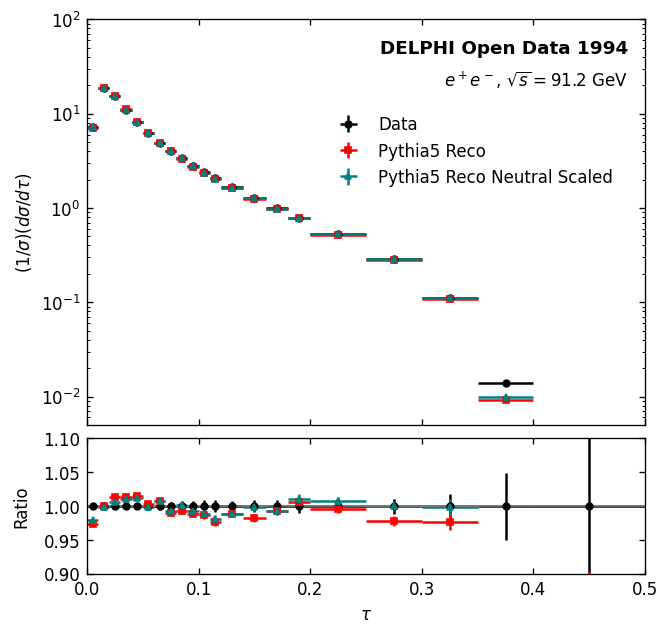}
    \includegraphics[width = 0.45\textwidth]{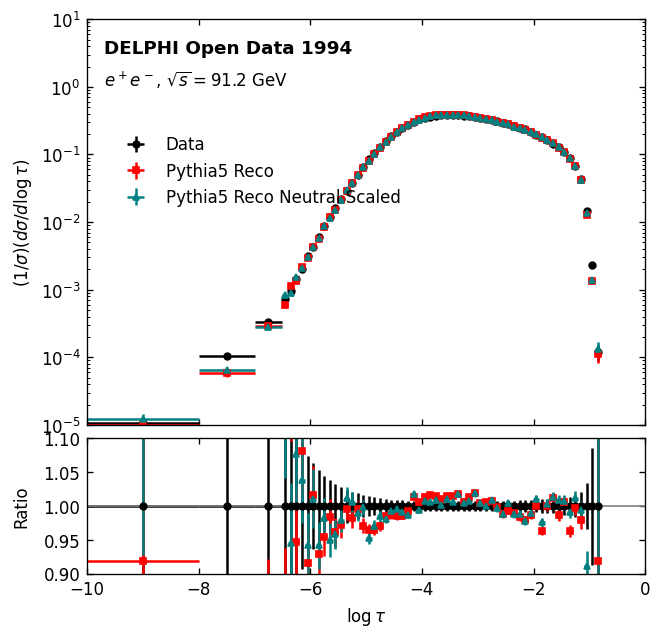}
    \caption{Effect of the neutral particle energy scale uncertainty on the reconstructed $\tau$ (left) and $\log\tau$ (right) distributions. The nominal MC (red) is compared to the varied MC (teal), where neutral particle energies have been shifted by 20\%. Data is shown in black for reference.}
    \label{fig:scaleNeutralSysThrust}
\end{figure}

\begin{enumerate}
    \item \textbf{Charged particle efficiency (Fig.~\ref{fig:effChargedSysThrust}):} To estimate the uncertainty from tracking efficiency, 2\% of reconstructed charged tracks are randomly removed from each MC event. This is inherited from the approach used in the DELPHI~\cite{DELPHI:2000uri} and CMS~\cite{CMS:2024mlf} analyses. The effect of this variation on the reconstructed thrust distribution is shown in the figure.

    \item \textbf{Charged particle momentum scale (Fig.~\ref{fig:scaleChargedSysThrust}):} To account for potential mis-calibration, the momentum of each charged track is shifted by a $p_{\rm T}$-dependent factor, ranging from 0.1\% at low $p_{\rm T}$ to 2.5\% at high $p_{\rm T}$. This is also inherited from the approach used in the DELPHI~\cite{DELPHI:2000uri} and CMS~\cite{CMS:2024mlf} analyses. The effect of this kinematic shift is shown in the figure.

    \item \textbf{High-$p_{\rm T}$ track modeling (Fig.~\ref{fig:tailChargedSysThrust}):} This systematic addresses the possibility that the data contains a source of high $p_{\rm T}$ fake tracks not modeled in the simulation. The potential impact of such a contamination is estimated by treating these tracks as a background and subtracting them from the data. Specifically, 10\% to 40\% of charged tracks with $p_{\rm T}>30$ GeV are randomly removed from the data sample. The difference between the nominal data distribution and this modified distribution is taken as a conservative estimate of this uncertainty.

    \item \textbf{Neutral particle efficiency (Fig.~\ref{fig:effNeutralSysThrust}):} In a similar procedure to the charged particle study, 2\% of reconstructed neutral particles are randomly removed from each MC event. The resulting change in the thrust distribution is shown in the figure. 

    \item \textbf{Neutral particle energy scale (Fig.~\ref{fig:scaleNeutralSysThrust}):} The energy of each neutral particle is shifted by a global factor of 20\%. This is an estimate, chosen to account for the larger intrinsic uncertainties associated with calorimeter measurements compared to tracking. The effect of this large energy shift is shown in the figure.
\end{enumerate}

\clearpage
\subsubsection{Simulation modeling systematics}
\label{sec:simUncert}
As shown in Section~\ref{sec:DataMCComp}, inconsistencies in neutral particle kinematics are observed between the 1994 and 1995 datasets and their corresponding simulations. To account for the potential impact of this detector-level mismodeling, three additional systematic uncertainties are evaluated.

\begin{figure}[ht!]
    \centering
    \includegraphics[width = 0.45\textwidth]{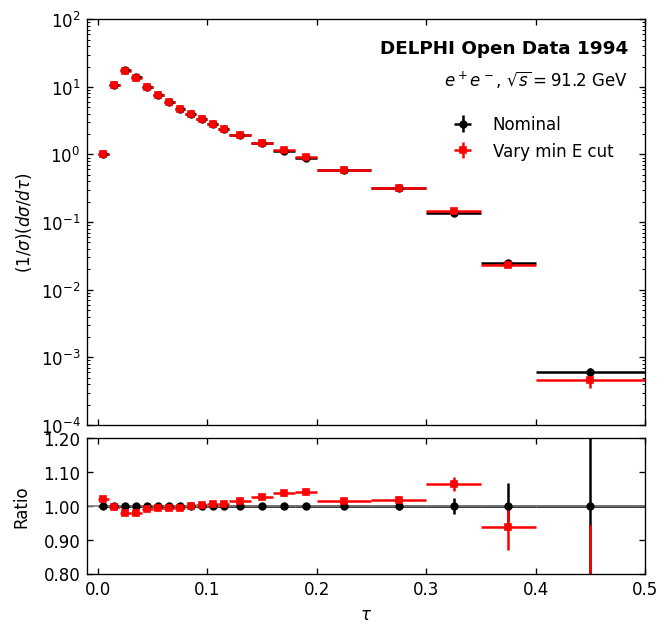}
    \includegraphics[width = 0.45\textwidth]{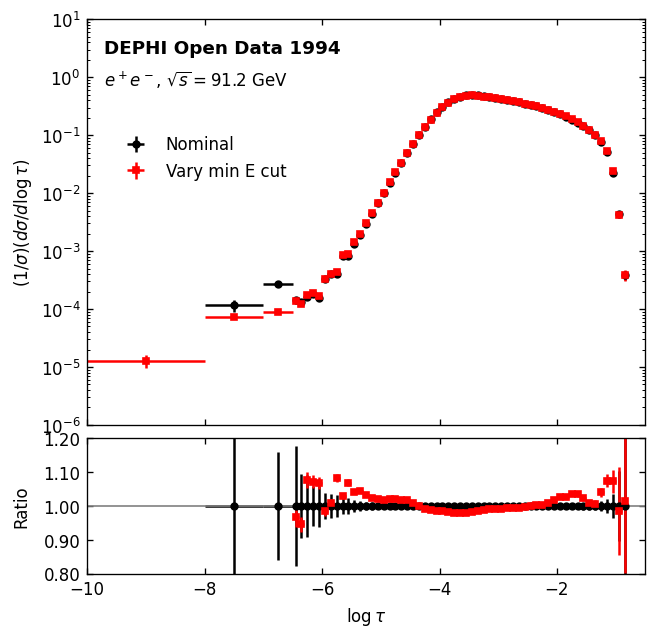}
    \caption{The relative uncertainty on the unfolded thrust distributions from varying the minimum neutral particle energy cut. The effect is shown as the ratio of the varied result (\(E_{\text{neutral}} > 5.0\)~GeV) (red) to the nominal result (\(E_{\text{neutral}} > 0.5\)~GeV) (black).}
    \label{fig:cutNeutral}
\end{figure}

\begin{figure}[ht!]
    \centering
    \includegraphics[width = 0.45\textwidth]{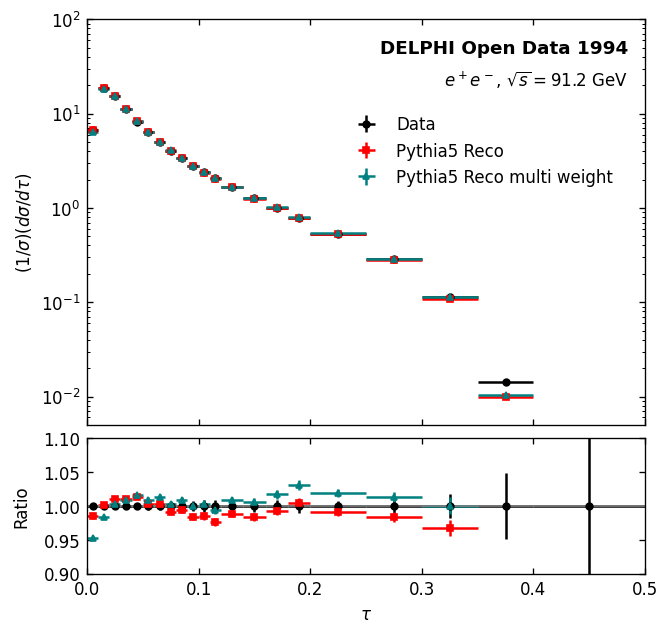}
    \includegraphics[width = 0.45\textwidth]{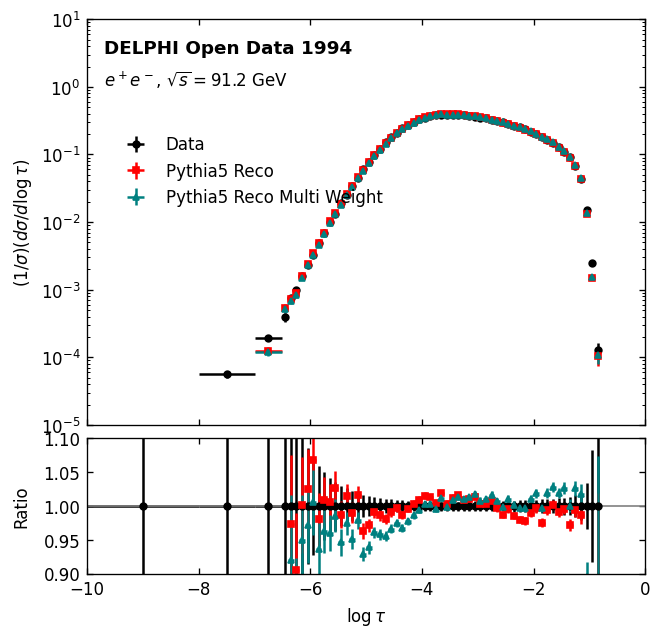}
    \caption{Effect of the event multiplicity reweighting on the reconstructed \(\tau\) (left) and \(\log\tau\) (right) distributions. The nominal 1994 MC (red) is compared to the varied 1994 MC (teal), which has been reweighted to match the multiplicity distribution of the 1995 data. The 1994 data is shown in black for reference.}
    \label{fig:multiCorr}
\end{figure}

\begin{figure}[ht!]
    \centering
    \includegraphics[width = 0.45\textwidth]{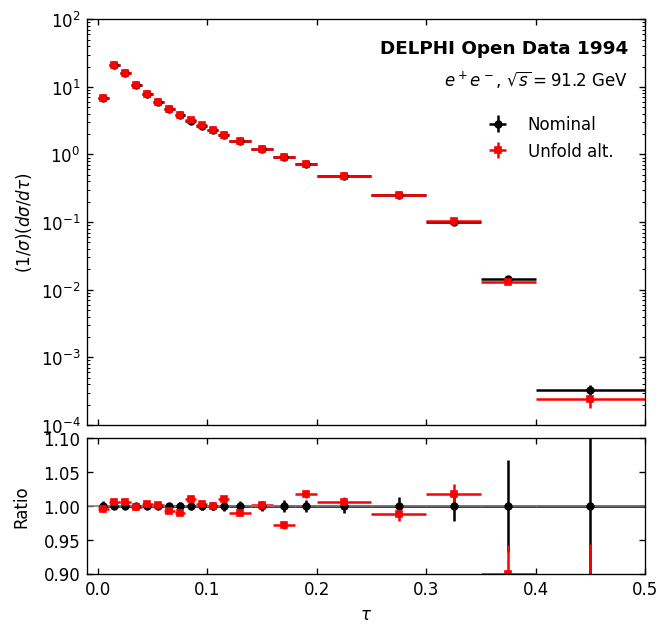}
    \includegraphics[width = 0.45\textwidth]{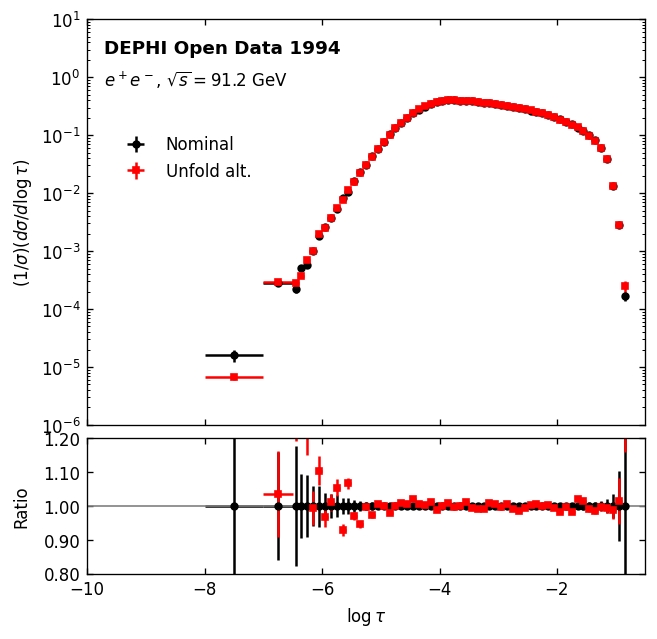}
    \caption{The relative uncertainty on the unfolded thrust distributions from using an alternative detector model. The effect is shown as the ratio of the result unfolded with the 1995 MC response matrix (red) to the nominal result unfolded with the 1994 MC matrix (black).}
    \label{fig:responseSys}
\end{figure}

\begin{enumerate}
    \item \textbf{Low-Energy neutral particles (Fig.~\ref{fig:cutNeutral}):} 
    To estimate the uncertainty from the modeling of low-energy neutral particles, the minimum energy cut is increased from 0.5~GeV to 5.0~GeV. The entire analysis chain, including the unfolding and phase-space correction, is then repeated. The full difference between this varied result and the nominal result is assigned as a systematic uncertainty.

    \item \textbf{Event multiplicity modeling (Fig.~\ref{fig:multiCorr}):} 
To investigate the impact of the observed year-to-year difference in the event multiplicity modeling, a reweighting is performed. An event-by-event weight is derived from the ratio of the 1995 data to the 1994 data multiplicity spectra. This weight is then applied to the PYTHIA 5 sample. The thrust variables are recalculated at the detector level with this reweighted MC sample, and the resulting change in the reconstructed thrust distributions, as shown in the figure, is taken as a conservative estimate of the systematic uncertainty associated with this population shift.

    \item \textbf{Alternative Detector Model (Fig.~\ref{fig:responseSys}):} 
    This study further addresses potential simulation inconsistencies by using the 1995 MC sample as an alternative model of the detector. The 1994 data is unfolded using a response matrix derived from the 1995 simulation (and vice versa for 1995). The difference between this result and the nominal unfolded distribution is taken as a systematic uncertainty. This directly quantifies the impact of the year-to-year changes in the official detector simulation.
\end{enumerate}

\clearpage
\subsubsection{Generation modeling systematics}
Systematic uncertainty sources can arise from a potential bias caused by the mis-modeling of data by the PYTHIA 5 MC used to derive the unfolding matrices and the acceptance and event selection corrections.

\begin{figure}[ht!]
    \centering
    \includegraphics[width = 0.45\textwidth]{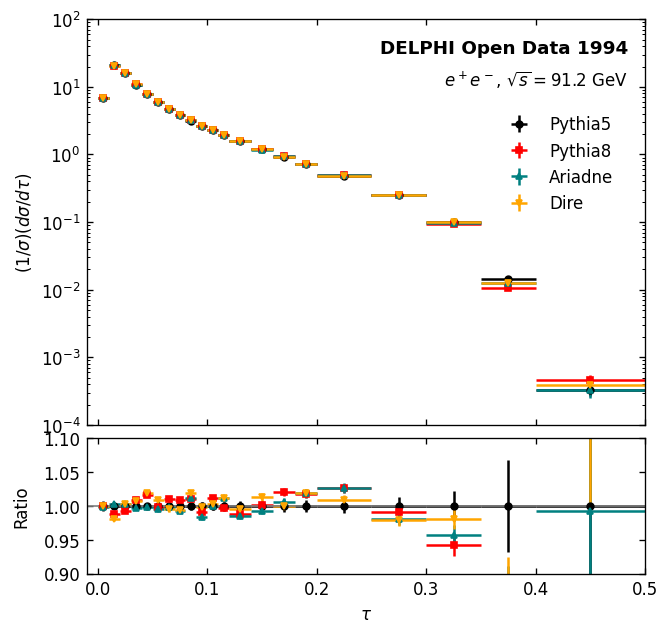}
    \includegraphics[width = 0.45\textwidth]{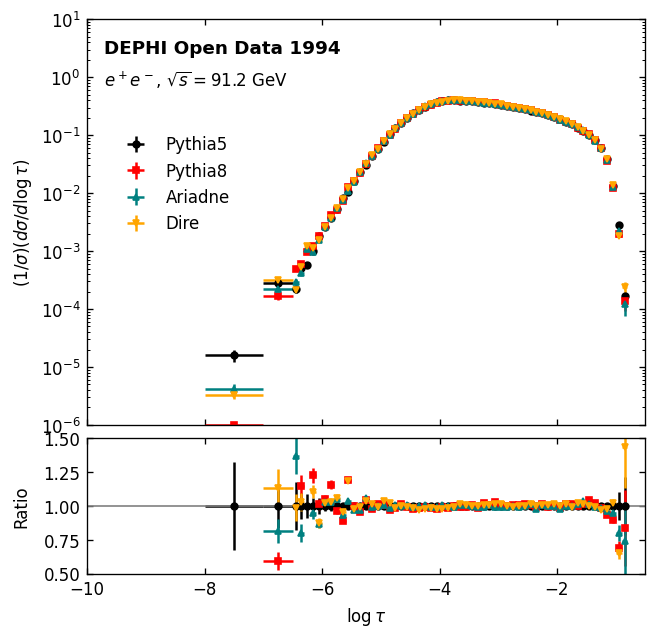}
    \caption{Unfolded thrust (left) and $\log\tau$ (right) distributions using response matrices derived from PYTHIA 5 (black), PYTHIA 8 (red), ARIADNE (teal), and PYTHIA 8 Dire (orange).}
    \label{fig:unfoldThrustModel}
\end{figure}

The modeling uncertainty associated with the \textbf{unfolding} is quantified by comparing the unfolded distribution using response matrices derived from the nominal PYTHIA 5, and alternative ARIADNE, PYTHIA 8, and PYTHIA 8 Dire generators. The comparison is shown in Fig.~\ref{fig:unfoldThrustModel}. The difference in results among these different MC is within 1\% in the perturbative region and grows to 50\% in the non-perturbative region.

\begin{figure}[ht!]
    \centering
    \includegraphics[width = 0.45\textwidth]{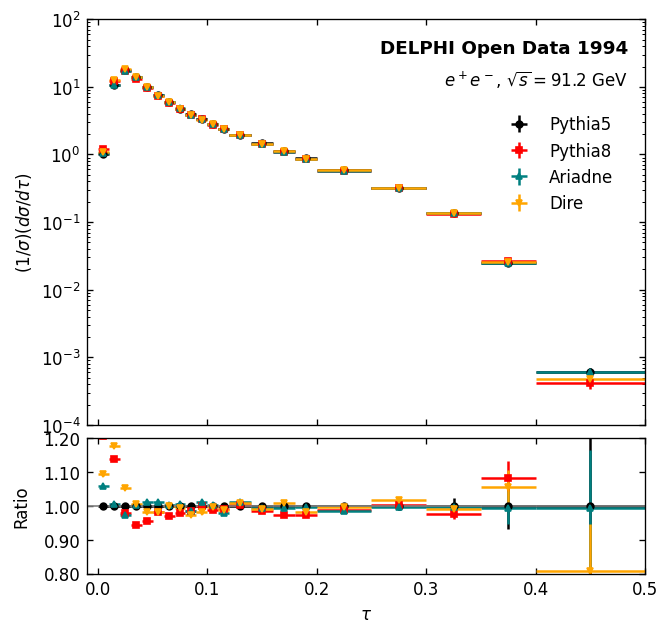}
    \includegraphics[width = 0.45\textwidth]{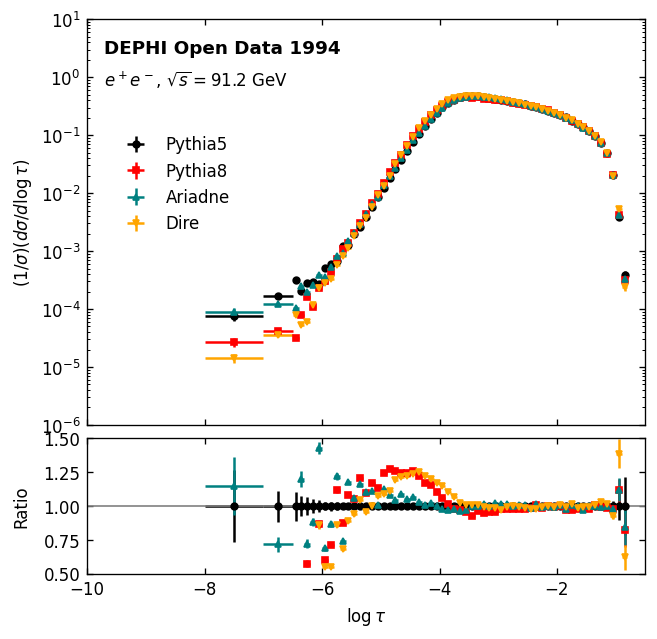}
    \caption{Unfolded $\tau$ (left) and $\log\tau$ (right) distributions corrected by acceptance correction factors derived from PYTHIA 5 (black), PYTHIA 8 (red), ARIADNE (teal), and PYTHIA 8 Dire (orange).}
    \label{fig:corrThrustModel}
\end{figure}

The modeling uncertainty associated with the \textbf{acceptance and event selection correction} is estimated by comparing corrected data using correction factors derived from PYTHIA 5, PYTHIA 8, ADRIANE, and PYTHIA 8 Dire generators, as shown in Fig.~\ref{fig:corrThrustModel}. 
The maximum deviations in these comparisons are assigned as the systematic uncertainties. 

\subsubsection{Summary of systematics}
\begin{figure}[ht!]
    \centering
    \includegraphics[width = 0.45\textwidth]{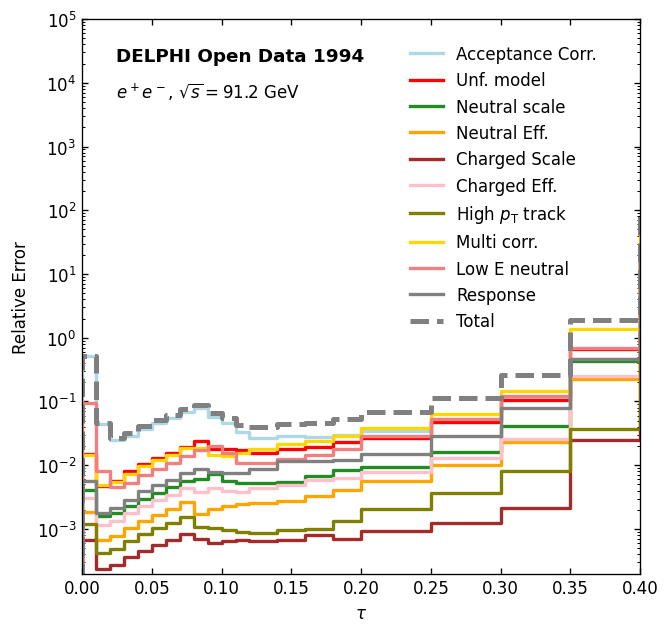}
    \includegraphics[width = 0.45\textwidth]{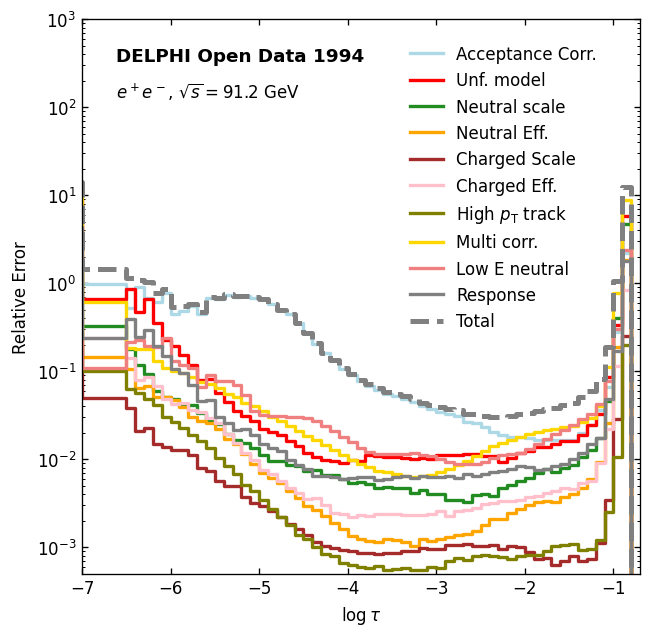}
    \caption{Summary of systematic uncertainties for $\tau$ (left) and $\log\tau$ (right) measurements.}
    \label{fig:sysSummaryThrust}
\end{figure}

A summary of the relative systematic uncertainties on the thrust measurement is shown in Figure~\ref{fig:sysSummaryThrust}. The total systematic uncertainty, shown as the gray dashed line, is calculated as the quadratic sum of each individual source.

As can be seen, the dominant contributions vary depending on the kinematic region. At small values of \(\tau\) (the non-perturbative, two-jet region), the uncertainty is dominated by the modeling of the acceptance correction. In the central and high-\(\tau\) regions, which are typically used for \(\alpha_s\) extractions, the largest uncertainties arise from the event multiplicity reweighting and the model dependence of the unfolding.
A detailed breakdown of the shape of each uncertainty component will be made available on HEPData after final paper publication. 

\clearpage
\section{Results}
\label{sec:results}

\begin{figure}[H]
    \centering
    \begin{overpic}[width=0.75\linewidth]{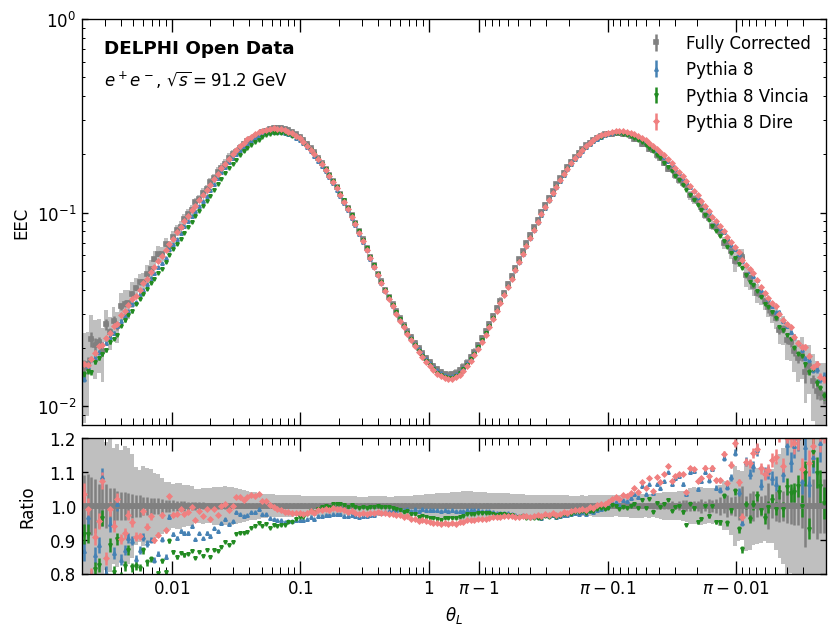}
      \put(36,64){\includegraphics[width=0.06\textwidth]{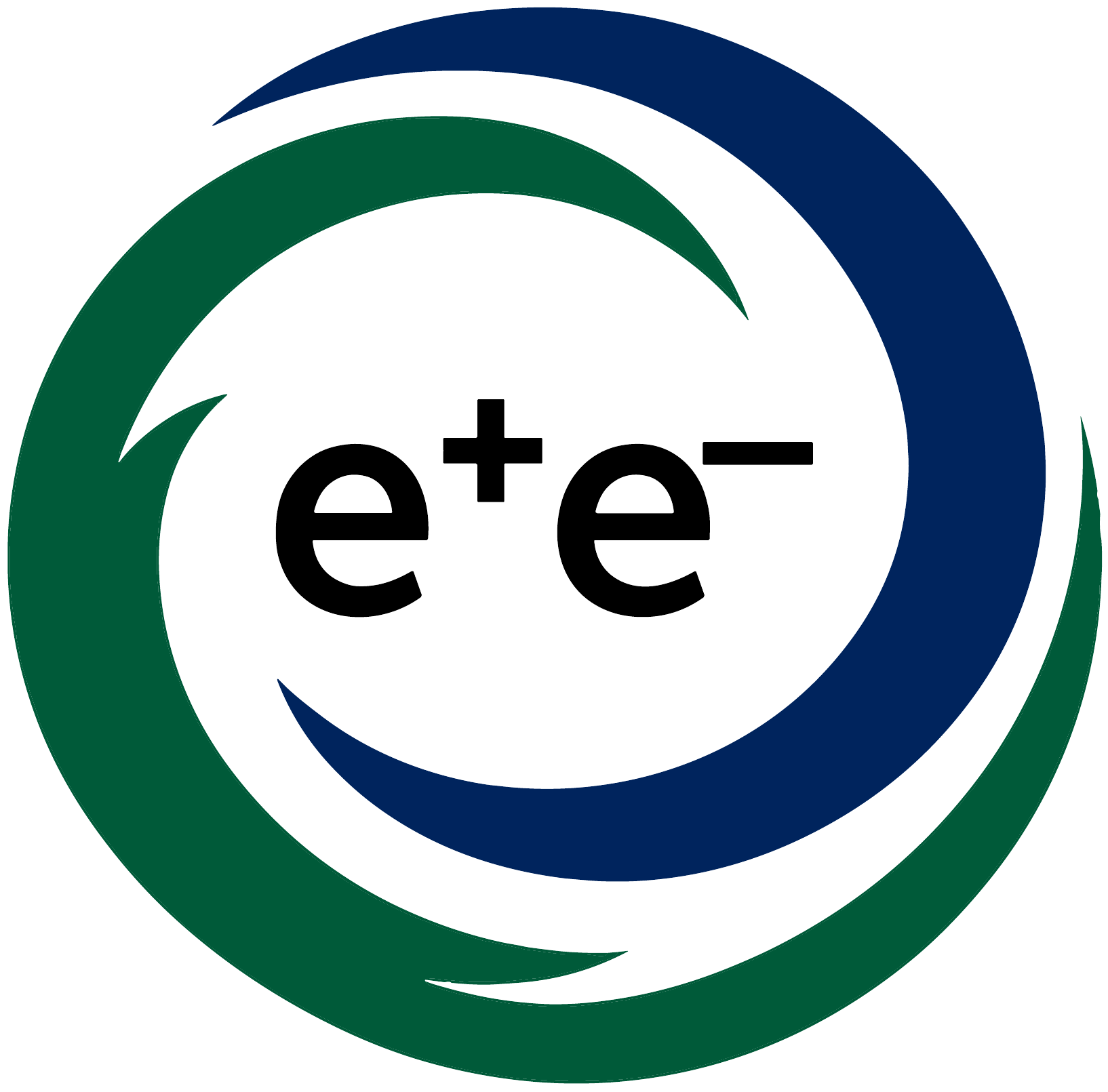}}
    \end{overpic}
    \caption{Fully-corrected EEC distributions from the DELPHI open data as a function of $\theta_{\rm L}$. Statistical error bars are displayed as vertical lines, and systematic error bars are shown in the gray boxes. Predictions from PYTHIA 8 default, Vincia, and Dire parton shower algorithms are shown in blue, green, and red, respectively. }
    \label{fig:resultsEECr}
\end{figure}

\begin{figure}[H]
    \centering
    \begin{overpic}[width=0.75\linewidth]{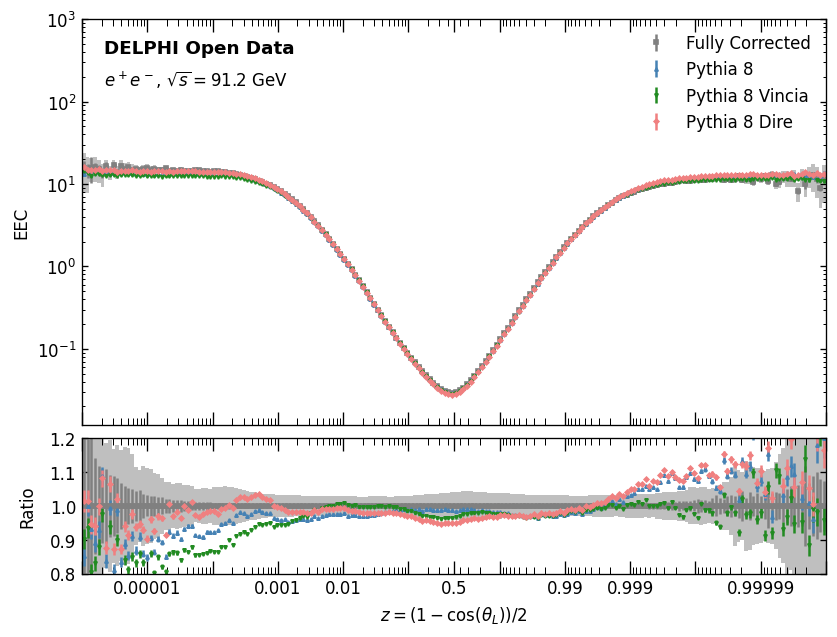}
    \put(36,64){\includegraphics[width=0.06\textwidth]{figures/Results/ee-logo.pdf}}
    \end{overpic}
    \caption{Fully-corrected EEC distributions from the DELPHI open data as a function of $z$. Statistical error bars are displayed as vertical lines, and systematic error bars are shown in the gray boxes. Predictions from PYTHIA 8 default, Vincia, and Dire parton shower algorithms are shown in blue, green, and red, respectively. }
    \label{fig:resultsEECz}
\end{figure}

The fully-corrected EEC distributions are presented as a function of \(\theta_\mathrm{L}\) and \(z\) in Figs.~\ref{fig:resultsEECr} and~\ref{fig:resultsEECz}, respectively. The results from the 1994 and 1995 datasets are combined, as they are found to show good consistency (see Figure~\ref{fig:compAleph}). Furthermore, the measurements are in good agreement with the recent re-analysis of ALEPH data~\cite{Bossi:2025xsi}, demonstrating strong cross-experiment validation (see Appendix~\ref{app:alephcomp}). While the ALEPH re-analysis was limited to angles above 0.006 ($\pi-0.006$) rad due to data format precision, this work extends the measurement's reach significantly further down to 0.002 ($\pi-0.002$) rad. The high-granularity of this measurement spans the full angular range from the collinear to the back-to-back regions. This provides a detailed view of the expected scaling behavior of QCD and the transition from the perturbative parton state to the non-perturbative confined hadron state.

These distributions are also compared to predictions from three modern parton shower algorithms in PYTHIA 8: the default, Vincia\cite{Skands:2016nry}, and Dire~\cite{Hoche:2015sya}. The level of agreement varies across the spectrum; for example, the Vincia shower provides a good description of the back-to-back region, while the Dire shower better captures the collinear region. For future detailed comparisons and the tuning of parton shower and hadronization models, the fully corrected distributions, with their covariance matrices, systematic uncertainties breakdowns, and a RIVET routine~\cite{BUCKLEY20132803} will be made available together with the journal publication.

\begin{figure}
    \centering
    \includegraphics[width=0.6\linewidth]{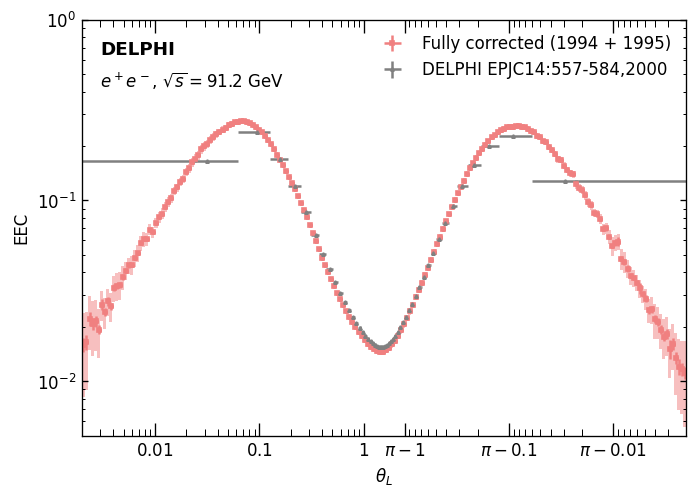}
    \caption{A comparison of the fully-corrected, track-based EEC distribution from this work (light coral) with the previous DELPHI measurement from Ref.~\cite{DELPHI:2000uri} (gray). For the measurement from this work, the statistical uncertainties are shown as vertical lines, and the total systematic uncertainties are shown as boxes. For the previous measurement, the error bars represent the statistical and systematic uncertainties added in quadrature. The definition of the EEC in this work differs from that of the legacy DELPHI analysis. Specifically, the legacy measurement normalizes the energy of charged track pairs by the visible event energy, while the current work uses the beam energy.}
    \label{fig:delphi_epjc}
\end{figure}

A comparison with the previous track-based DELPHI measurement~\cite{DELPHI:2000uri} is shown in Figure~\ref{fig:delphi_epjc}. This new measurement offers unprecedented resolution, particularly revealing the shape of the distribution in the collinear and back-to-back limits. This enhanced precision is crucial for detailed investigations of hadronization in the collinear limit and of non-perturbative power corrections and the Collins-Soper kernel in the back-to-back region. The definition of the EEC in this work differs from that of the legacy DELPHI analysis. Specifically, the legacy measurement normalizes the energy of charged track pairs by the visible event energy, while the current work uses the beam energy. Further physics interpretations, including detailed comparisons to analytical resummed predictions, will be presented in a forthcoming journal publication.

\begin{figure}[ht!]
    \centering
    \begin{minipage}{0.48\textwidth}
        \begin{overpic}[width=\textwidth]{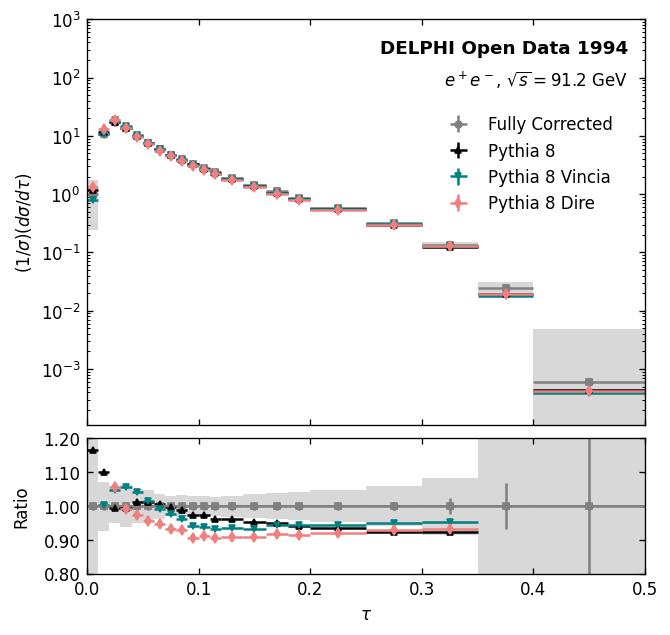}
         \put(16,82){\includegraphics[width=0.08\textwidth]{figures/Results/ee-logo.pdf}}
        \end{overpic}
        \centering
        \text{(a)}\\
    \end{minipage}
    \begin{minipage}{0.48\textwidth}
        \begin{overpic}[width=\textwidth]{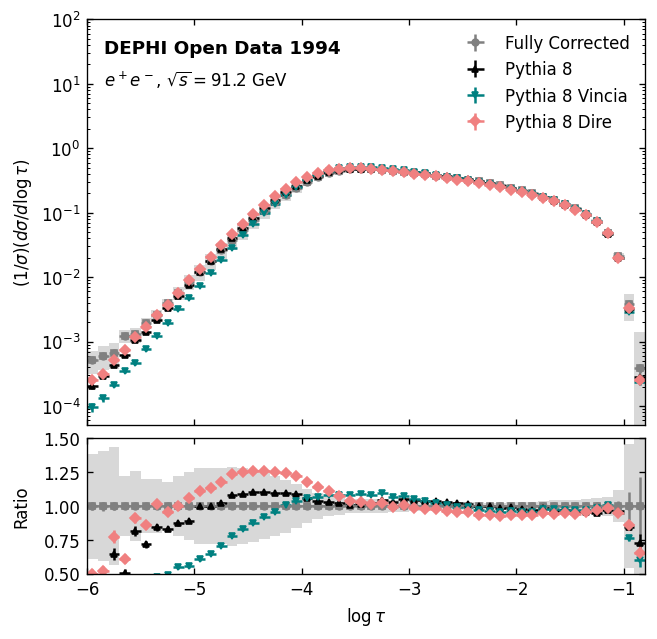}
         \put(16,74){\includegraphics[width=0.08\textwidth]{figures/Results/ee-logo.pdf}}
        \end{overpic}
        \centering
        \text{(d)}\\
    \end{minipage}\\
    \begin{minipage}{0.48\textwidth}
        \begin{overpic}[width=\textwidth]{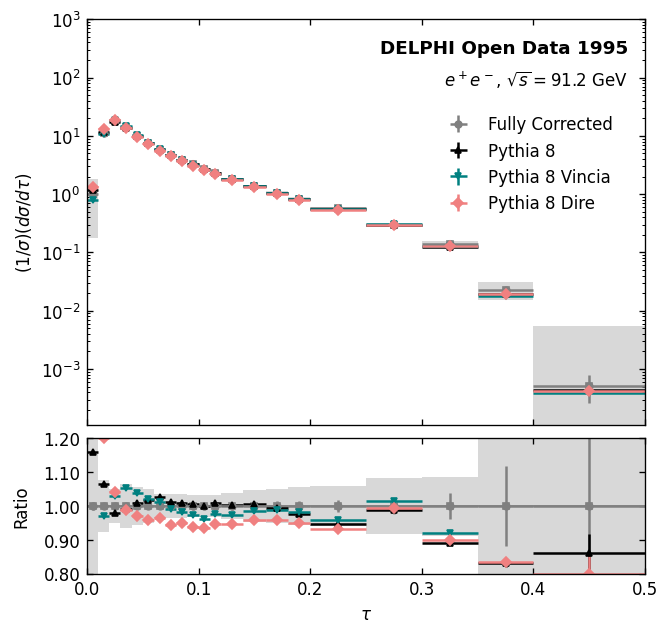}
        \put(16,82){\includegraphics[width=0.08\textwidth]{figures/Results/ee-logo.pdf}}
        \end{overpic}
        \centering
        \text{(c)}\\
    \end{minipage}
    \begin{minipage}{0.48\textwidth}
        \begin{overpic}[width=\textwidth]{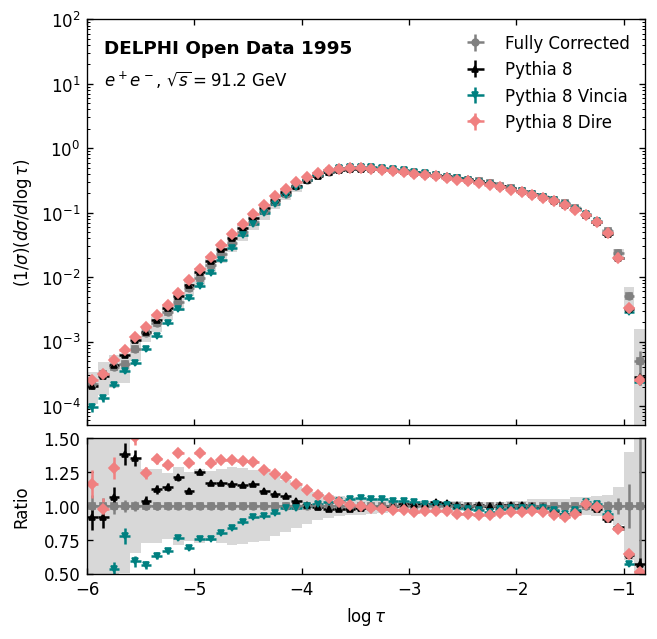}
         \put(16,74){\includegraphics[width=0.08\textwidth]{figures/Results/ee-logo.pdf}}
        \end{overpic}
        \centering
        \text{(d)}\\
    \end{minipage}\\
    \caption{Fully-corrected thrust distributions for the 1994 and 1995 datasets. Statistical uncertainties are displayed as vertical error bars, and the total systematic uncertainties are shown as gray boxes. The data are compared to predictions from three PYTHIA~8 parton shower models: default (blue), Vincia (green), and Dire (red).}
    \label{fig:resultsThrust}
\end{figure}

The fully-corrected \(\tau = 1-T\) and \(\log\tau\) distributions for the 1994 and 1995 data-taking years are presented separately in Figure~\ref{fig:resultsThrust}. The results are not combined in order to preserve information about potential year-to-year variations in detector performance. These measurements are compared to predictions from the three PYTHIA 8 parton shower models: the default, Vincia, and Dire. While the default PYTHIA 8 model provides the best overall description, both the Vincia and Dire showers exhibit noticeable deviations from the data, particularly in the non-perturbative, two-jet region at small \(\tau\). To facilitate more detailed comparisons and precision constraints on \(\alpha_s\) and non-perturbative models, a full breakdown of all systematic uncertainty components will be made publicly available on HEPData.

\begin{figure}
    \centering
    \begin{overpic}[width=0.75\linewidth]{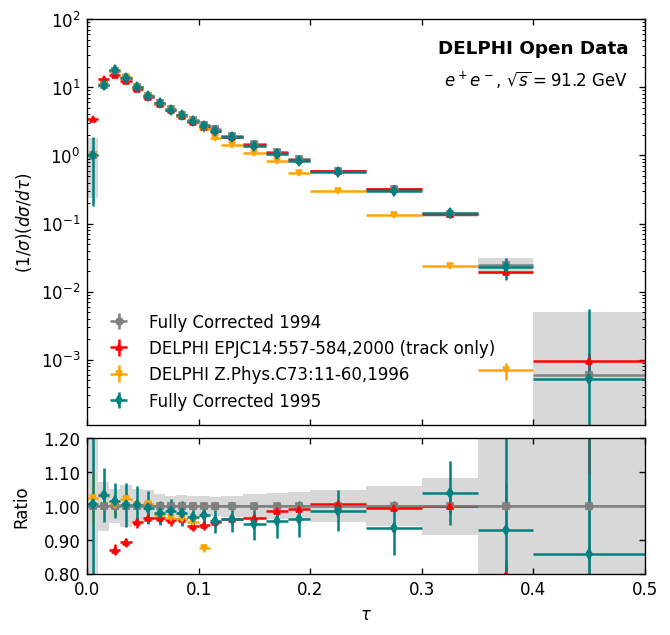}
    \put(85,70){\includegraphics[width=0.07\textwidth]{figures/Results/ee-logo.pdf}}
    \end{overpic}
    \caption{Comparison of fully-corrected thrust (\(\tau\)) distributions. The new all-particle measurement from this work for 1994 (gray) is shown with statistical uncertainties as vertical lines and total uncertainties as gray boxes. This is compared with the 1995 result from this work (teal), a legacy track-only measurement from 1994~\cite{DELPHI:2000uri} (red), and a legacy all-particle measurement from 1991-1993~\cite{DELPHI:1996sen} (orange). For these latter three datasets, the error bars represent the statistical and systematic uncertainties added in quadrature.}
    \label{fig:resultsCompThrust}
\end{figure}

Figure~\ref{fig:resultsCompThrust} presents a comparison of this thrust measurement with previous DELPHI results that are publicly available on HEPData. The legacy results include a track-only measurement from 1994 data~\cite{DELPHI:2000uri} (red points) and an all-particle measurement using 1991-1993 data~\cite{DELPHI:1996sen} (orange points). Other DELPHI measurements are not shown as their data is not publicly available. Significant differences are observed between this work and the legacy all-particle thrust measurements. A cross-check of the track-only thrust can be seen in Appendix~\ref{app:trackThrust}, showing good agreement between this and the legacy results. 

\section{Summary and outlook}
\label{sec:summary}
In this note, new measurements of the thrust and EEC observables are presented, using the DELPHI open data from \(e^+e^-\) collisions at \(\sqrt{s} = 91.2\)~GeV. This analysis represents the first comprehensive physics study using these legacy datasets, establishing a benchmark for future investigations. A defining feature of this measurement is a rigorous and detailed evaluation of systematic uncertainties, performed to modern standards.

The track-based EEC measurement achieves a significantly improved resolution over previous DELPHI results, enabling detailed studies of QCD in the collinear (\(\theta_{\mathrm{L}} \to 0\)) and back-to-back (\(\theta_{\mathrm{L}} \to \pi\)) regions. The all-particle thrust measurement focuses on systematic uncertainty breakdown, provides a high-precision result for new constraints on the strong coupling constant (\(\alpha_s\)) and non-perturbative physics models. Crucially, by establishing this binned result, this work serves as an essential benchmark for future unbinned measurements of these and other event shapes with the same data~\cite{Andreassen:2019cjw,Badea:2025wzd}.

This work marks the beginning of a renewed program of precision QCD with LEP data, revisiting classic observables to address contemporary questions. Natural extensions include measurements of the EEC's energy evolution, higher-point correlators~\cite{Budhraja:2024xiq,Alipour-fard:2024szj}, and flavor-tagged event shapes~\cite{Craft:2022kdo}. Excitingly, a new extraction of \(\alpha_s\) from these distributions is particularly timely, as results from \(e^+e^-\) event shapes with analytical hadronization models were recently excluded from the world average~\cite{ParticleDataGroup:2020ssz,ParticleDataGroup:2022pth}. This work also has the potential to shape the future, serving as a catalyst to inspire and inform studies at the proposed $e^{+}e^{-}$ colliders~\cite{Benedikt:2651299, CEPCStudyGroup:2018rmc, CEPCStudyGroup:2025kmw}.

\clearpage
\section*{Acknowledgments}
This work would not have been possible without the decades of effort by the DELPHI collaboration in designing, building, and operating the detector, nor without the foresight of its data preservation team. We are profoundly grateful for their monumental effort in making these pristine datasets and simulation codes publicly available.

We would especially like to thank Ulrich Schwickerath and Dietrich Liko for their guidance on the DELPHI open data. For many theoretical insights that shaped this work, we are also grateful to Miguel Benitez, Hao Chen, Max Jaarsma, Kyle Lee, Yibei Li, Ian Moult, Iain Stewart, Jesse Thaler, Wouter Waalewijn, Xiaoyuan Zhang, and HuaXing Zhu. We would like to extend special thanks to Ian Moult and HuaXing Zhu for their invaluable discussions on the analysis strategy.

\clearpage

\bibliographystyle{JHEP}
\typeout{}
\bibliography{EEC}

\clearpage
\appendix

\section{Analysis code}
\label{app:analysiscode}
The various software bases used for this analysis are listed below. 

The code used to convert the DELPHI open data in its original ZEBRA format to the ROOT format is available publicly at \url{https://github.com/jingyucms/delphi-nanoaod}. 

The code used for the analysis is available publicly at \url{https://github.com/jingyucms/delphi-nanoaod/delphi-analysis/python}. 

The code used for DELPHI detector simulation is available publicly at \url{https://github.com/jingyucms/Delphi-Sim-Pipeline}. 

\section{Binnings for the thrust variables}
\label{app:thrustbinning}

\begin{table}[ht!]
    \centering
    \caption{Bin boundaries for the \(\tau = 1-T\) distribution.}
    \label{tab:tauBinning}
    \begin{tabular}{c}
        \toprule
        \(\tau\) bin boundaries \\
        \midrule
        \texttt{0.00, 0.01, 0.02, 0.03, 0.04, 0.05, 0.06} \\
        \texttt{0.07, 0.08, 0.09, 0.10, 0.11, 0.12, 0.14} \\
        \texttt{0.16, 0.18, 0.20, 0.25, 0.30, 0.35, 0.40} \\
        \texttt{0.50, 1.00} \\
        \bottomrule
    \end{tabular}
\end{table}

\begin{table}[ht!]
    \centering
    \caption{Bin boundaries for the \(\log\tau\) distribution.}
    \label{tab:logTauBinning} 
    \begin{tabular}{c}
        \toprule
        \(\log\tau\) bin boundaries \\
        \midrule
        \texttt{-10.0,  -8.0,  -7.0,  -6.5,  -6.4,  -6.3,  -6.2,  -6.1,  -6.0} \\
        \texttt{ -5.9,  -5.8,  -5.7,  -5.6,  -5.5,  -5.4,  -5.3,  -5.2,  -5.1} \\
        \texttt{ -5.0,  -4.9,  -4.8,  -4.7,  -4.6,  -4.5,  -4.4,  -4.3,  -4.2} \\
        \texttt{ -4.1,  -4.0,  -3.9,  -3.8,  -3.7,  -3.6,  -3.5,  -3.4,  -3.3} \\
        \texttt{ -3.2,  -3.1,  -3.0,  -2.9,  -2.8,  -2.7,  -2.6,  -2.5,  -2.4} \\
        \texttt{ -2.3,  -2.2,  -2.1,  -2.0,  -1.9,  -1.8,  -1.7,  -1.6,  -1.5} \\
        \texttt{ -1.4,  -1.3,  -1.2,  -1.1,  -1.0,  -0.9,  -0.8,  -0.7,  -0.6} \\
        \texttt{ -0.5,  -0.4,  -0.3,  -0.2,  -0.1,   0.0} \\
        \bottomrule
    \end{tabular}
\end{table}

\section{Derivation of the covariance matrix formula}
\label{app:eq5.1}
\begin{align*}
S &= \frac{1}{N-1} \sum_{i=1}^{N} (y_i - \bar{y})(y_i - \bar{y})^T \\
&= \frac{1}{N-1} \sum_{i=1}^{N} (y_i y_i^T - y_i \bar{y}^T - \bar{y} y_i^T + \bar{y}\bar{y}^T) \\
&= \frac{1}{N-1} \left( \sum_{i=1}^{N} y_i y_i^T - \sum_{i=1}^{N} y_i \bar{y}^T - \sum_{i=1}^{N} \bar{y} y_i^T + \sum_{i=1}^{N} \bar{y}\bar{y}^T \right) \\
&= \frac{1}{N-1} \left( \sum_{i=1}^{N} y_i y_i^T - \left(\sum_{i=1}^{N} y_i\right) \bar{y}^T - \bar{y} \left(\sum_{i=1}^{N} y_i^T\right) + N\bar{y}\bar{y}^T \right) \\
&= \frac{1}{N-1} \left( \sum_{i=1}^{N} y_i y_i^T - (N\bar{y})\bar{y}^T - \bar{y}(N\bar{y}^T) + N\bar{y}\bar{y}^T \right) \\
&= \frac{1}{N-1} \left( \sum_{i=1}^{N} y_i y_i^T - N\bar{y}\bar{y}^T - N\bar{y}\bar{y}^T + N\bar{y}\bar{y}^T \right) \\
&= \frac{1}{N-1} \left( \sum_{i=1}^{N} y_i y_i^T - N\bar{y}\bar{y}^T \right) \\
V_y &= \frac{1}{N} S \\
&= \frac{1}{N} \left[ \frac{1}{N-1} \left( \sum_{i=1}^{N} y_i y_i^T - N\bar{y}\bar{y}^T \right) \right] \\
&= \frac{1}{N(N-1)} \left( \sum_{i=1}^{N} y_i y_i^T - N\bar{y}\bar{y}^T \right) \\
&= \frac{1}{N(N-1)} \left( \sum_{i=1}^{N} y_i y_i^T - N \left(\frac{1}{N}\sum_{i=1}^{N} y_i\right) \left(\frac{1}{N}\sum_{j=1}^{N} y_j\right)^T \right) \\
V_y &= \frac{1}{N(N-1)} \left( \sum_{i=1}^{N} y_i y_i^T - \frac{1}{N} \left(\sum_{i=1}^{N} y_i\right) \left(\sum_{j=1}^{N} y_j\right)^T \right)
\end{align*}

\clearpage
\section{Full analysis chain closure test}
\label{app:CorrectionClosure}
A closure test using the simulated PYTHIA 5 sample is performed on all the EEC analysis steps, including fake subtraction, unfolding, matching efficiency correction, 2D projection correction, and event and track selection correction, before unblinding using data. 
The result, shown in~Fig.~\ref{fig:fullclosure}, demonstrates the perfect closure of the analysis. 
\begin{figure}[ht!]
\centering
\includegraphics[width=0.6\linewidth]{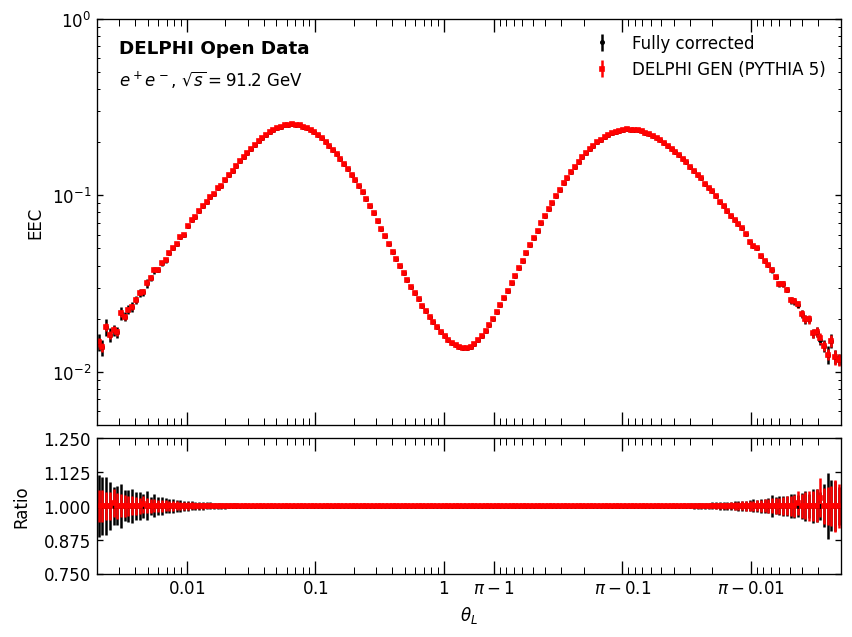}
\caption{EEC from fully corrected simulated PYTHIA 5 sample (black) and the generator-level PYTHIA 5 sample (red). The fully corrected sample is lying on top of the simulated one, showing perfect closure. }
\label{fig:fullclosure}
\end{figure}

\begin{figure}[ht!]
\centering
\includegraphics[width=0.4\linewidth]{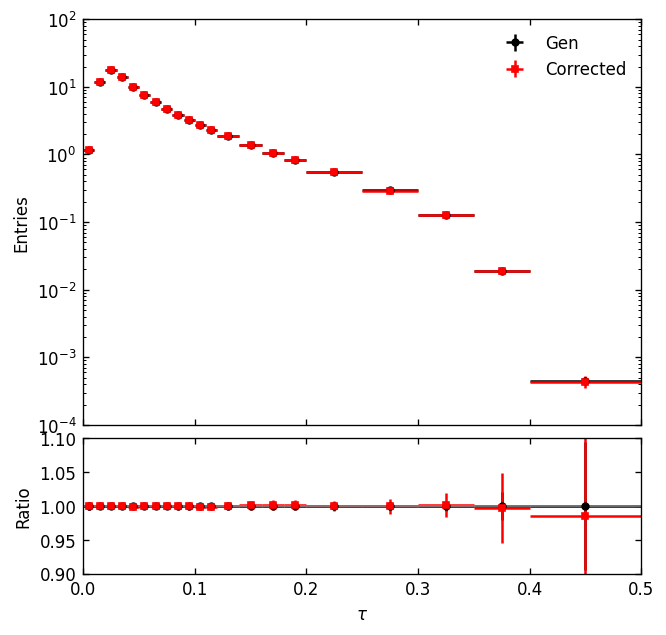}
\caption{Thrust from fully corrected simulated PYTHIA 5 sample (black) and the generator-level PYTHIA 5 sample (red). The fully corrected sample is lying on top of the simulated one, showing perfect closure. }
\label{fig:fullclosure}
\end{figure}

\clearpage
\section{PYTHIA 8 simulation efficiency and fake rate}
\label{app:pythia8perf}
\begin{figure}[ht!]
    \centering
    \includegraphics[width=0.45\linewidth]{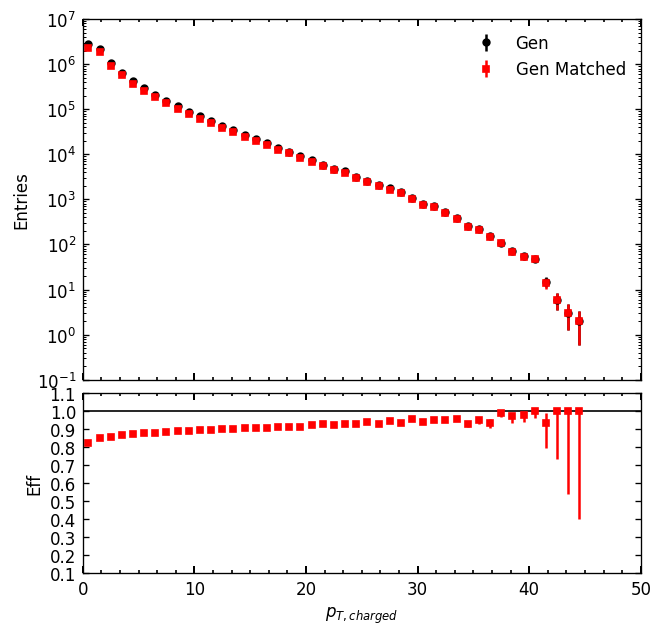}
    \includegraphics[width=0.45\linewidth]{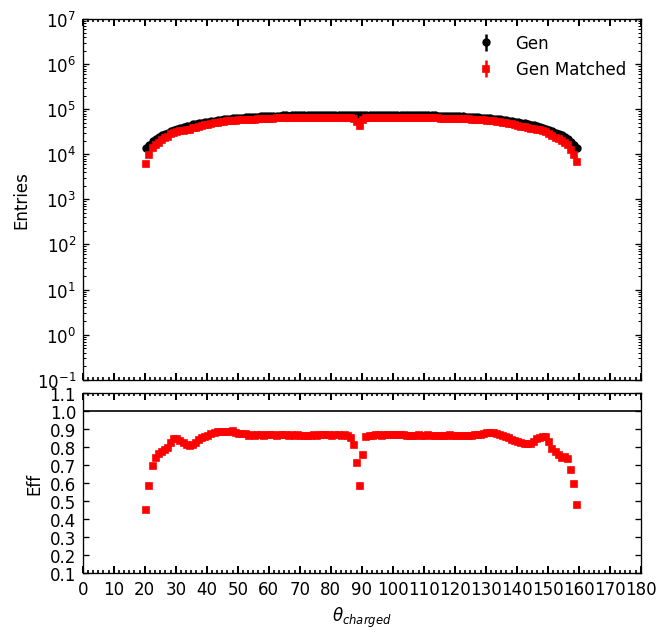}
    \caption{Single track matching efficiency as a function of PYTHIA 8 generator-level $p_{\rm T}$ (left) and $\theta$ (right). The upper panels show the distributions of matched (red) and all (black) generator-level tracks, while the lower panels show the efficiency ratio.}
    \label{fig:trk_eff_pythia}
\end{figure}

\begin{figure}[ht!]
    \centering
    \includegraphics[width=0.45\linewidth]{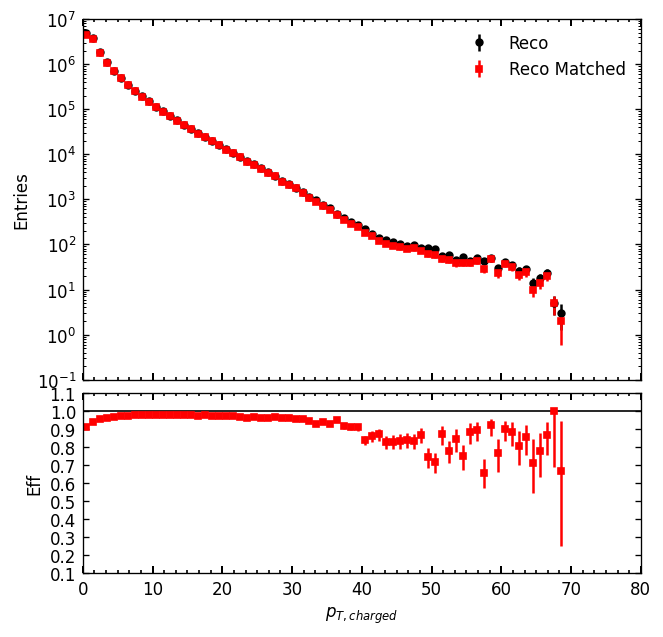}
    \includegraphics[width=0.45\linewidth]{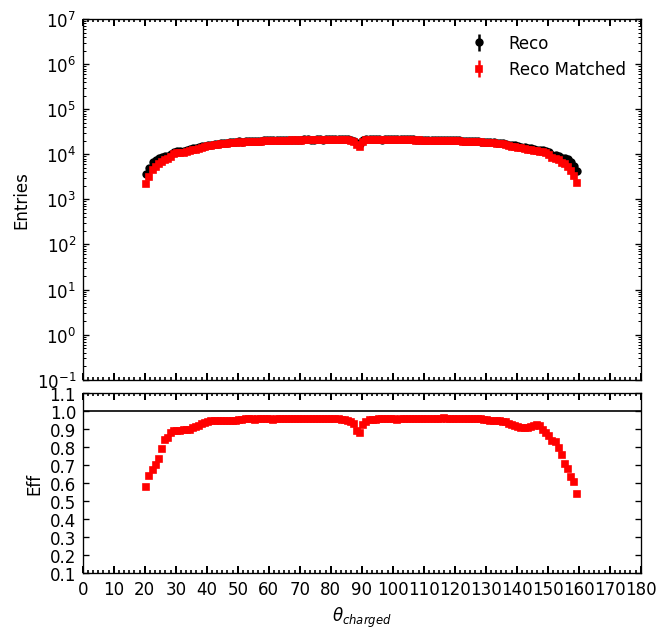}
    \caption{Single track fake rate as a function of reconstructed PYTHIA 8 $p_{\rm T}$ (left) and $\theta$ (right). The upper panels show the distributions of matched (red) and all (black) reconstruction-level tracks, while the lower panels show the fake rate ratio.}
    \label{fig:trk_fake_pythia}
\end{figure}

To cross-check the correctness of the self-developed PYTHIA-to-DELSIM workflow, several cross-checks are made on the track reconstruction performance. 
Figs.~\ref{fig:trk_eff_pythia} and \ref{fig:trk_fake_pythia} show the track efficiencies and fakes derived from the PYTHIA 8 sample. 
The results show good agreement with those from the DELPHI legacy PYTHIA 5 sample shown in Section~\ref{sec:matching}, demonstrating the validity of the simulation pipeline.  

\clearpage
\section{Treatment of V0 decays in the DELPHI simulation}
\label{app:Ks}

A unique feature of the DELPHI simulation chain is the handling of long-lived neutral particles (V0's), primarily $K^{0}_{\rm S}$ and $\Lambda$. 
In the nominal PYTHIA~5.7/JETSET~7.4 simulation, these particles (identified with status code 4) are not decayed by the event generator. 
Instead, their decays are handled within the GEANT3-based detector simulation to more accurately model interactions with detector material. 
This procedure, confirmed by the DELPHI data preservation team, differs from the approach used in many other experiments.

\begin{figure}[ht!]
    \centering
    \includegraphics[width=0.6\linewidth]{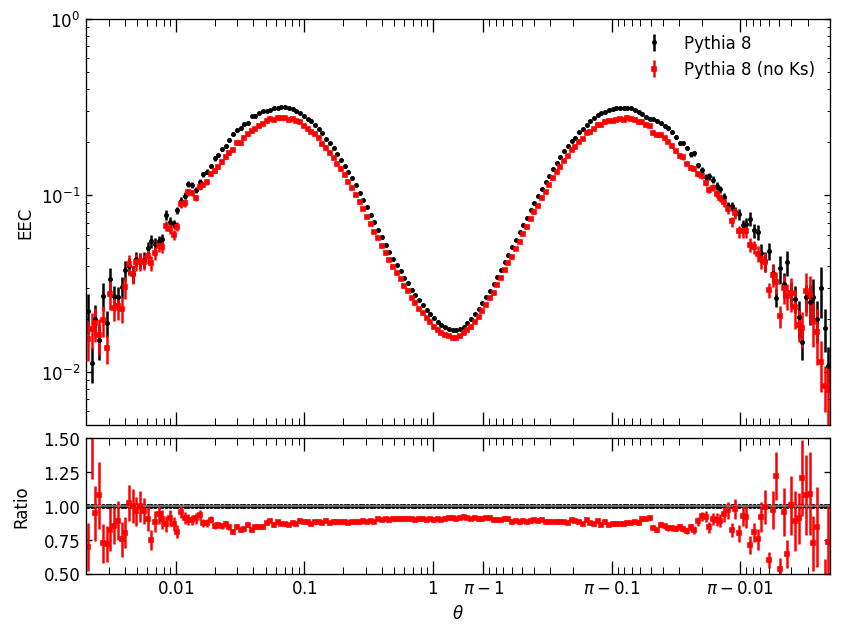}
    \caption{A generator-level comparison of the EEC distribution from a standard PYTHIA~8 sample (black) and a sample where $K^{0}_{\rm S}$ decays have been disabled (red).}
    \label{fig:KsEEC}
\end{figure}

\begin{figure}[ht!]
    \centering
    \includegraphics[width=0.6\linewidth]{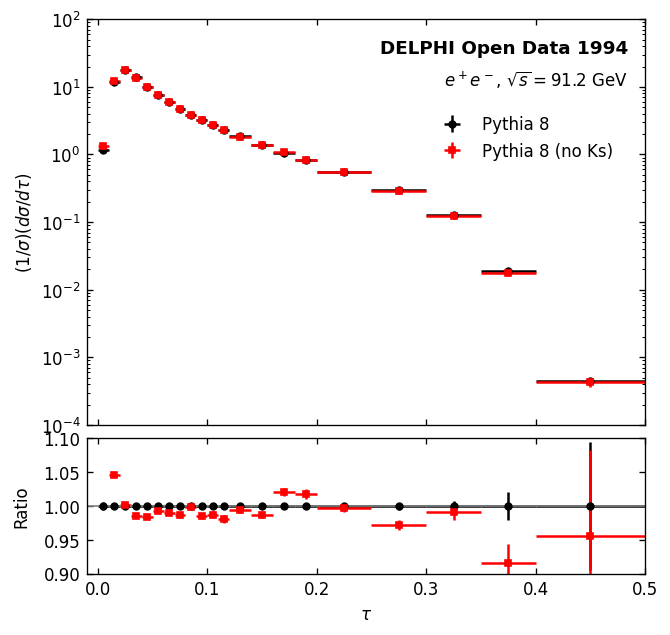}
    \caption{A generator-level comparison of the thrust distribution from a standard PYTHIA~8 sample (black) and a sample where $K^{0}_{\rm S}$ decays have been disabled (red).}
    \label{fig:KsThrust}
\end{figure}

This simulation choice has a direct impact on the generator-level distributions used for corrections. 
To illustrate the effect, the decay of $K^{0}_{\rm S}$ particles was disabled in a sample of PYTHIA~8 events. 
Figures~\ref{fig:KsEEC} and~\ref{fig:KsThrust} show the resulting generator-level EEC and thrust distributions, comparing the nominal PYTHIA~8 (with decays) to the sample without $K^{0}_{\rm S}$ decays, revealing a significant difference in the observables.

To construct a particle-level definition consistent with modern theoretical predictions, the decay products of these V0's must be retrieved from the simulation output. A complication arises if a V0 particle interacts with the detector material before it decays, as this can initiate a particle shower. In such cases, the summed four-momentum of the simulated shower particles will not match the initial V0's four-momentum. This occurs in approximately 0.2\% of \(Z \to q\bar{q}\) events in the PYTHIA 5 sample. To handle these cases while preserving the full event sample, the original generator-level V0 is kept as a single stable neutral particle, and its shower-level decay products are discarded.


\clearpage
\section{Comparison of EEC between data-taking years and with ALEPH}
\label{app:alephcomp}
\begin{figure}[H]
    \centering
    \includegraphics[width=0.6\linewidth]{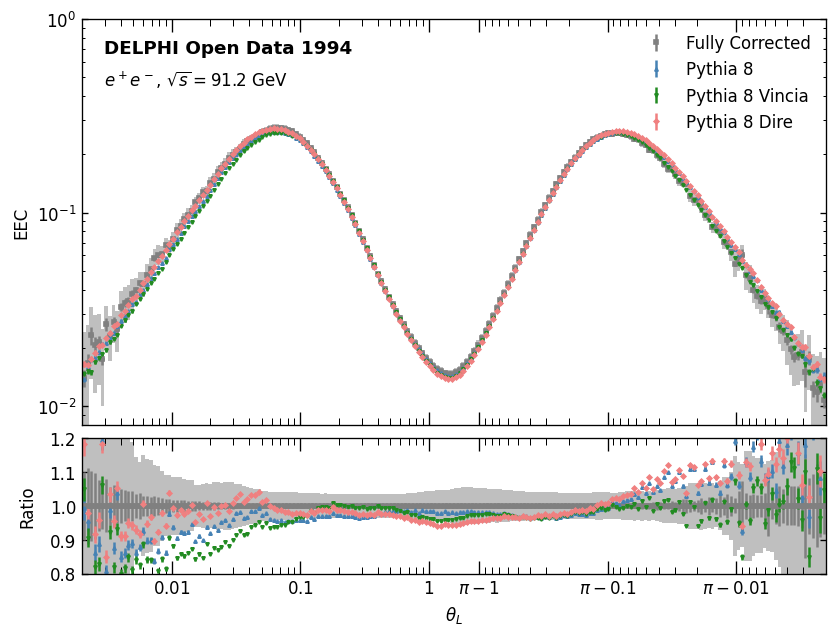}
    \caption{Fully-corrected EEC distributions from the 1994 DELPHI open data as a function of $\theta_{\rm L}$. Statistical error bars are displayed as vertical lines, and systematic error bars are shown in the gray boxes. Predictions from PYTHIA 8 default, Vincia, and Dire parton shower algorithms are shown in blue, green, and red, respectively. }
    \label{fig:resultsEECr94}
\end{figure}

\begin{figure}[H]
    \centering
    \includegraphics[width=0.6\linewidth]{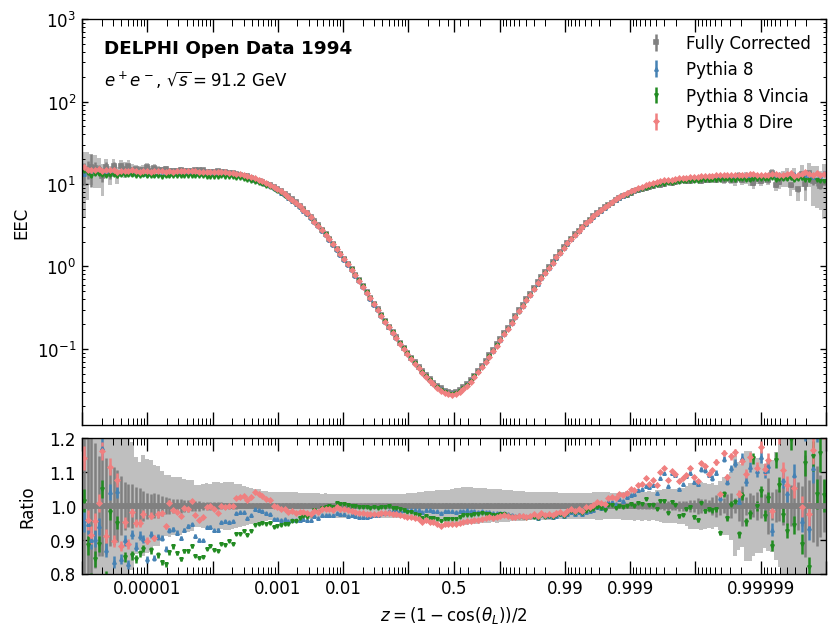}
    \caption{Fully-corrected EEC distributions from the 1994 DELPHI open data as a function of $z$. Statistical error bars are displayed as vertical lines, and systematic error bars are shown in the gray boxes. Predictions from PYTHIA 8 default, Vincia, and Dire parton shower algorithms are shown in blue, green, and red, respectively. }
    \label{fig:resultsEECz94}
\end{figure}

\begin{figure}[H]
    \centering
    \includegraphics[width=0.6\linewidth]{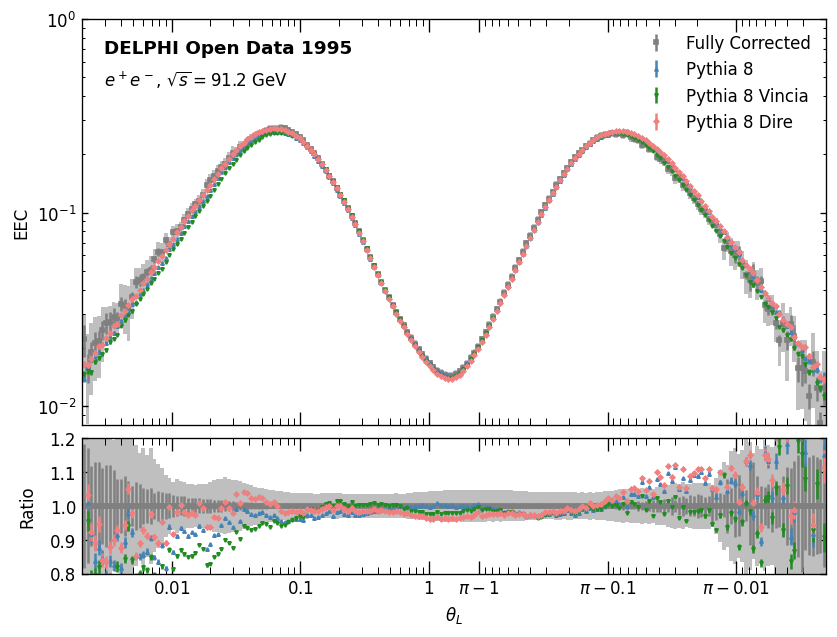}
    \caption{Fully-corrected EEC distributions from the 1995 DELPHI open data as a function of $\theta_{\rm L}$. Statistical error bars are displayed as vertical lines, and systematic error bars are shown in the gray boxes. Predictions from PYTHIA 8 default, Vincia, and Dire parton shower algorithms are shown in blue, green, and red, respectively. }
    \label{fig:resultsEECr95}
\end{figure}

\begin{figure}[H]
    \centering
    \includegraphics[width=0.6\linewidth]{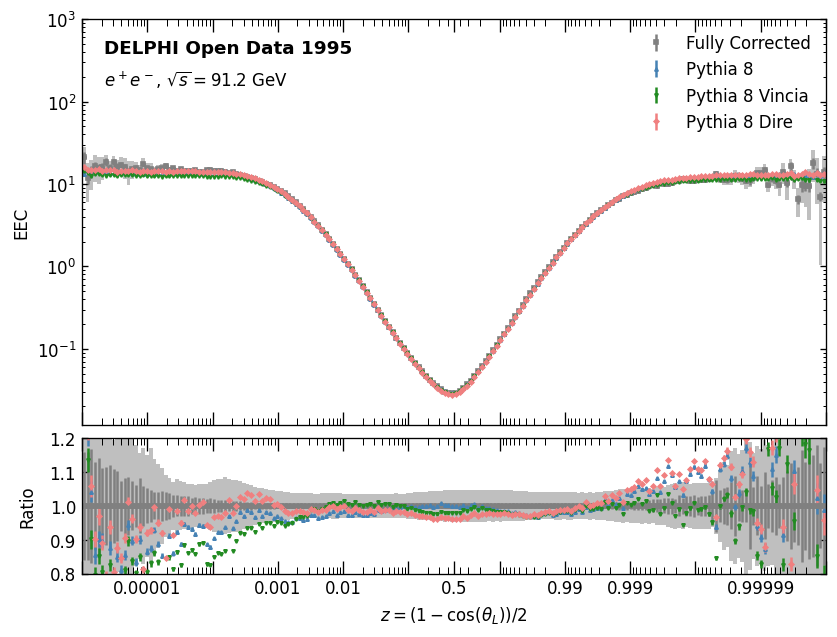}
    \caption{Fully-corrected EEC distributions from the 1995 DELPHI open data as a function of $z$. Statistical error bars are displayed as vertical lines, and systematic error bars are shown in the gray boxes. Predictions from PYTHIA 8 default, Vincia, and Dire parton shower algorithms are shown in blue, green, and red, respectively. }
    \label{fig:resultsEECz95}
\end{figure}

The fully corrected EEC distribution from the 1994 and 1995 DELPHI datasets is shown in Figs~\ref{fig:resultsEECr94},~\ref{fig:resultsEECz94},~\ref{fig:resultsEECr95}, and~\ref{fig:resultsEECz95}. 

\clearpage
The results are also compared to the recent re-analysis of ALEPH data~\cite{Bossi:2025xsi}.
The DELPHI results from the two different data-taking years show good agreement, indicating a high level of stability in the final corrected measurement. Furthermore, the DELPHI distributions are in good agreement with the ALEPH re-analysis result across the entire angular range. This cross-experiment consistency provides a strong validation for both of the measurement procedures. Noted that the ALEPH re-analysis did not provide results for angles below 0.006 ($\pi-0.002$) rad due to precision limitations in its data format.
\begin{figure}[ht!]
\centering
\includegraphics[width=0.75\linewidth]{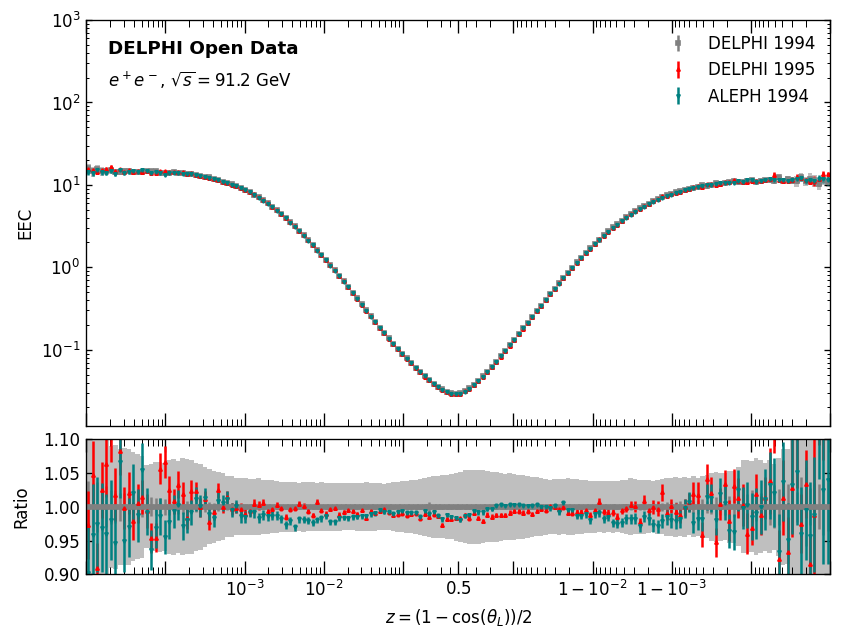}
\caption{Comparison of the fully corrected EEC distributions from the 1994 (light coral) and 1995 (black) DELPHI datasets with the recent ALEPH re-analysis result~\cite{Bossi:2025xsi} (blue). The ALEPH points include statistical uncertainties only.}
\label{fig:compAleph}
\end{figure}


\clearpage
\section{Additional figures for 1995 track matching performance}\label{app:1995}
\begin{figure}[ht!]
    \centering
    \includegraphics[width=\linewidth]{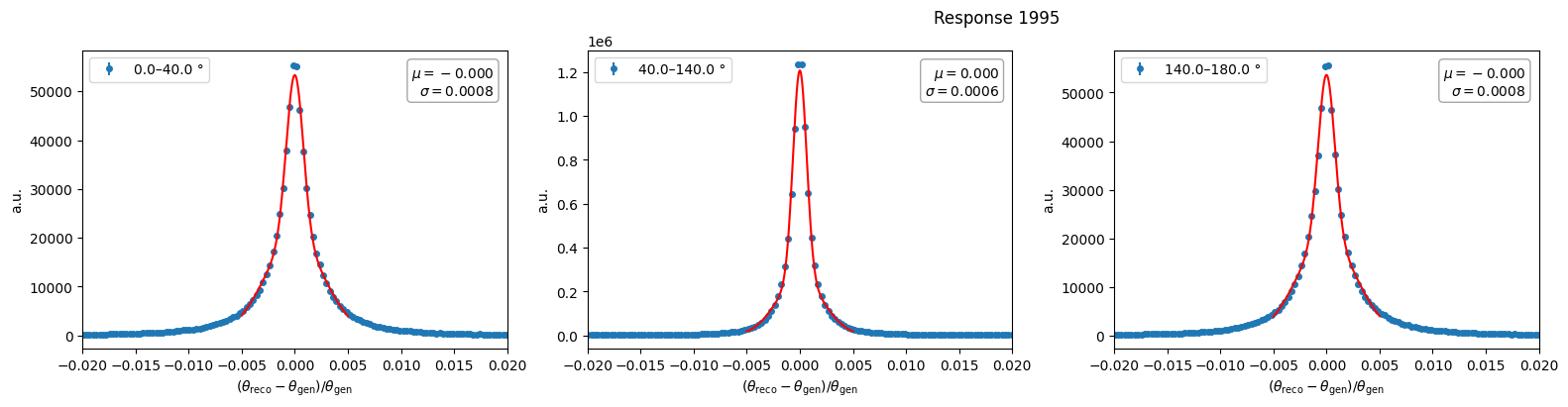}
    \caption{Track $\theta$ resolution distributions for 1995 in different detector regions: forward endcap (left), barrel (middle), and backward endcap (right). The $\theta$ response is calculated as $(\theta_{\rm reco} - \theta_{\rm gen}) / \theta_{\rm gen}$. Double Gaussian fits are overlaid to extract the resolution parameters.}
    \label{fig:theta_res_1995}
\end{figure}

\begin{figure}[ht!]
    \centering
    \includegraphics[width=\linewidth]{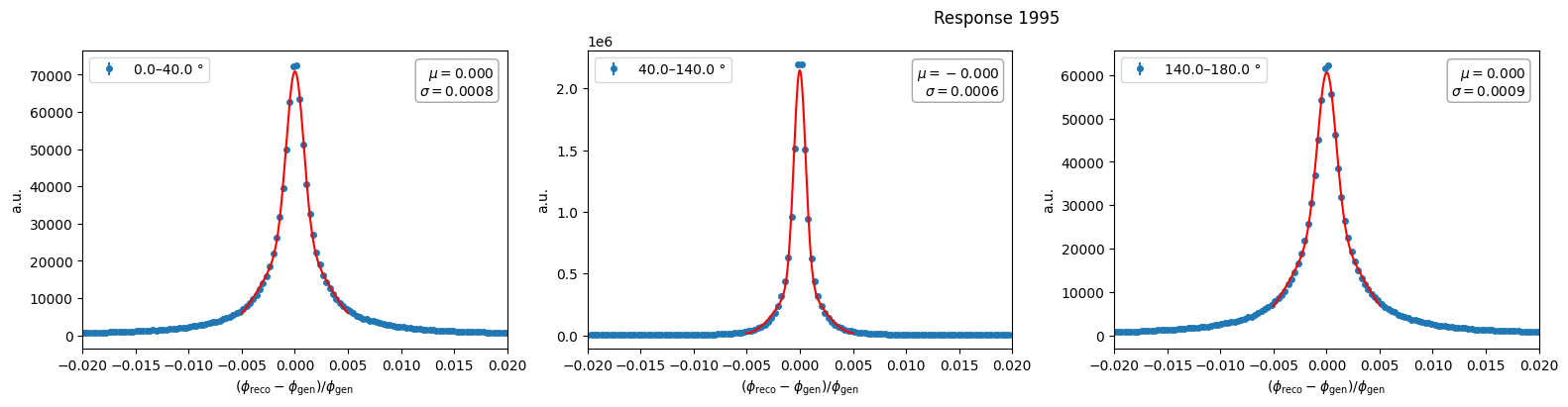}
    \caption{Track $\phi$ resolution distributions for 1995 in different detector regions: forward endcap (left), barrel (middle), and backward endcap (right). The $\phi$ response is calculated as $(\phi_{\rm reco} - \phi_{\rm gen}) / \phi_{\rm gen}$. Double Gaussian fits are overlaid to extract the resolution parameters.}
    \label{fig:phi_res_1995}
\end{figure}

\begin{figure}[ht!]
    \centering
    \includegraphics[width=\linewidth]{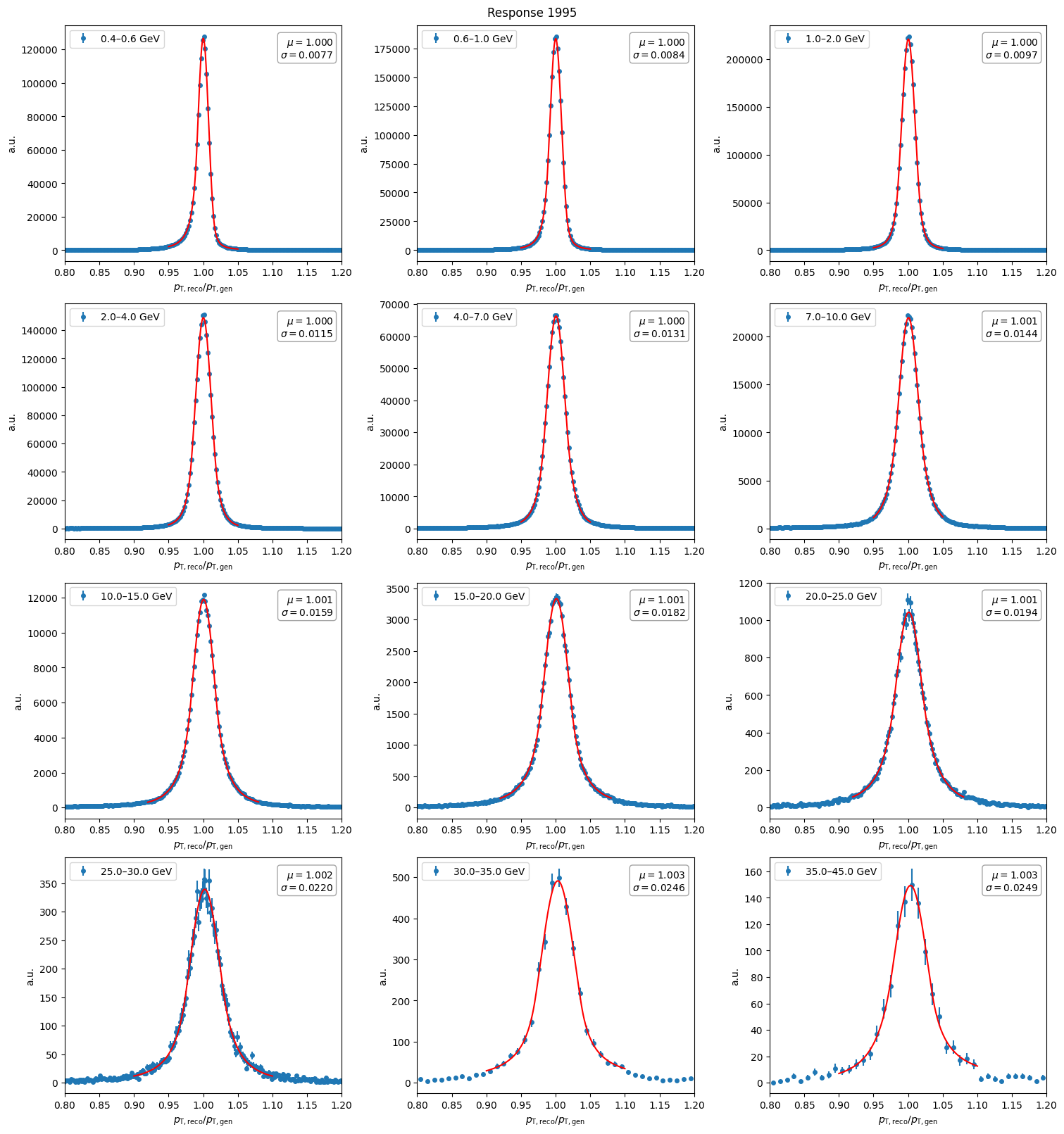}
    \caption{Track $p_{\rm T}$ resolution distributions for 1995 in different $p_{\rm T}$ ranges. The $p_{\rm T}$ response is calculated as $p_{\rm T, reco} / p_{\rm T, gen}$. Double-sided crystal ball fits are overlaid to extract the response and resolution parameters.}
    \label{fig:phi_pt_1995}
\end{figure}

\clearpage
\section{Additional figures for 1995 EEC analysis}\label{app:1995EEC}
This section shows some key plots for the 1995 analysis. 

\begin{figure}[ht!]
\centering
\includegraphics[width=0.6\linewidth]{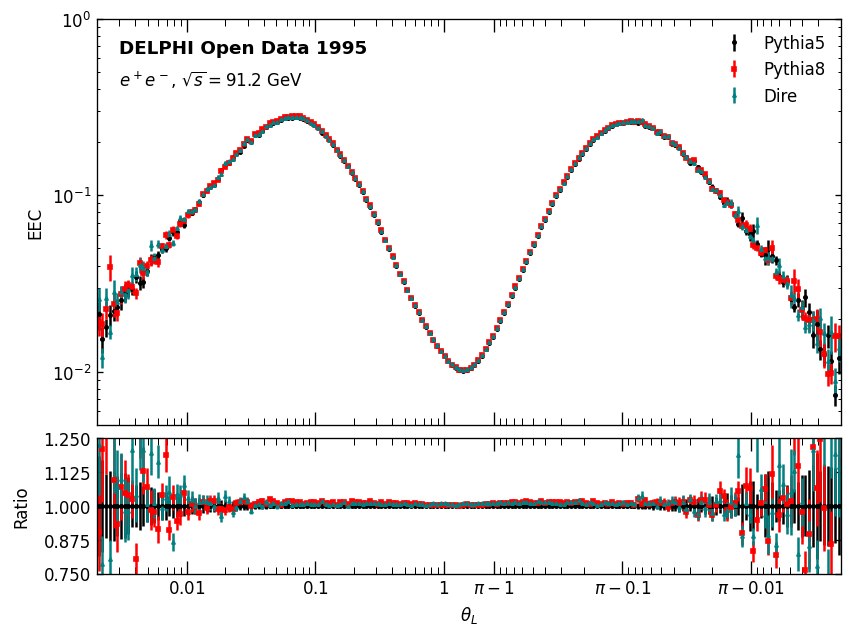}
\caption{Major systematic uncertainty: comparison of the unfolded EEC distributions obtained using response matrices derived from four different generators: PYTHIA 5 (black), PYTHIA 8 (red), and PYTHIA 8 Dire (teal).}
\label{fig:prior}
\end{figure}

\begin{figure}[ht!]
\centering
\includegraphics[width=0.6\linewidth]{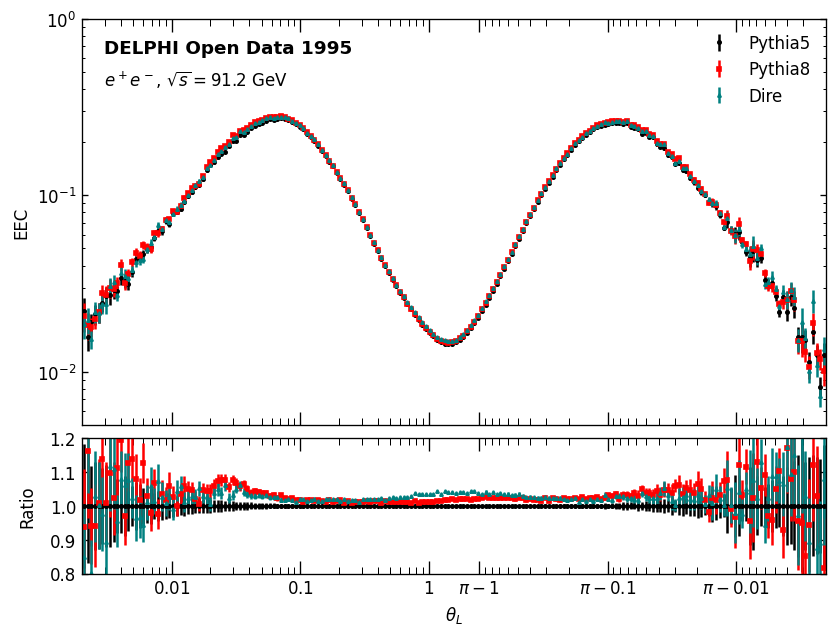}
\caption{Major systematic uncertainty: unfolded data distributions corrected by track and event selection correction factors derived from PYTHIA 5 (black), PYTHIA 8 (red), and PYTHIA 8 Dire (teal).}
\label{fig:acceptcorr}
\end{figure}

\begin{figure}[ht!]
    \centering
    \includegraphics[width=0.9\linewidth]{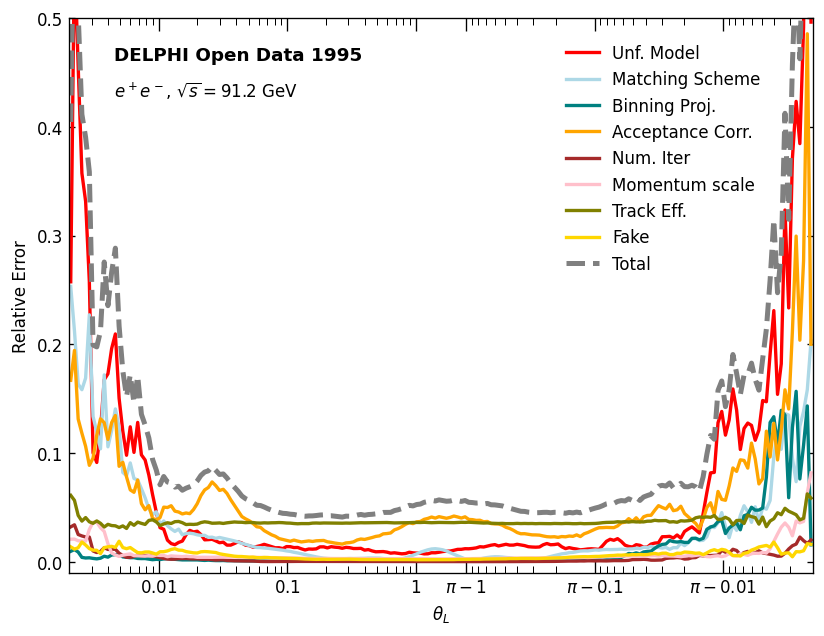}
    \caption{Summary of the systematic uncertainties for 1995. The total uncertainty (gray dashed) is calculated as the quadratic sum of the individual uncertainty sources, including event and track selection corrections~\ref{sec:selectionUncert} (orange), unfolding model dependence~\ref{sec:unfoldingUncert} (red), matching scheme~\ref{sec:matchingUncert} (blue), 2D binning projection~\ref{sec:projectionUncert} (teal), unfolding regularization strength~\ref{sec:unfoldingUncert} (brown), track efficiency (olive) and momentum scale (pink)~\ref{sec:trkUncert}, and fake~\ref{sec:trackTail} (gold). }
    \label{fig:systematicSummary}
\end{figure}

\clearpage
\begin{figure}[ht!]
    \centering
    \includegraphics[width=0.45\linewidth]{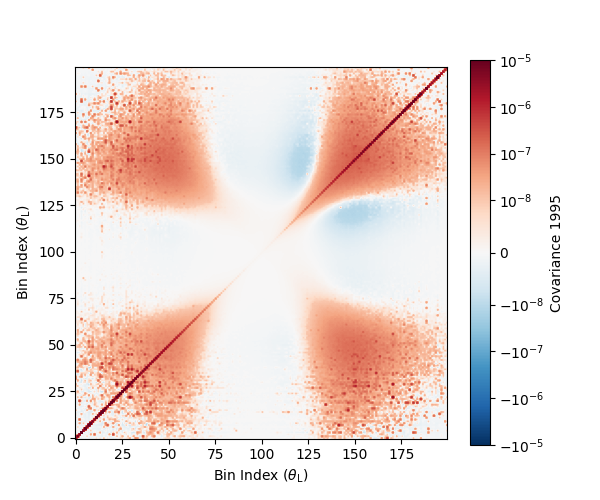}
    \includegraphics[width=0.45\linewidth]{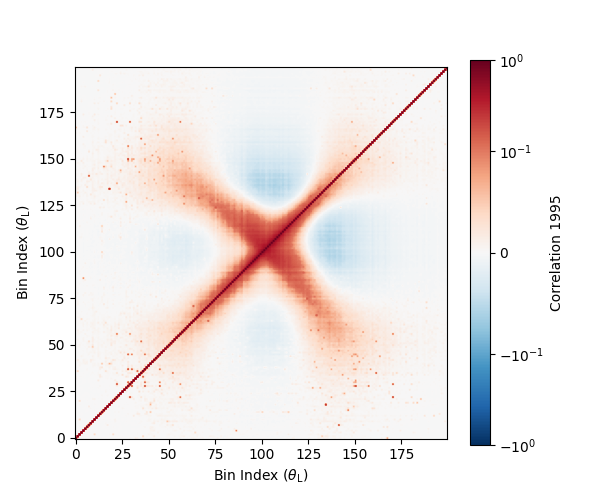}
    \caption{Covariance (left) and correlation (right) matrices of the one-dimensional EEC distribution as a function of $\theta_{\rm L}$ from 1995 data.}
    \label{fig:correlation_1995}
\end{figure}

\begin{figure}[ht!]
    \centering
    \includegraphics[width=0.45\linewidth]{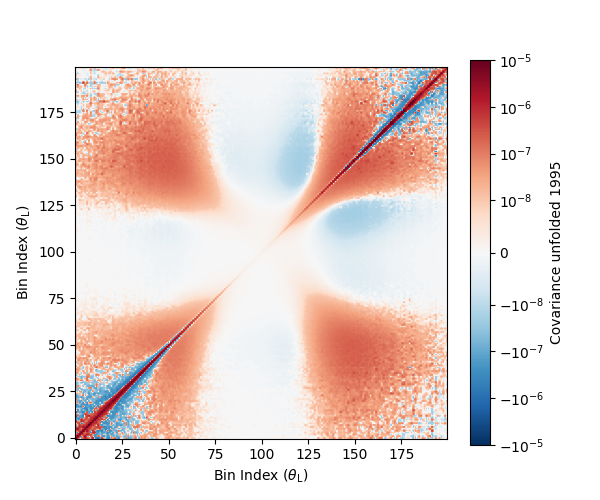}
    \includegraphics[width=0.45\linewidth]{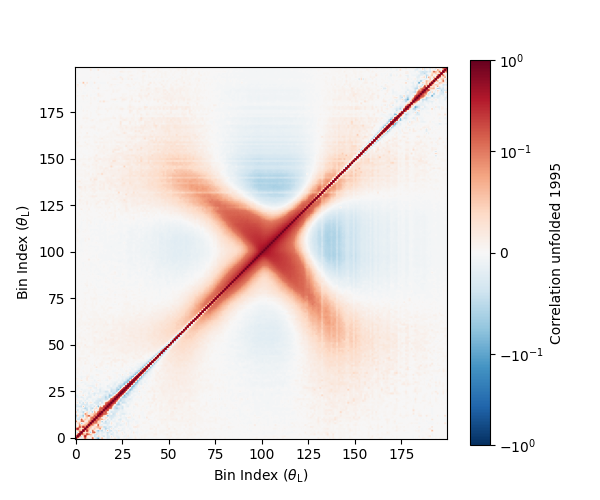}
    \caption{Covariance (left) and correlation (right) matrices of the one-dimensional EEC distribution as a function of $\theta_{\rm L}$ after unfolding for 1995.}
    \label{fig:covUnfold_1995}
\end{figure}

\clearpage
\begin{figure}[ht!]
    \centering
    \includegraphics[width=0.45\linewidth]{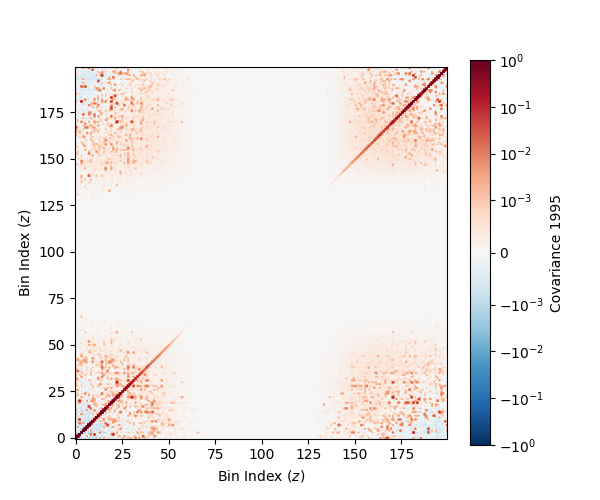}
    \includegraphics[width=0.45\linewidth]{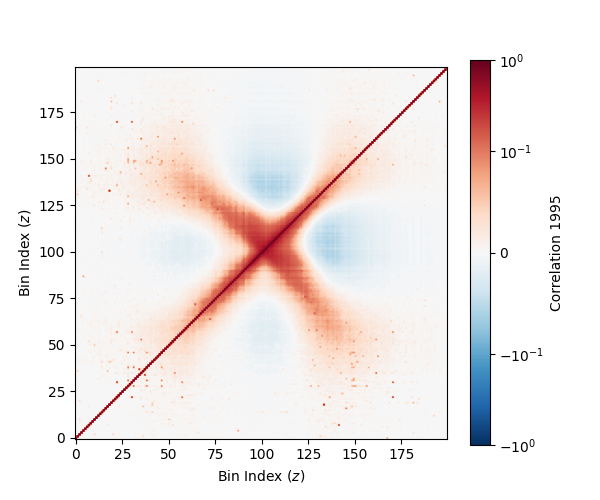}
    \caption{Covariance (left) and correlation (right) matrices of the one-dimensional EEC distribution as a function of $z$ from 1995 data.}
    \label{fig:correlation_z_1995}
\end{figure}

\begin{figure}[ht!]
    \centering
    \includegraphics[width=0.45\linewidth]{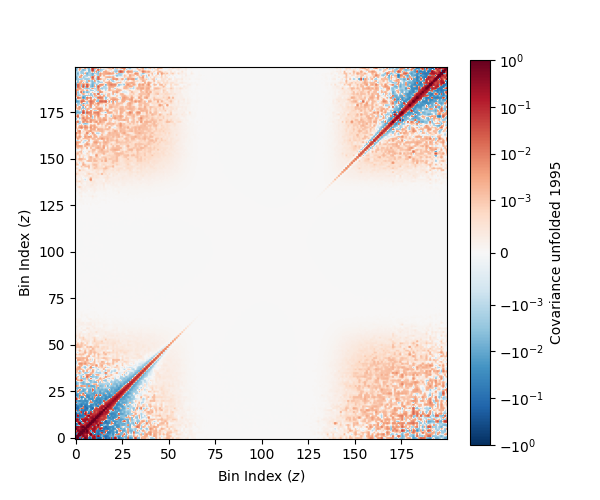}
    \includegraphics[width=0.45\linewidth]{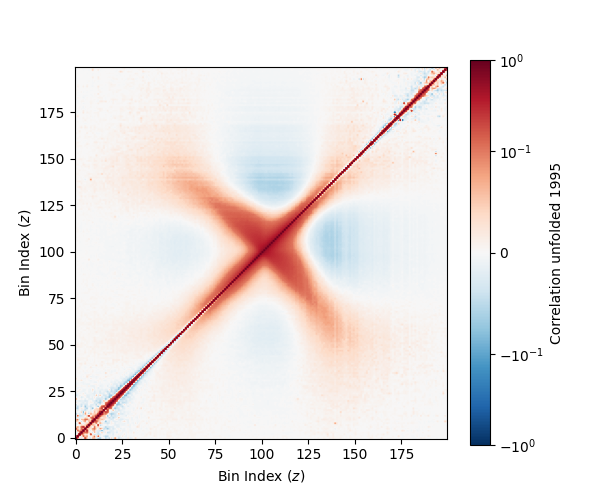}
    \caption{Covariance (left) and correlation (right) matrices of the one-dimensional EEC distribution as a function of $z$ after unfolding for 1995.}
    \label{fig:covUnfold_z_1995}
\end{figure}

\clearpage
\section{Additional figures for 1995 thrust analysis}\label{app:1995Thrust}
This section shows some key plots for the 1995 analysis. 

\begin{figure}[ht!]
    \centering
    \includegraphics[width = 0.45\textwidth]{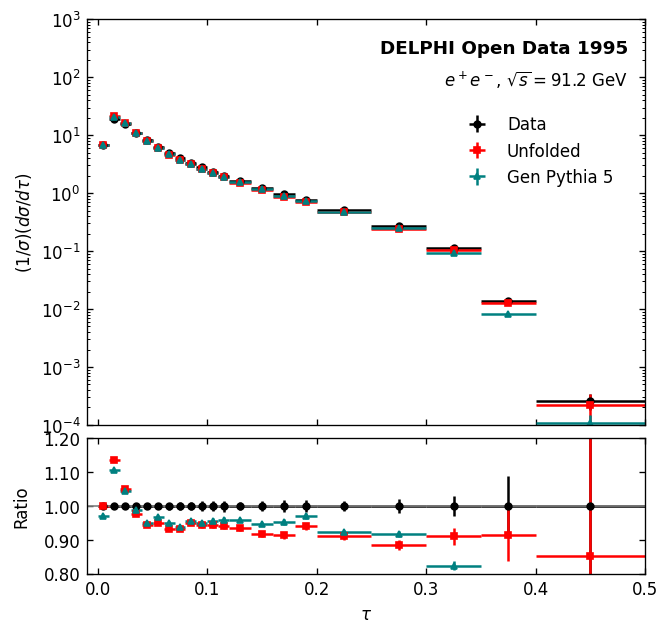}
    \includegraphics[width = 0.45\textwidth]{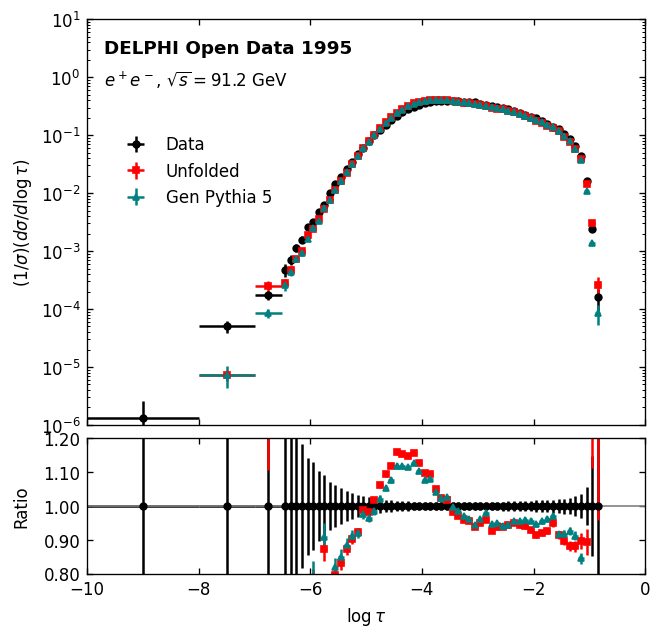}
    \caption{Size of the unfolding correction: raw (black) and unfolded (red) data distributions for $\tau$ (left) and $\log\tau$ (right). Also shown are the generator-level distributions from PYTHIA 5 (teal).}
    \label{fig:unfoldThrust1995}
\end{figure}

\begin{figure}[ht!]
    \centering
    \includegraphics[width = 0.45\textwidth]{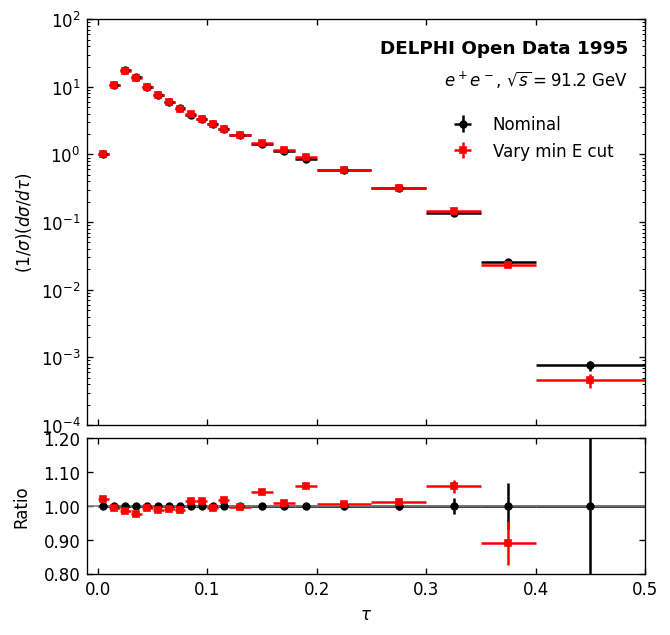}
    \includegraphics[width = 0.45\textwidth]{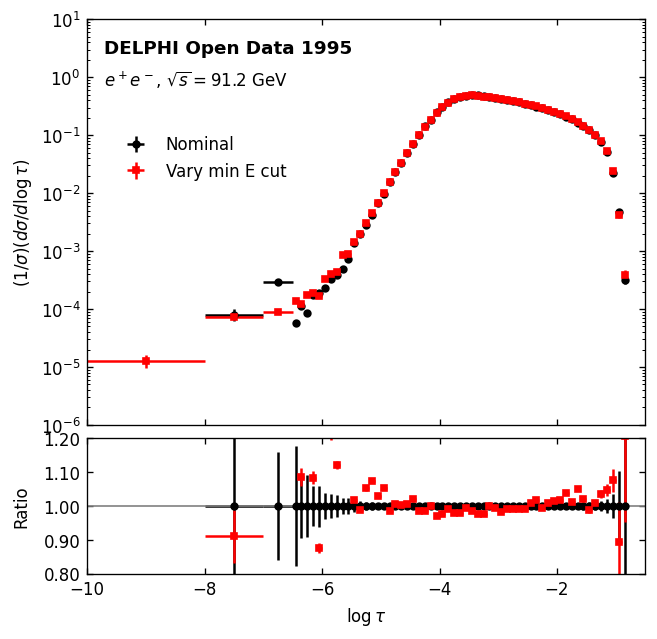}
    \caption{Major systematic uncertainty: the relative uncertainty on the unfolded thrust distributions from varying the minimum neutral particle energy cut. The effect is shown as the ratio of the varied result (\(E_{\text{neutral}} > 5.0\)~GeV) (red) to the nominal result (\(E_{\text{neutral}} > 0.5\)~GeV) (black).}
    \label{fig:cutNeutral1995}
\end{figure}

\begin{figure}[ht!]
    \centering
    \includegraphics[width = 0.45\textwidth]{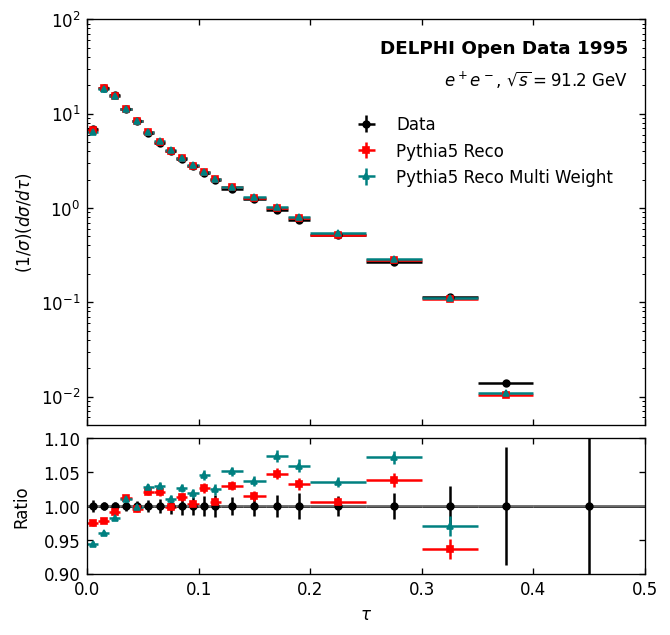}
    \includegraphics[width = 0.45\textwidth]{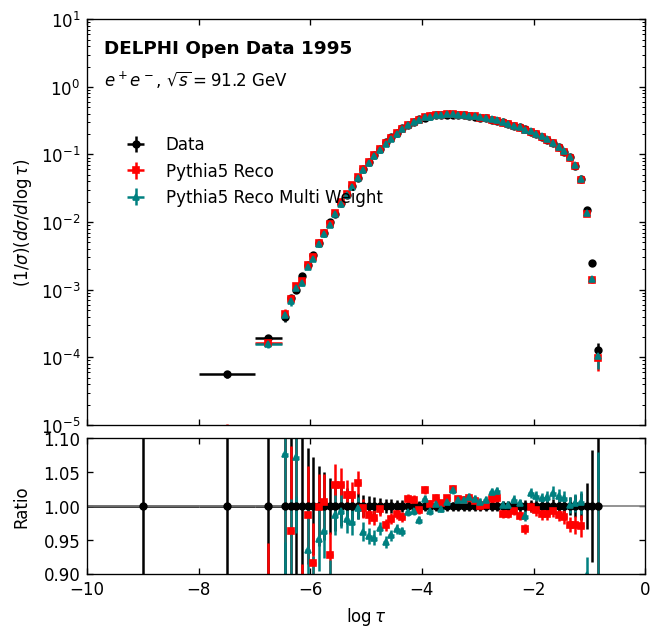}
    \caption{Major systematic uncertainty: effect of the event multiplicity reweighting on the reconstructed \(\tau\) (left) and \(\log\tau\) (right) distributions. The nominal 1995 MC (red) is compared to the re-weighted 1995 MC (teal). The 1995 data is shown in black for reference.}
    \label{fig:multiCorr1995}
\end{figure}

\begin{figure}[ht!]
    \centering
    \includegraphics[width = 0.45\textwidth]{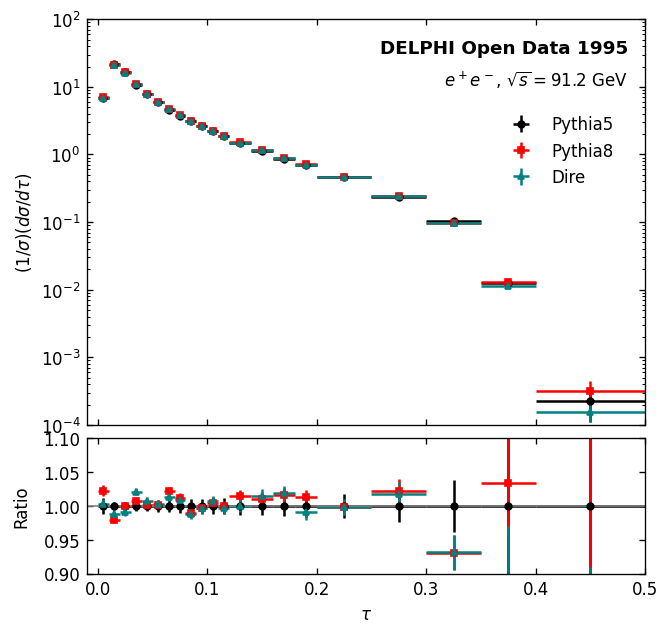}
    \includegraphics[width = 0.45\textwidth]{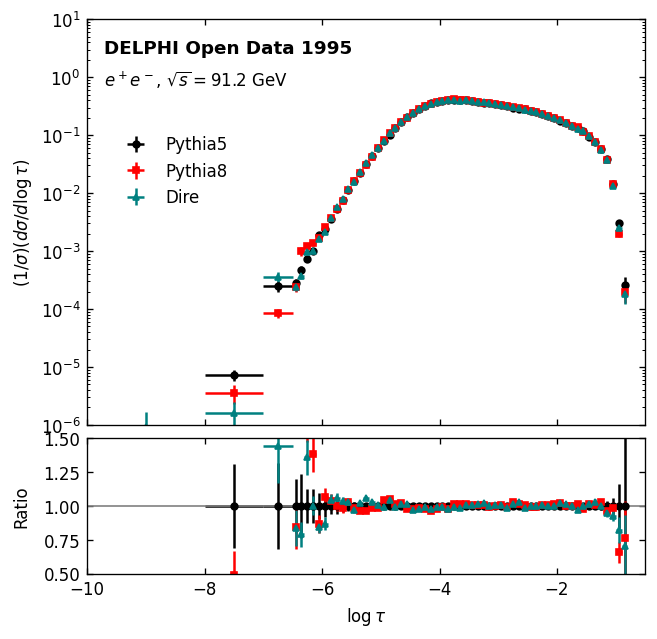}
    \caption{Major systematic uncertainty: unfolded thrust (left) and $\log\tau$ (right) distributions using response matrices derived from PYTHIA 5 (black), PYTHIA 8 (red), and PYTHIA 8 Dire (teal).}
    \label{fig:unfoldThrustModel1995}
\end{figure}

\begin{figure}[ht!]
    \centering
    \includegraphics[width = 0.45\textwidth]{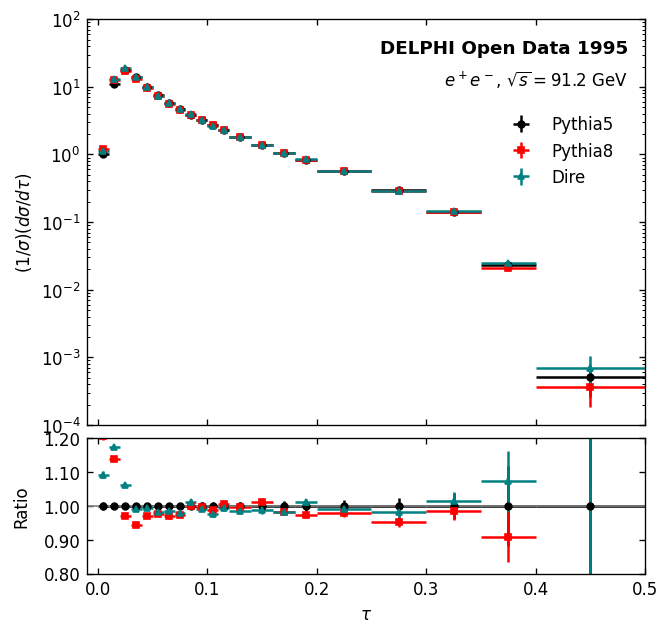}
    \includegraphics[width = 0.45\textwidth]{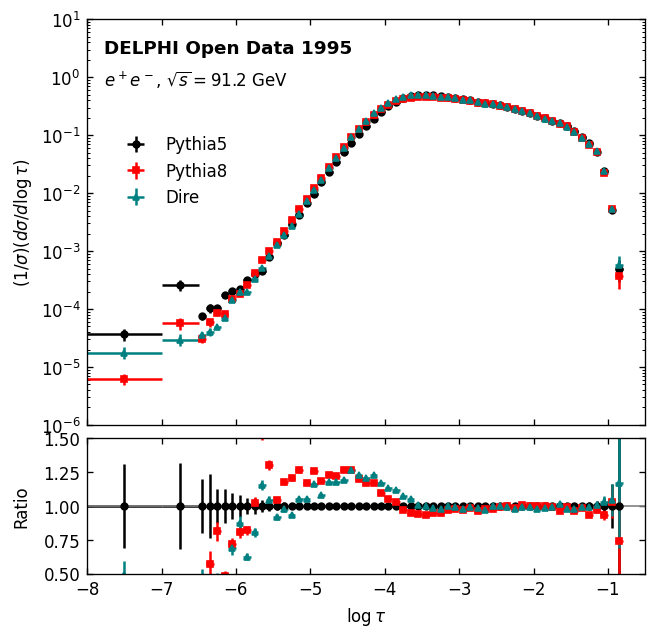}
    \caption{Major systematic uncertainty: unfolded $\tau$ (left) and $\log\tau$ (right) distributions corrected by acceptance correction factors derived from PYTHIA 5 (black), PYTHIA 8 (red), and PYTHIA 8 Dire (teal).}
    \label{fig:corrThrustModel1995}
\end{figure}

\begin{figure}[ht!]
    \centering
    \includegraphics[width = 0.45\textwidth]{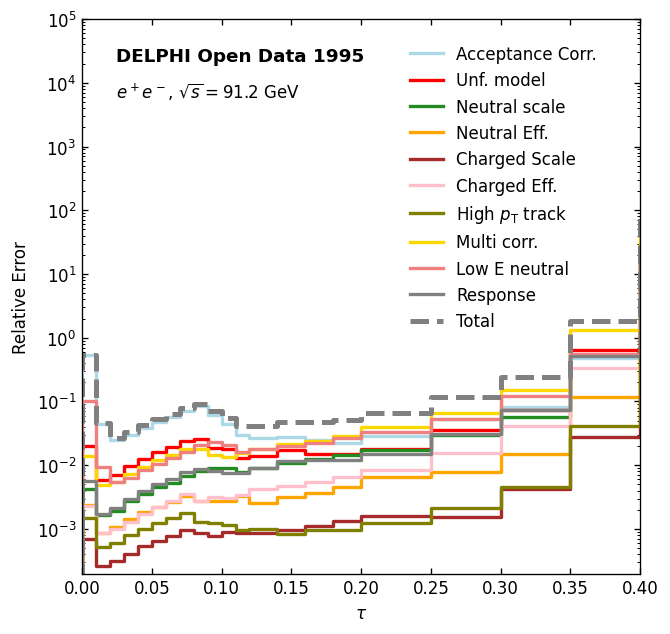}
    \includegraphics[width = 0.45\textwidth]{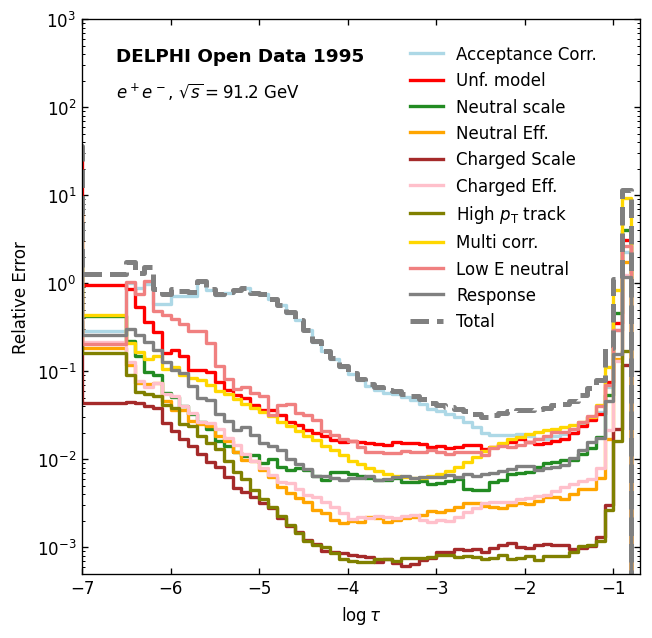}
    \caption{Summary of systematic uncertainties for $\tau$ (left) and $\log\tau$ (right) measurements for 1995.}
    \label{fig:sysSummaryThrust1995}
\end{figure}

\clearpage
\section{Track-only thrust cross-check}\label{app:trackThrust}
A cross-check is performed using the track-only thrust distribution. The result, unfolded and corrected for detector effects, is compared to a legacy measurement from the DELPHI collaboration~\cite{DELPHI:2000uri}. 

\begin{figure}[H]
    \centering
    \includegraphics[width=0.6\linewidth]{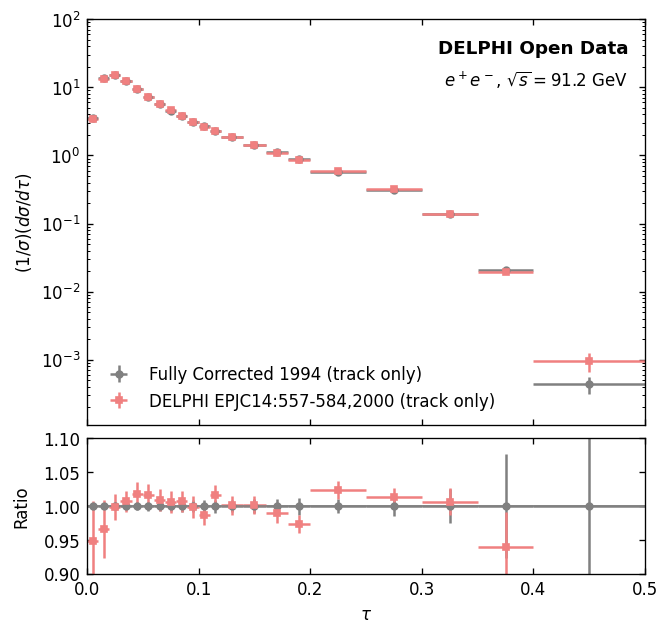}
    \caption{Comparison of the fully corrected track-only thrust from this analysis (using DELPHI open data) and the legacy DELPHI measurement~\cite{DELPHI:2000uri}. Both results are based on 1994 data.}
    \label{fig:trackThrust}
\end{figure}

The comparison is presented in Figure~\ref{fig:trackThrust}, which shows agreement to within 2\% in the primary region of interest for the extraction of $\alpha_{\rm s}$. The error bars shown for the cross-check represent statistical uncertainties only. Residual discrepancies between the two measurements can be attributed to differences in the underlying algorithms used for thrust calculation and the different methods used for detector corrections.

\end{document}